\documentclass[11pt]{article}

\usepackage{cite}
 \usepackage{subfigure}
\usepackage{multirow}
\usepackage{helvet}
\usepackage{amsmath}
\usepackage{amssymb}
\usepackage{setspace}
\usepackage{graphicx}
\usepackage{empheq}
\usepackage{soul}
\usepackage{tabularx}
\usepackage{array}
\usepackage{float}
\usepackage{braket}
\usepackage[utf8]{inputenc}
\usepackage{url}
\usepackage{hyperref}
\usepackage{slashed}
\usepackage[font=footnotesize,labelfont=bf]{caption}

\def\be{\begin{equation}}
\def\ee{\end{equation}}
\def\non{\nonumber}

\usepackage[a4paper,top=3cm,bottom=3cm,left=2.5cm,right=2.5cm,bindingoffset=0mm]{geometry}

\begin{document}

\begin{center}
{\Large \bf Modelling $W^+ W^-$ production with rapidity gaps at the LHC}

\vspace*{1cm}
                                                   
S. Bailey, L. A. Harland--Lang \\                                                 
                                                   
\vspace*{0.5cm}
Rudolf Peierls Centre, Beecroft Building, Parks Road, Oxford, OX1 3PU                                          
                                                    
\vspace*{1cm}

\begin{abstract}
\noindent
We present a new calculation of $W^+ W^-$ production in the semi--exclusive channel, that is either with intact outgoing protons or rapidity gaps present in the final state, and with no colour flow between the colliding protons. This study provides the first complete prediction of the $W^+ W^-$  semi--exclusive cross section, as well as the breakdown between elastic and proton dissociative channels. It combines the structure function calculation for a precise modelling of the region of low momentum transfers with a parton--level calculation in the region of high momentum transfers. The survival factor probability of no additional proton--proton interactions is fully accounted for, including its  kinematic and process dependence.  We analyse in detail the role that the pure photon--initiated ($\gamma\gamma \to W^+ W^-$) subprocess plays, a comparison that is only viable by working in the electroweak axial gauge. In this way, we find that the dominance of this is not complete in the proton dissociative cases, although once $Z$--initiated production is included a significantly better matching to the complete calculation is achieved. A direct consequence of this is that the relative elastic, single and double dissociative fractions are in general  different in comparison to the case of lepton pair production. We present a direct comparison to the recent ATLAS data on semi--exclusive $W^+ W^-$ production, finding excellent agreement within uncertainties. Our calculation is provided in the publicly available  \texttt{SuperChic 4.1} Monte Carlo (MC) generator, and can be passed to a general purpose MC for showering and hadronization of the final state. 
  
\end{abstract}

\end{center}

\section{Introduction}

The production of vector boson pairs via vector--boson scattering (VBS) is a broad class of process that provides unique sensitivity to the gauge structure of Standard Model (SM) and of BSM effects that may modify it, see~\cite{Ballestrero:2018anz,Buarque:2021dji,Covarelli:2021gyz} for recent reviews. The effective isolation of this process generally requires that VBS cuts are imposed, such that two jets with a sufficiently large rapidity separation and invariant mass be present in the detector. However, an alternative way in which to select such VBS events is to consider the exclusive channel, where both colliding protons remain intact, and/or the semi--exclusive channel where one or both protons dissociate but there is nonetheless no colour flow between the protons, and hence rapidity gaps are present. In particular, for the case of $ZZ$ and $W^+ W^-$ production, one can expect the photon--initiated (PI) production channel, that is due to $\gamma \gamma \to W^+ W^-/ZZ$ scattering, to play a significant role; indeed in the purely exclusive case it is to very good approximation the only channel. As discussed in~\cite{Chapon:2009hh,LHCForwardPhysicsWorkingGroup:2016ote} this can provide unique sensitivity to this sector of the SM, and of anomalous gauge couplings in particular. The $W^+ W^-$ channel is highly topical in light of the recent  first observation by ATLAS  of semi--exclusive $W^+ W^-$ production~\cite{ATLAS:2020iwi}, at 13 TeV.

More generally, the production of electroweak (EW) particles with intact protons and/or rapidity gaps in the final state is a key ingredient in the LHC precision physics programme, with unique sensitivity to physics within and beyond the SM, see e.g.~\cite{Harland-Lang:2020veo} for further discussion and references, and~\cite{LHCForwardPhysicsWorkingGroup:2016ote,Hagiwara:2017fye,Bruce:2018yzs,Harland-Lang:2018hmi,Vysotsky:2018slo,Schoeffel:2020svx,Goncalves:2020saa} for reviews and studies. A particularly promising avenue is to measure the outgoing intact protons using dedicated forward proton detectors, namely the AFP~\cite{AFP,Tasevsky:2015xya} and CT--PPS~\cite{CT-PPS}  detectors, which have been installed in association with both ATLAS and CMS, respectively. These  have most recently been used in the first measurement of lepton pair production with a single proton tag by ATLAS~\cite{Aad:2020glb} (evidence for which, but not a cross section measurement, was presented by CMS--TOTEM in~\cite{Cms:2018het}) and to place limits on anomalous gauge couplings in the diphoton final state with both protons tagged by CMS--TOTEM~\cite{CMS:2020rzi}. Both experiments are equipped with time--of--flight detectors, which serve to suppress pile--up backgrounds, see~\cite{Cerny:2020rvp} for a recent study. Moreover, as described in detail in~\cite{CMS:2021ncv}, an exciting and broad range of measurements is also possible during HL--LHC running. 

The key element in the above measurements is that by selecting events with intact protons, we can effectively isolate the photon--initiated (PI) production mechanism for EW particles. This  is rather well understood and hence can provide a very clean probe of BSM effects in this sector. In particular, the colour singlet nature of the initial--state photons naturally leads to exclusive events with intact protons\footnote{There are many interesting possibilities in the context of ultraperipheral production in heavy ions collisions~\cite{Bruce:2018yzs}, although this is not the focus of the current paper.} in the final state. However, even without tagged protons, one can still select events due to PI production by requiring that rapidity gaps are  present in the final state. More commonly, in the high pile--up environment of the LHC, one requires that no additional   tracks associated with the primary vertex be present. Indeed, a range of data on PI lepton pair production has been taken at the LHC using this method, by ATLAS~\cite{Aad:2015bwa,Aaboud:2017oiq} and CMS~\cite{Chatrchyan:2011ci}. Most recently, the first observation of semi--exclusive $W^+ W^-$ production was reported by ATLAS~\cite{ATLAS:2020iwi} at 13 TeV, following previous evidence from both ATLAS~\cite{Aaboud:2016dkv} and CMS~\cite{Chatrchyan:2013foa,Khachatryan:2016mud}, in the two lepton decay channel. 

In general, for events selected via a rapidity veto alone both exclusive and semi--exclusive channels will contribute, and  therefore  a unified theoretical treatment of these is required. The first complete treatment of this was presented in~\cite{Harland-Lang:2020veo}, for the case of lepton pair production. This combined the precise treatment of the underlying PI production process provided by the structure function (SF) approach presented in~\cite{Harland-Lang:2019eai,Harland-Lang:2021zvr} with a fully differential modelling of the survival factor probability of no additional particle production due to 
multi--parton interactions (MPI). This was implemented in the  \texttt{SuperChic 4} Monte Carlo (MC) generator~\cite{SuperCHIC} in a form that could then subsequently be passed to a general purpose MC such as \texttt{Pythia}~\cite{Sjostrand:2014zea} for showering and hadronization of the proton dissociation products.

Purely exclusive $W^+ W^-$ production has been previously studied extensively, see e.g.~\cite{Chapon:2009hh,Harland-Lang:2011mlc}, and indeed has been implemented for some time in \texttt{SuperChic}~\cite{Harland-Lang:2015cta}, while the semi--exclusive case has been considered in~\cite{Luszczak:2014mta,Luszczak:2018ntp,Forthomme:2018sxa} in the on--shell approximation, and without the survival factor accounted for. However a unified treatment
has so far been lacking, and hence in this paper we present this for the first time. The basic framework follows directly from that applied in the case of lepton pair production, with however some key differences. As we will discuss, the particular sensitivity of $W^+ W^-$ production to the EW symmetry breaking sector of the SM requires a more careful treatment of the semi--exclusive channel once one goes beyond the on--shell approximation for the initial--state photons. In particular, as we go away from this limit gauge invariance dictates that we include diagrams where the $W$ bosons are emitted from the quark legs that generate the initial--state photons in the pure PI diagrams. In principle, one might expect these diagrams to be kinematically suppressed in comparison to pure PI diagrams, due to the $t$--channel enhancement, $\sim 1/Q_i^2$, of the photon propagators in the latter case; indeed, precisely this effect leads to the equivalent diagrams in the lepton pair production case being suppressed, as we will discuss. However, this argument dramatically fails in the EW unitary gauge, due to the well known unitary violating effects that are present here, leading to amplitudes that  grow indefinitely with energy when  gauge dependent subsets of diagrams are considered in isolation, and hence a breakdown in the expectations that come from naive counting in powers of $1/Q_i^2$. This issue is well known in the context of $WW$ scattering, see e.g.~\cite{Kleiss:1986xp,Kunszt:1987tk,Accomando:2006mc,Borel:2012by}. 

A  route out of the above issue is, as discussed in~\cite{Kunszt:1987tk,Accomando:2006iz,Accomando:2006mc,Borel:2012by} to work instead in the EW axial gauge (see e.g.~\cite{Dams:2004vi}), where such unitarity violating effects are explicitly absent. In such a gauge, the dominance (or not) of the PI process may be more appropriately analysed.
We therefore examine the impact of working in the axial gauge on the current case, which is to our knowledge the first time this has been applied in context of $\gamma \gamma \to W^+ W^-$ (or more properly, $\gamma/Z \gamma/Z \to W^+ W^-$) scattering. Once a rapidity veto is imposed, we find that the contribution from pure PI diagrams is $\sim 50\, (75)\%$ of the overall cross section for the DD (SD) cases, within the ATLAS fiducial region~\cite{ATLAS:2020iwi}. This is therefore significant, but not overwhelmingly so. On the other hand, once we include $Z$--initiated production the matching is significantly improved, with agreement at the $\sim 10\%$ level or less achieved. This demonstrates that, once an appropriate gauge is chosen, the semi--exclusive $W^+ W^-$ signal can to this level of precision be viewed as proceeding  via the $\gamma/Z \gamma/Z \to W^+ W^-$ channel, but not the purely PI one.

Nonetheless, a precise and gauge invariant treatment of course requires that all relevant diagrams are included. Hence in this paper we take a hybrid approach. In the region of low photon $Q^2$ and or proton dissociation system $W^2$, where one cannot reliably apply the parton model to calculate the underlying $p \to \gamma^* X$ vertex, but where as discussed in~\cite{Manohar:2016nzj,Harland-Lang:2019eai} precise experimental determination of the corresponding proton structure functions are available, we apply the SF approach of~\cite{Harland-Lang:2019eai,Harland-Lang:2021zvr}. The key observation here is that in this region the pure PI diagrams are indeed completely dominant, as we will show. Away from this region, i.e. at higher photon $Q^2$, we instead apply LO perturbation theory to calculate the full set of $qq' \to W^+ W^- qq'$ and $\gamma^* q  \to  W^+ W^- q$ diagrams (where $q, q'$ denote arbitrary quark/antiquarks). We have implemented this in the \texttt{SuperChic 4.1} MC generator, and in this paper present a detailed study of the results of this approach, the uncertainties in it, and their implications for the LHC. 

We will in particular use these results to compare directly with the recent ATLAS measurement~\cite{ATLAS:2020iwi}, finding excellent agreement. For such data, there are three channels that contribute, namely the purely elastic (EL) case, where both protons remain intact, the single dissociative (SD) case, where one protons breaks up, and the double dissociative (DD) case, where both break up. The relative contributions from these are in general sensitive to the particular process under consideration, the final--state kinematics, and the appropriate modelling of the soft survival factor. The predicted contributions from EL, SD and DD production are in particular found to be rather different from the case of lepton pair production, which therefore provides an obstacle to using the measured relative components in this case to derive an effective exclusive signal in the $W^+ W^-$ case, as is done in~\cite{ATLAS:2020iwi} and in earlier analyses~\cite{Aaboud:2016dkv,Khachatryan:2016mud}. While the difference is relatively mild, it is non negligible, and this issue may in particular be crucial if the aim is to use such data to look for small deviations from the SM, for example in the context of an EFT analysis. Of course, if such data are taken with single or double proton tags this would enable the relative components to be directly measured, and bypass this issue as well as  providing a more fine grained analysis of the overall signal.

The outline of this paper is as follows. In Section~\ref{sec:SF} we outline the basic formalism behind the SF approach. In Section~\ref{sec:firstlook} we examine the issues inherent in a naive application of this approach, within the unitary gauge. In Section~\ref{sec:ax} we discuss the EW axial gauge, and demonstrate how this allows the appropriate power counting in $Q_i^2/M^2$ to be uncovered, for the pure PI diagrams with respect to the fully gauge invariant set of contributing diagrams. In Section~\ref{sec:hybrid} we present the new hybrid approach. In Section~\ref{sec:hybridres} we present results for this in the context of semi--exclusive production, and compare with more approximate approaches. In Section~\ref{sec:vbs} we briefly discuss the implications of our study for the cases where VBS cuts are instead imposed. In Section~\ref{sec:lep} we revisit the case of lepton pair production. In Section~\ref{sec:S2} we describe how the soft survival factor can be evaluated. In Section~\ref{sec:mc} we describe how the calculation is implemented in the \texttt{SuperChic 4.1} MC. In Section~\ref{sec:theorunc} we discuss the theoretical uncertainties in our results. In Section~\ref{sec:ATLAS} we compare to the ATLAS 13 TeV analysis. Finally, in Section~\ref{sec:conc} we conclude.

\section{Modelling $W^+ W^-$ pair production}

\subsection{The Structure Function Approach}\label{sec:SF}

A key ingredient in our calculation of $W^- W^-$ pair production is the structure function (SF) approach for calculating PI production, as discussed in~\cite{Harland-Lang:2019eai,Harland-Lang:2021zvr}, and which we summarise here. The basic idea comes from the the analysis of~\cite{Budnev:1974de} (see also~\cite{Chen:1973mv}), namely that in the high--energy limit ($\sqrt{s} \gg m_p$) the PI cross section in proton--proton collisions can be written in the general form
  \be\label{eq:sighh}
  \sigma_{pp} = \frac{1}{2s} \int \frac{{\rm d}^3 p_1' {\rm d}^3 p_2' {\rm d}\Gamma}{E_1' E_2'}   \alpha(Q_1^2)\alpha(Q_2^2)
  \frac{\rho_1^{\mu\mu'}\rho_2^{\nu\nu'} M^*_{\mu'\nu'}M_{\mu\nu}}{Q_1^2 Q_2^2}\delta^{(4)}(q_1+q_2 - k)\;.
 \ee
 Here the outgoing hadronic systems have momenta $p_{1,2}'$ and the photons have momenta $q_{1,2}$, with $q_{1,2}^2 = -Q_{1,2}^2$. We consider the production of a system of 4--momentum $k = q_1 + q_2 = \sum_{j=1}^N k_j$ of $N$ particles, where ${\rm d}\Gamma = \prod_{j=1}^N {\rm d}^3 k_j / 2 E_j (2\pi)^3$ is the standard phase space volume. $M^{\mu\nu}$ corresponds to the $\gamma\gamma \to X(k)$ production amplitude, with arbitrary photon virtualities. The generalisation to include $Z$--initiated production is straightforward~\cite{Harland-Lang:2021zvr}, and will be considered at the end of this section.
 
This result is the basis of the equivalent photon approximation~\cite{Budnev:1974de}, as well as being precisely the formulation used in the structure function approach~\cite{Han:1992hr} applied to the calculation of Higgs Boson production via VBF. It was applied in~\cite{Harland-Lang:2019eai} to the case of lepton pair production at the LHC, while in~\cite{Harland-Lang:2021zvr} this was extended to include initial--state $Z$ and mixed $\gamma / Z+ q $ contributions. In~\cite{Harland-Lang:2021zvr} this approach was also applied for the first time to the production of a back--to--back same--sign lepton pair of the same flavour, or a lepton pair of differing flavours and arbitrary signs, that is via lepton--lepton scattering. This has subsequently been extended in~\cite{Buonocore:2021bsf} to include further kinematically subleading contributions\footnote{The additional diagram included in~\cite{Buonocore:2021bsf} is suppressed by the back--to--back requirement, as described in~\cite{Harland-Lang:2021zvr}. We in addition note that our formulation of the SF approach, as given  in terms of the photon density matrix $\rho$, is taken for consistency with the original work of~\cite{Budnev:1974de} (see in particular (5.1)  and Appendix D); however, as discussed below \eqref{eq:rho} the integral over $M_i$ is understood as being performed simultaneously with the phase space integral in \eqref{eq:sighh}, i.e. is not factorized from it. This stipulation has not  been accounted for in~\cite{Buonocore:2021bsf}, where it is incorrectly stated that the formulation of the SF approach as presented here and in previous papers (as well as~\cite{Budnev:1974de}) is incorrect. Finally, for the avoidance of confusion, we note that in~\cite{Buonocore:2021bsf} the SF approach is instead labelled the `hadronic tensor' (`HT') approach; however these are the same.}.

 In the above expression, $\rho$ is the density matrix of the virtual photon, which is given in terms of the well known proton structure functions:
 \be\label{eq:rho}
 \rho_i^{\alpha\beta}=2\int \frac{{\rm d}M_i^2}{Q_i^2}  \bigg[-\left(g^{\alpha\beta}+\frac{q_i^\alpha q_i^\beta}{Q_i^2}\right) F_1(x_{B,i},Q_i^2)+ \frac{(2p_i^\alpha-\frac{q_i^\alpha}{x_{B,i}})(2p_i^\beta-\frac{q_i^\beta}{x_{B,i}})}{Q_i^2}\frac{ x_{B,i} }{2}F_2(x_{B,i},Q_i^2)\bigg]\;,
 \ee
where $x_{B,i} = Q^2_i/(Q_i^2 + M_{i}^2 - m_p^2)$ for a hadronic system of mass $M_i$ and we note that the definition of the photon momentum $q_i$ as outgoing from the hadronic vertex is opposite to the usual DIS convention. Here, the integral over $M_i^2$ is understood as being performed simultaneously with the phase space integral over $p_{i}'$, i.e. is not fully factorized from it (the energy $E_i'$ in particular depends on $M_i$).

The input for the proton structure functions comes from noting that the same density matrix $\rho$ appears in the cross section for lepton--proton scattering. One can therefore make use of the wealth of data for this process to constrain the structure functions, and hence the photon--initiated cross section, to high precision. In more detail, the structure function receives contributions from: elastic photon emission, for which we use the A1 collaboration~\cite{Bernauer:2013tpr}  fit to the elastic proton form factors;  CLAS data on inelastic  structure functions in the resonance $W^2 < 3.5$ ${\rm GeV}^2$ region, primarily concentrated at lower $Q^2$ due to the $W^2$ kinematic requirement; the HERMES fit~\cite{Airapetian:2011nu} to the inelastic low $Q^2 < 1$ ${\rm GeV}^2$ structure functions in the continuum $W^2 > 3.5$ ${\rm GeV}^2$ region; inelastic high $Q^2 > 1$ ${\rm GeV}^2$ structure functions for which the pQCD prediction in combination with PDFs determined from a global fit  provide the strongest constraint (we take the ZM--VFNS at NNLO in QCD  predictions for the structure functions as implemented in~\texttt{APFEL}~\cite{Bertone:2013vaa}, with the \texttt{MSHT20qed\_nnlo} PDFs~\cite{Cridge:2021pxm} throughout). The inputs we take are as discussed in the MMHT15 and MSHT20 photon PDF determinations~\cite{Harland-Lang:2019pla,Cridge:2021pxm}, which are closely based on that described in~\cite{Manohar:2016nzj,Manohar:2017eqh} for the \texttt{LUXqed} set. 

As will compare with this later, we recall that \eqref{eq:sighh} can straightforwardly be connected to the result of the equivalent photon approximation (EPA)~\cite{Budnev:1974de}. As in~\cite{Harland-Lang:2019zur} we can write
 \be\label{eq:sighhf}
 \sigma_{pp} = \frac{1}{2s}  \int  {\rm d}x_1 {\rm d}x_2\,{\rm d}^2 q_{1_\perp}{\rm d}^2 q_{2_\perp
} {\rm d \Gamma} \,\alpha(Q_1^2)\alpha(Q_2^2)\frac{1}{\tilde{\beta}} \frac{\rho_1^{\mu\mu'}\rho_2^{\nu\nu'} M^*_{\mu'\nu'}M_{\mu\nu}}{Q_1^2Q_2^2}\delta^{(4)}(q_1+q_2 - p_X)\;,
 \ee
 where
 \be
 x_{1,2} = \frac{1}{\sqrt{s}}\left(E_{X} \pm p_{X,z}\right) = \frac{m_{X_\perp}}{\sqrt{s}} e^{\pm y_{X}}\;,
 \ee
 with $X=W^+ W^-$ indicating the kinematics of the centrally produced system (we will keep the results in terms of $X$ for generality), and $q_{i\perp}$ are the photon transverse momenta, while $\tilde{\beta}$ is defined in~\cite{Harland-Lang:2019zur}. The amplitude squared $M^*_{\mu'\nu'}M_{\mu\nu}$ permits a general expansion~\cite{Budnev:1974de}
\be\label{eq:mtran}
M^*_{\mu'\nu'}M_{\mu\nu} = R_{\mu\mu'} R_{\nu\nu'}  \,\frac{1}{4}\sum_{\lambda_1 \lambda_2} |M_{\lambda_1 \lambda_2}|^2 + \cdots\;,
\ee
where we omit various terms that vanish when taking the $Q_{1,2}^2 \ll M_X^2$ limit, or after integration over the photon azimuthal angle. Here $\lambda_i=\pm $ are the transverse photon helicities, and $R$ is the metric tensor that is transverse to the photon momenta $q_{1,2}$:
\be
R^{\mu\nu} = -g^{\mu\nu} + \frac{(q_1 q_2)(q_1^\mu q_2^\nu+q_1^\nu q_2^\mu)+Q_1^2 q_2^\mu q_2^\nu+Q_2^2 q_1^\mu q_1^\nu}{(q_1 q_2)^2-Q_1^2 Q_2^2}\;.
\ee
In the $Q_{1,2}^2 \ll M_X^2$ limit we have
\be\label{eq:rhodr}
\rho_{\mu\nu}^i R^{\mu\nu} \approx 2\int \frac{{\rm d}M_i^2}{Q_i^2} \frac{x_{B,i}}{x^2_{i}} \left[\left(z_i p_{\gamma q}(z_i)+\frac{2x_i^2 m_p^2}{Q_i^2}\right)F_2(x_i/z_i,Q^2_i)-z_i^2 F_L(x_i/z_i,Q^2_i)\right]\;,
\ee
where $z_i=x_i/x_{B_i}$ with $i={1,2}$, and as before the integral over $M_i$  is understood as being performed simultaneously with the phase space integral in \eqref{eq:sighhf}, i.e. is not factorized from it (as e.g. the photon $Q_i^2$ depend on $M_i$ at fixed $q_{i_\perp}$). At this point, we can evaluate the helicity amplitudes  $M_{\lambda_1 \lambda_2}$ in the on--shell limit to give the `on-shell' approximation to the full result, i.e. giving the leading contribution in $Q_i^2/M_X^2$. There is no unique way to make the on--shell projection for the initial--state photons, but a straightforward way is to simply evaluate the helicity amplitudes in the $W^+ W^-$ rest frame, as generated according to \eqref{eq:sighhf}, and assuming the initial--state photons are on--shell and collinear. 

Alternatively, noting that  applying \eqref{eq:mtran} and \eqref{eq:rhodr} leaves \eqref{eq:sighhf} independent of the azimuthal angle of the photons, we can integrate over this and change variables to give
 \be\label{eq:sighhfq}
 \sigma_{pp} = \frac{\pi^2}{2s}  \int  {\rm d}\xi_1 {\rm d}\xi_2\,{\rm d}Q_1^2{\rm d}Q_2^2 {\rm d \Gamma} \,\alpha(Q_1^2)\alpha(Q_2^2) \frac{\rho_1^{\mu\mu'}\rho_2^{\nu\nu'} M^*_{\mu'\nu'}M_{\mu\nu}}{Q_1^2Q_2^2}\delta^{(4)}(q_1+q_2 - p_X)\;,
 \ee
where $\xi_i$ are the photon momentum fractions with respect to the parent proton momenta. We can then rewrite \eqref{eq:rhodr} by changing variables from $M_i$ to $x_{B,i}$ (at fixed $Q_i^2$) to give
 \be\label{eq:rhophot}
 \frac{1}{\alpha(\mu^2)}\int  \frac{{\rm d} Q^2}{Q^2}\alpha(Q^2)^2\rho_{\mu\nu}^i R^{\mu\nu} \approx  \frac{4\pi}{\xi_i}   f_{\gamma/p}^{\rm PF}(\xi_i,\mu^2)\;,
\ee
where `PF' indicates this is the physical photon PDF, following the notation of~\cite{Manohar:2017eqh}, i.e.
\begin{align}\nonumber
  x f_{\gamma/p}^{\rm PF}(x,\mu^2) &= 
  \frac{1}{2\pi \alpha(\mu^2)} \!
  \int_x^1
  \frac{dz}{z}
  \int^{\frac{\mu^2}{1-z}}_{\frac{x^2 m_p^2}{1-z}} 
  \frac{dQ^2}{Q^2} \alpha^2(Q^2)
  \\  \label{eq:xfgamma-phys}
  &\cdot\Bigg[\!
  \left(
    zp_{\gamma q}(z)
    + \frac{2 x ^2 m_p^2}{Q^2}
  \right)\! F_2(x/z,Q^2)
    -z ^2
  F_L\!\left(\frac{x}{z},Q^2\right)
  \Bigg]\,.
\end{align}
We note that \eqref{eq:xfgamma-phys} only differs from the result one gets from evaluating the LHS of \eqref{eq:rhophot} by the upper limit on the $Q^2$ integral, which in \eqref{eq:sighhfq} is set by the kinematic endpoint, whereas in \eqref{eq:xfgamma-phys} an artificial `factorization' scale has been introduced. As we have made the $Q_{1,2}^2 \ll M_X^2$ approximation, the sensitivity to this choice is beyond the accuracy at which we calculate, and hence can be viewed as a natural parameterisation of our sensitivity to this, see~\cite{Harland-Lang:2019eai} for further discussion. Beyond LO this interpretation can be made more precise, and a proper $\overline{\rm MS}$ matching can be achieved, as shown in~\cite{Manohar:2017eqh}. This requires modification of \eqref{eq:xfgamma-phys} to include a  $\overline{\rm MS}$ matching term, but we note that this is only relevant beyond LO in the parton--level calculation, which the above discussion corresponds to.

Putting the above together we arrive at
\be\label{eq:colllo}
\sigma_{pp} \approx \int  {\rm d}\xi_1 {\rm d}\xi_2\,f_{\gamma/p}(\xi_1,\mu^2)f_{\gamma/p}(\xi_2,\mu^2)\hat{\sigma}(\gamma\gamma \to X)\;.
\ee
Here, we have substituted for the full photon PDF (i.e. including $\overline{\rm MS}$ matching), which we are free to do at this level of precision, and absorbed a factor of $\alpha^2(\mu^2)/\alpha(Q_1^2)\alpha(Q_2^2)$ in the $\gamma\gamma \to X$ cross section. The latter procedure simply corresponds to evaluating the scale of the couplings entering this process, due to the initial--state photons, at $\mu$ rather than $Q_i$ (there is in particular an implicit factor of $\alpha(Q_1^2)\alpha(Q_2^2)$ in \eqref{eq:mtran}), which as discussed in \cite{Harland-Lang:2016lhw,Kallweit:2017khh} is the appropriate choice in the collinear calculation. 

The above expressions give two approximate results for the case of e.g. $W^+ W^-$ production, in both cases treating the initial--state photons as on--shell in the $\gamma \gamma \to W^+ W^-$ matrix elements. In the former case, i.e. applying \eqref{eq:rhodr} in \eqref{eq:sighhf}, the full photon kinematics are still included at the phase space level, which corresponds to a form of $k_\perp$ factorization, as in e.g. the treatment of.~\cite{Luszczak:2018ntp}. The latter approach corresponds to the usual LO result within collinear factorization. In both cases the underlying production process is fully gauge invariant, and so we will not see any of the potential issues that become apparent when we try to extend this, as in the following sections. However, they remain approximations to the full results, which in particular only include the pure PI diagrams; we will comment on the possible extension of the LO collinear result to NLO (and beyond) below. 

Finally, the  expression \eqref{eq:sighh} in the presence of $Z$--initiated production is the same but with the replacement
\be
M^*_{\mu'\nu'}M_{\mu\nu} \to \sum_{Y_1,Y_2} \rho_{Y_1,1}^{\mu\mu'}\rho_{Y_2,2}^{\nu\nu'} \left[M^*_{\mu'\nu'}M_{\mu\nu}\right]_{Y_1,Y_2}
\ee
 where $Y_i= \gamma\gamma, Z\gamma, ZZ$ are the PI, $\gamma Z$ interference and  $Z$--initiated production amplitudes, respectively. That is, $M^{\mu\nu}$ is the relevant $VV \to X(k)$ production amplitude, with $V=\gamma,Z$ and the `$Y_1, Y_2$' subscript indicating that  pure $Z$, $\gamma$ or $Z/\gamma$ interference be included in the appropriate way. We then have
   \begin{align}\nonumber
 \rho_{Y_i,i}^{\alpha\beta}=2\eta_{Y_i}'\int \frac{{\rm d}M_i^2}{Q_i^2} &\bigg[-\left(g^{\alpha\beta}+\frac{q_i^\alpha q_i^\beta}{Q_i^2}\right) F_1^{Y_i}(x_{B,i},Q_i^2)+ \frac{(2p_i^\alpha-\frac{q_i^\alpha}{x_{B,i}})(2p_i^\beta-\frac{q_i^\beta}{x_{B,i}})}{Q_i^2}\frac{ x_{B,i} }{2}F_2^{Y_i}(x_{B,i},Q_i^2)\\ \label{eq:rhoZ}
 &-i\epsilon^{\alpha\beta\mu\nu}q_{i,\mu} p_{i,\nu} \frac{ x_{B,i} }{Q_i^2}F_3^{Y_i}(x_{B,i},Q_i^2)\bigg]\;,
 \end{align}
The prefactors $\eta'$ are given by:
\be
\eta_{\gamma\gamma,i}' = 1\;,\qquad \eta_{\gamma Z, i }' = \left(\frac{G_F M_Z^2}{2 \sqrt{2}\pi \alpha}\right)^{1/2}\frac{Q^2_i}{Q_i^2 + M_Z^2}\;,\qquad \eta_{ZZ, i}' = (\eta_{\gamma Z, i }')^2\;.
\ee
We note that in~\cite{Harland-Lang:2021zvr} the prefactors $\eta_{Y_i}$ were defined to be consistent with the standard DIS convention, i.e. with $\eta_{\gamma Z, i}$ absorbing the factor of $g_w/(2e \cos\theta_W)$ from the $Z ll $ vs. $\gamma ll$ coupling in the lepton pair production amplitude $M^{\mu\nu}$, and a similar factor absorbed on the proton side.  However, we now move beyond the case of lepton pair production,  with a correspondingly different $Z WW$ coupling entering the amplitude $M$. For clarity, we therefore now apply a the  standard normalization for $M^{\mu\nu}$, with no factors from this absorbed into the definition of $\eta$. The priming above signifies that the modified convention for the $\eta$ factors is now used, with the difference simply amounting to the square root in $\eta_{\gamma Z, i }' $.

\subsection{A first look: the SF approach alone}\label{sec:firstlook}

\begin{figure}[t]
\begin{center}
\includegraphics[scale=0.63]{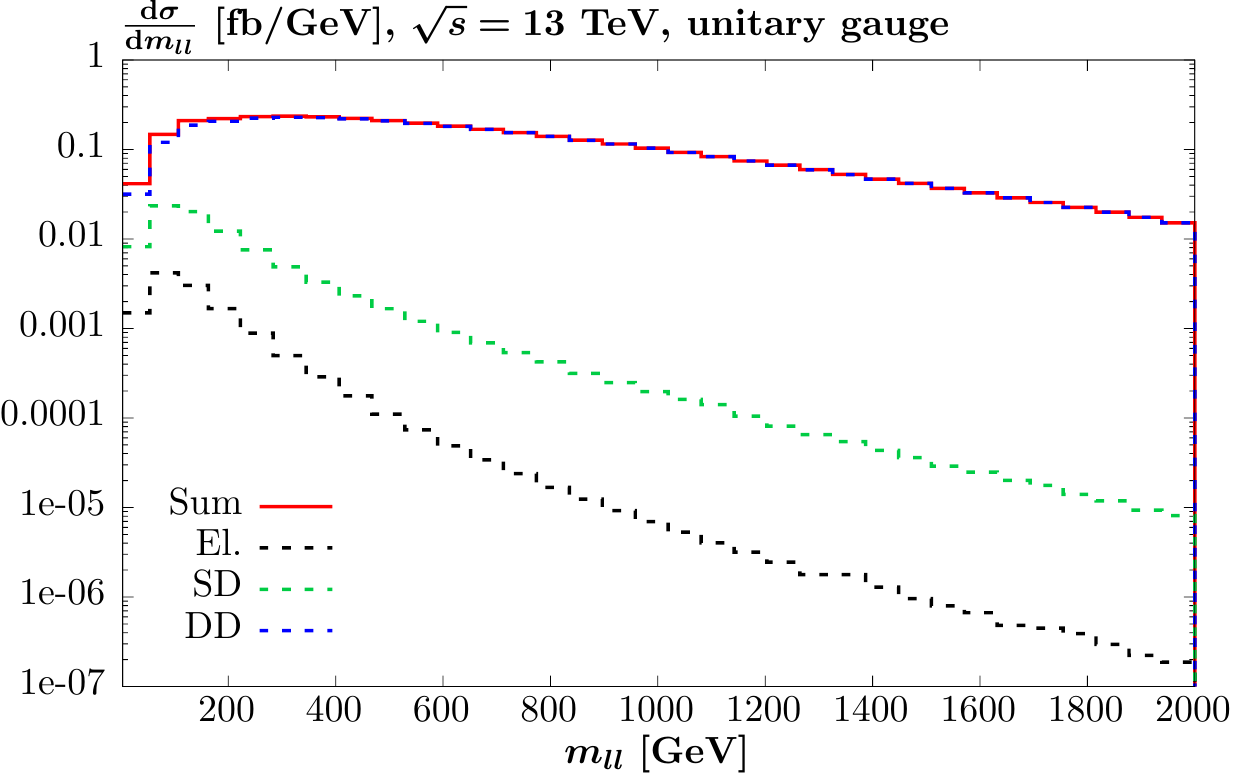}
\includegraphics[scale=0.63]{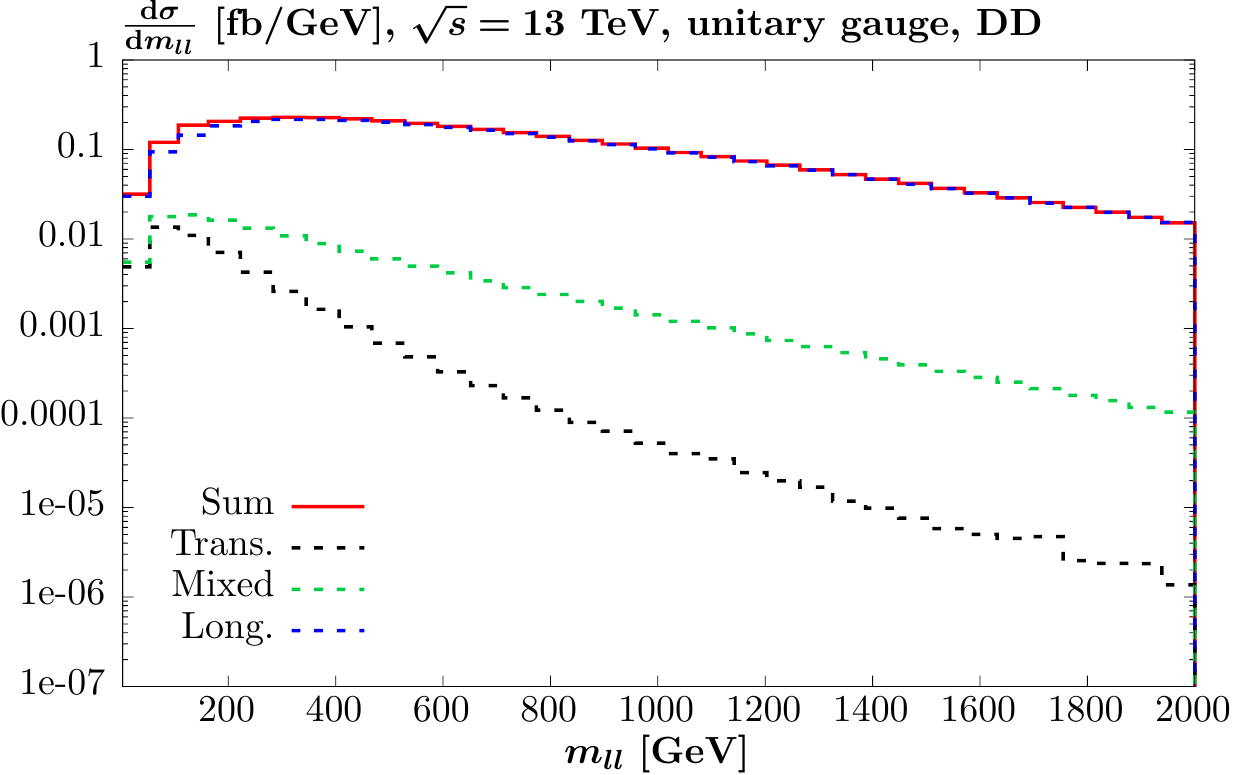}
\includegraphics[scale=0.63]{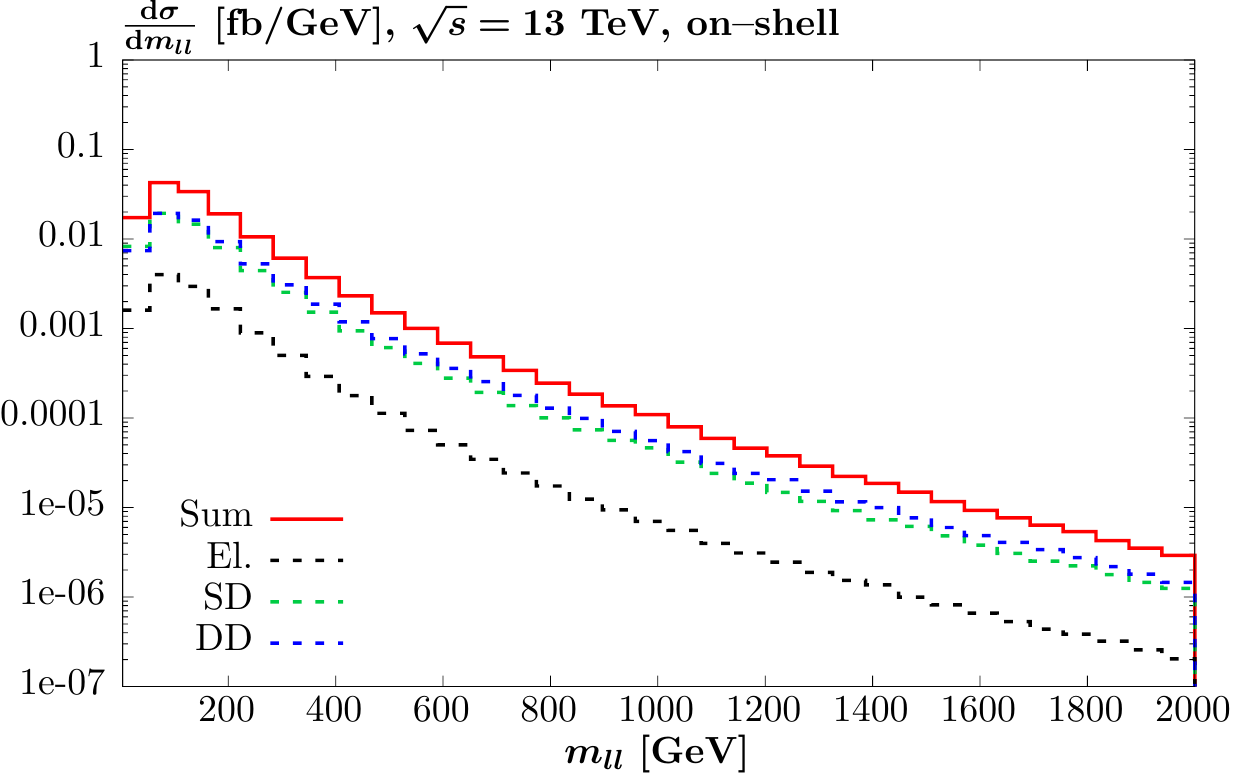}
\caption{Differential cross section with respect to the dilepton invariant mass, $m_{ll}$, for $W^+ W^- \to e^\pm \nu \mu^\mp \nu$ production at the 13 TeV LHC, within the event selection of the ATLAS measurement~\cite{ATLAS:2020iwi}. Cross sections calculated in (top) SF approach in the unitary gauge and (bottom) the on--shell approximation. (top left) and (bottom): The breakdown between elastic (El.), single dissociative (SD) and double dissociative (DD) is given, as well as the sum of the three. (top right): For double dissociative production, the breakdown between purely transverse, purely longitudinal and mixed $W$ polarizations is given.}
\label{fig:unitary_nvbf}
\end{center}
\end{figure}

We begin by simply applying the formula \eqref{eq:sighh} directly to the case of $W^+ W^-$ production in the PI channel. All results which follow are shown at 13 TeV, and correspond to the $e^\pm \nu \mu^\mp \nu$ decay channel, within the ATLAS 13 TeV~\cite{ATLAS:2020iwi} event selection. That is:
\begin{align}\label{eq:atcuts}
&|\eta_l| < 2.5\;,\\ \non
&p_{l,\perp}^{\rm min}> 20\, {\rm GeV},\; p_{l,\perp}^{\rm max}> 27\, {\rm GeV}\;,\\ \non
&m_{ll} > 20\, {\rm GeV}\;,\\ \non
&p_\perp^{e \mu} > 30 \,{\rm GeV}\;.
\end{align}
In~\cite{ATLAS:2020iwi} a veto requiring no additional charged particles with
\be\label{eq:veto}
p_\perp > 500\,{\rm MeV}, \, |\eta | < 2.5\;,
\ee
is also imposed. We will consider for comparison results without this imposed and with it imposed, either approximately or via a full MC implementation; we will discuss this further below.

We first consider the result of working in the unitary gauge for the $\gamma \gamma \to W^+ W^-$ amplitudes in \eqref{eq:sighh}, where here and in what follows the initial--state photon may be off--shell, depending on the context. Omitting the rapidity veto for now, we show in Fig.~\ref{fig:unitary_nvbf} the distribution with respect to the dilepton invariant mass, $m_{ll}$. This is strongly correlated with the (unobservable) $W$ pair invariant mass, $\sqrt{\hat{s}}= m_{WW}$, and indeed qualitatively very similar results are found if we instead consider this quantity directly. In the top left and bottom figures we show the breakdown between elastic (EL), single dissociative (SD, where a single proton dissociates) and double dissociative (DD, where both dissociate) contributions, while the top right figure shows the DD contribution alone, but broken down into purely transverse, purely longitudinal and mixed $W$ polarizations. These are of course not directly observable quantities, and moreover without any rapidity veto imposed the total signal itself will not be isolated from $s$--channel production, which is not shown. Hence, these plots are for illustration purposes only. The top two figures apply \eqref{eq:sighhf}, with the off--shell $\gamma \gamma \to W^+ W^-$ amplitudes calculated in the unitary gauge, while for the bottom figure we for comparison show results using the on--shell approximation described in the previous section, broken down into EL, SD and DD components. The corresponding cross section values for the top left and bottom plots are shown in Table~\ref{tab:cs_unitons}. 

We can immediately see in Fig.~\ref{fig:unitary_nvbf} (top left) that the overall PI cross section (`Sum') falls very slowly with $m_{ll}$ out to rather large $m_{ll} \sim 2$ TeV. Indeed  the $m_{WW}$ distribution itself (not shown) is found to be essentially flat. This is clearly unphysical behaviour that is not seen in the corresponding on-shell case. From the top right plot we can see that the dominant enhancement comes from purely longitudinal $W$ polarizations, with the mixed case also somewhat enhanced. This sort of behaviour is of course exactly what we would expect: in the gauges such as the unitary, individual diagrams exhibit unphysical growth with powers of $E_{W^\pm}/M_W$ which only cancel in the complete gauge invariant combination of diagrams. Once we allow the initial--state photons to be off--shell, the pure PI $\gamma\gamma \to W^+ W^-$ diagrams are no longer individually gauge invariant and hence this behaviour is no longer tamed. With this in mind, we would expect the case of elastic scattering, where the photons are close to being on--shell ($Q_i^2 \ll M_{WW}^2$), to not exhibit any observable sensitivity to this effect, while the DD case, where both photons can be far off--shell, should be the most sensitive to it.  Of course in the on--shell case the pure PI diagrams are gauge invariant, and hence no such issue is observed. All of these effects are observed in the figures, and confirmed quantitatively in Table~\ref{tab:cs_unitons}. The significance of the effect relative to the on--shell case grows with $m_{ll}$ (and hence $E_{W^\pm} \sim M_{WW}$), as we would expect, although in the DD case this enhancement persists even at lower $m_{ll}$. We will discuss this further below. 

As noted above, without applying any further veto (or requirements on additional jets for VBS cuts) the VBS signal cross section we have calculated will not be isolated from the $s$--channel QCD production mode, and hence such a comparison is of limited phenomenological relevance. We therefore now consider the same comparison as above, but effectively accounting for the veto \eqref{eq:veto} on additional charged particles. To do this, for inelastic photon emission from the proton, we evaluate the kinematics of the outgoing quark in LO $q\to q\gamma$ emission so that it matches the photon momentum; at LO this corresponds to the jet kinematics in the standard VBS case. We then require that this passes the veto\footnote{More precisely, to ensure no colour flow between the two proton beams, we require that these outgoing quarks have the same sign of their rapidity as the initiating beam; this emulates the effect of a full MC treatment, where the opposite topology would lead to additional showering in the central detector. In reality the impact of imposing this additional requirement is at the percent level of less, given such configurations are in general kinematically suppressed.}. For elastic emission, this is to very good approximation automatically passed, and hence no veto need be applied in this case. We note that this is clearly only an approximation, with a full evaluation requiring a MC implementation to account for showering/hadronization effects, in particular at the low $p_\perp$ values at which the veto enters. Moreover, this does not account for the impact of MPI. Both of these effects will be addressed in the sections which follow. However, the current comparison will illustrate some of the key issues.

\begin{figure}
\begin{center}
\includegraphics[scale=0.63]{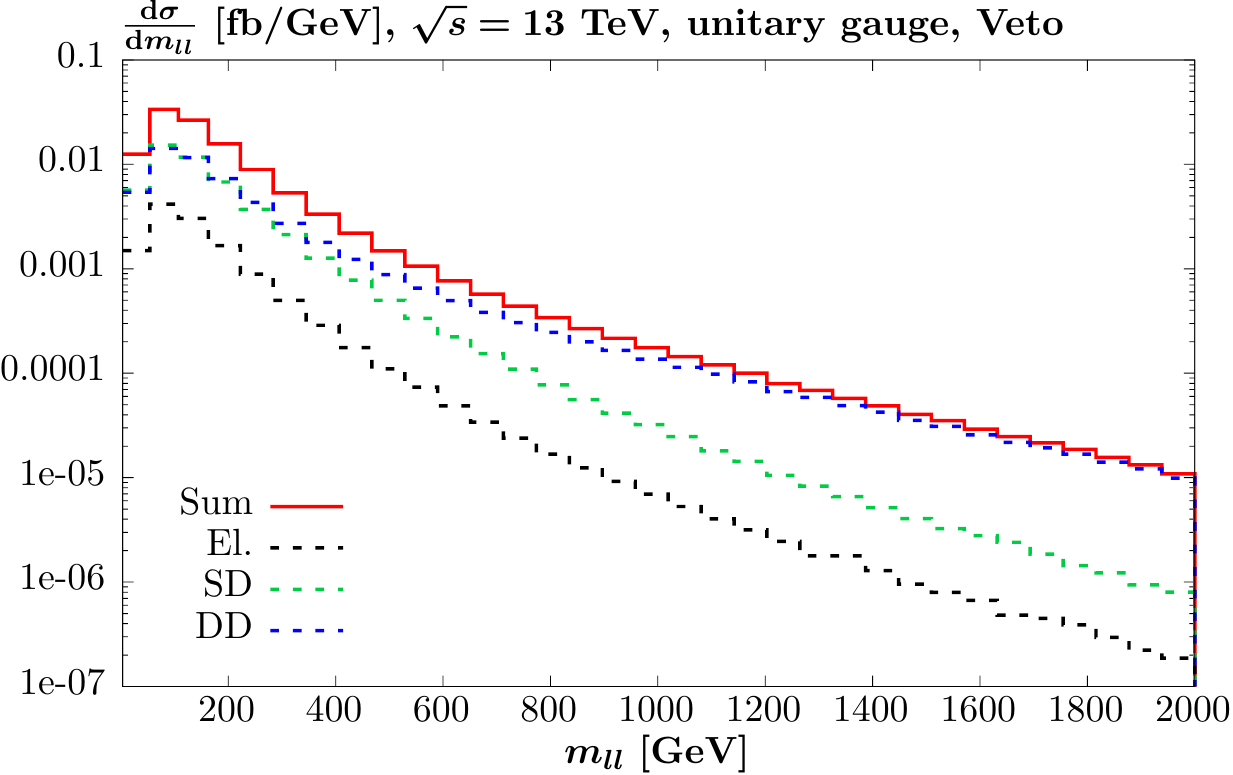}
\includegraphics[scale=0.63]{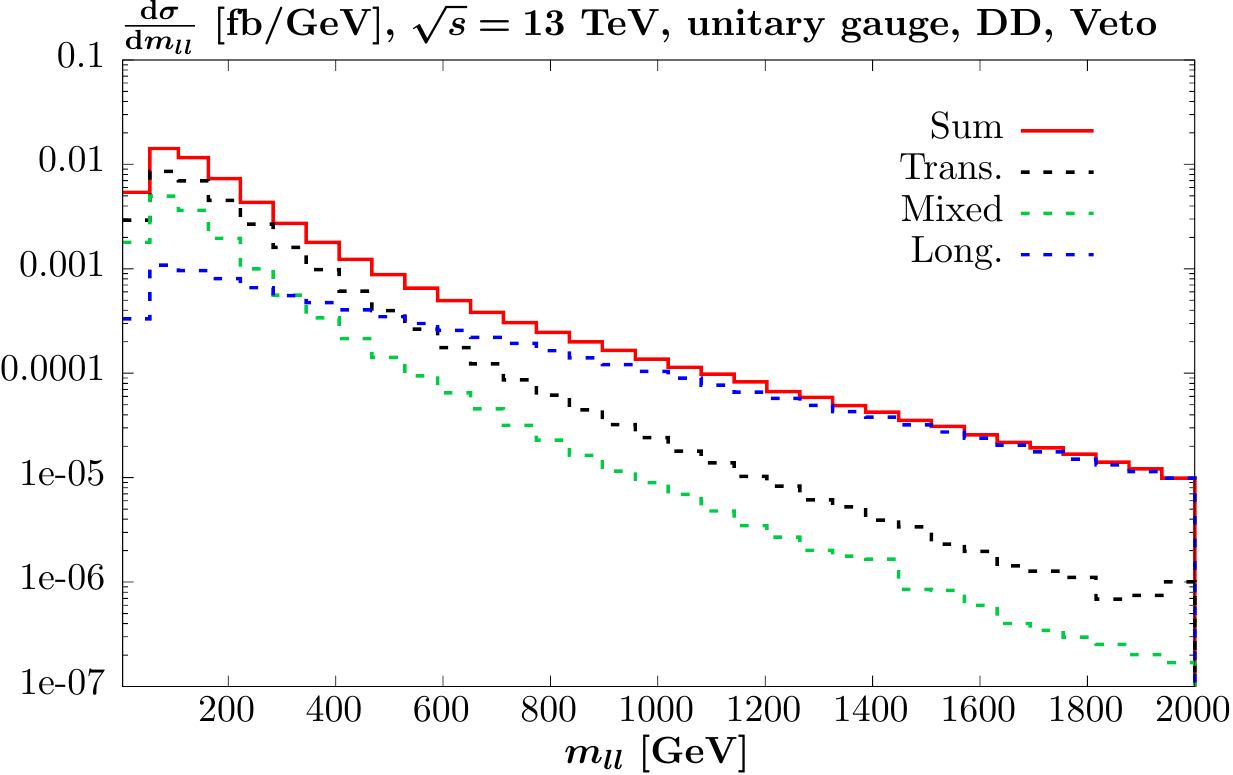}
\includegraphics[scale=0.63]{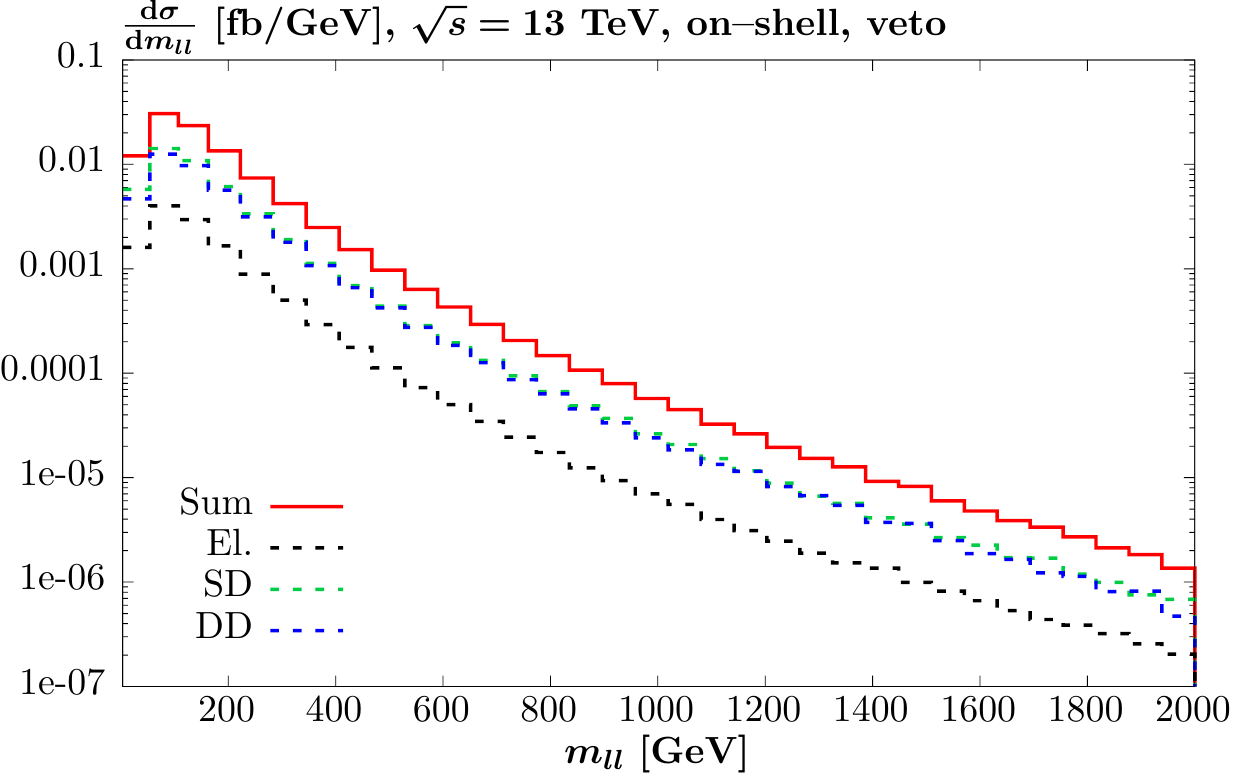}
\caption{As in Fig.~\ref{fig:unitary_nvbf}, but now with the veto \eqref{eq:veto} imposed at the parton level (no survival factor included), as described in the text.}
\label{fig:unitary_vbf2}
\end{center}
\end{figure}

The corresponding distributions are shown in Fig.~\ref{fig:unitary_vbf2}, for the same breakdowns as in Fig.~\ref{fig:unitary_nvbf}, while the cross section values are again given in Table~\ref{tab:cs_unitons}. We can see that the enhancement in the unitary case is significantly reduced, in particular at the level of the total cross sections shown in Table~\ref{tab:cs_unitons}. However it is not absent entirely, and at larger $m_{ll}$ we can still observe in the top left plot a significant enhancement in the DD case, which in the top right plot we can see is driven by the case of purely longitudinal $W$ polarizations, i.e. precisely the unitarity breaking effects discussed above.

\renewcommand{\arraystretch}{1.5}
\begin{table}
\begin{center}
\begin{tabular}{|c||c|c|c||c|c|c|}
\hline
 $\sigma$ [fb]& \multicolumn{3}{c||}{Unitary} & \multicolumn{3}{c|}{On--shell}\\
 \cline{2-7}
 &EL  & SD& DD&EL& SD& DD\\
\hline 
No veto &0.704& 5.01& 222&0.696& 3.31& 3.81\\
\hline
Veto &0.704  & 2.76& 3.03&0.696& 2.53& 2.30\\
\hline
\end{tabular}
\end{center}
\caption{Cross sections (in fb) corresponding to Figs.~\ref{fig:unitary_nvbf} (left) and~\ref{fig:unitary_vbf2} (left). See figure captions for definitions.}  \label{tab:cs_unitons}
\end{table}

In fact, in this case we can derive some rather simple analytic expectations for the impact of unitarity breaking effects. In particular, the effect of the veto is to suppress larger values of the photon $Q_i^2$ (see~\cite{Harland-Lang:2016apc} for exact expressions), such that these effects are driven by the large $M_{WW}$ behaviour of the $\gamma \gamma \to W^+ W^-$ amplitudes at fixed $Q_i^2 \ll M_{WW}^2$. We can in particular expand in terms of the helicity amplitudes:
\be
\mathcal{M}^{\mu\nu} =  \sum_{\lambda_1 \lambda_2} \mathcal{M}_{\lambda_1 \lambda_2}  \epsilon_{\lambda_1}^\mu \epsilon_{\lambda_2}^\nu \;
\ee
where $\mathcal{M}$ is the $\gamma(q_1) \gamma(q_2) \to W^+(p_+) W^- (p_-)$ amplitude, with $q_i^2 = - Q_i^2 \neq 0$ in general. Here the sum is over the photon polarization vectors $\epsilon_{\lambda_i}$, while the $W$ polarizations are left implicit. As we have the usual Ward identity relation $q_1^\mu M_{\mu \nu}=q_2^\nu M_{\mu \nu}=0$ there are three independent polarizations, namely the two standard transverse vectors, and the scalar polarization vector, which we can write as
\be
s_1^\mu = \sqrt{\frac{Q_1^2}{(q_1 q_2)^2 - Q_1^2 Q_2^2}} \left( q_2^\mu  + q_1^\mu \frac{(q_1 q_2)}{Q_1^2}\right)\;, 
\ee
with $s_2$ given by interchanging $1\leftrightarrow 2$. Given the Ward identity relation, we can drop the second term in the above expression, which makes explicit the vanishing  of the contribution from longitudinal polarizations  in the on--shell $Q^2_i \to 0$ limit, as is well known.

Using this, we can study  the sensitivity of the corresponding helicity amplitudes to unitarity breaking effects, namely by determining whether they do indeed grow with $M_{WW}$ (recalling that unitarity dictates that here the amplitudes should at most be constant with energy). Although on dimensional grounds the amplitude with at least one longitudinal $W$ polarization could behave in this way, we find that it is only for purely longitudinal  $W$ bosons that this occurs. This in particular only happens when both photons have scalar polarizations. The high energy, unitarity breaking, behaviour in the amplitude then takes the simple form
\be\label{eq:Munitb}
\mathcal{M}_{ss} \ni \frac{\sqrt{Q_1^2 Q_2^2}}{4 M_W^2} \cdot M_{WW}^2\;.
\ee 
That is, we only expect unitarity breaking effects when both photons are off--shell. This is confirmed in Fig.~\ref{fig:unitary_vbf2} (top left), where the DD channel shows significant growth with increasing $M_{WW}$, while the EL and SD channels do not exhibit this. It is therefore in the DD channel that we will expect particular sensitivity to these effects, and therefore to non--PI diagrams; we will confirm this later on. We note that in principle for SD, and even EL, production the elastic photons are not completely on--shell, and hence at sufficiently high $M_{WW}$ we will still expect to in principle see unitarity breaking effects when working in the unitary gauge. This will however be parametrically suppressed by a factor of $\sim \sqrt{Q_i^2}/M_W \lesssim 10^{-2}$ for each elastic photon, and hence is of  limited phenomenological relevance.

Jumping ahead a little, it is interesting to observe how the behaviour of \eqref{eq:Munitb} is cured when the full gauge invariant set of diagrams is considered. This is particularly simple if we consider for illustration the case of right handed quarks in Fig.~\ref{fig:wwfig}, in which case all diagrams where $W$ bosons attach to the quark legs to do not enter. Then we should include $Z$--initiated production as well, which excluding the $s$--channel Higgs contribution is simply achieved by replacing:
\be
\frac{1}{Q_i^2} \to \frac{1}{Q_i^2}\left(1-\frac{Q_i^2}{Q_i^2 + M_Z^2}\right) =  \frac{1}{Q_i^2}\frac{M_Z^2}{Q_i^2 + M_Z^2}\;.
\ee
Hence \eqref{eq:Munitb} effectively becomes
\be
\mathcal{M}_{ss,Z/\gamma} = \mathcal{M}_{ss} \cdot \frac{M_Z^4}{(Q_1^2 + M_Z^2)(Q_2^2 + M_Z^2)}\;.
\ee
It is then straightforward to show that this is exactly equal and opposite to the high energy behaviour given by the corresponding Higgs diagram. 

The above discussion explains why the DD case is in particular so sensitive to unitarity breaking effects, but it only provide a rough guide. That is, the photon $Q_i^2$ are of course integrated over according to \eqref{eq:sighhf} and hence the assumption of fixed $Q_i^2 \ll M_{WW}^2$ is not necessarily justified.  Indeed, once this requirement is dropped, unitarity breaking effects are found to enter beyond the DD, purely longitudinal case, as  is evident from Fig.~\ref{fig:unitary_nvbf}. Indeed, although the impact is highest at high mass, we can see that down to low $m_{ll}$ the cross section in the unitary gauge is artificially enhanced with respect to the on--shell case,  due to unphysical positive scaling with the photon $Q_i^2$ that occurs for longitudinal $W$ polarizations.

\subsection{The axial gauge}\label{sec:ax}

 \begin{figure}
\begin{center}
\includegraphics[scale=0.63]{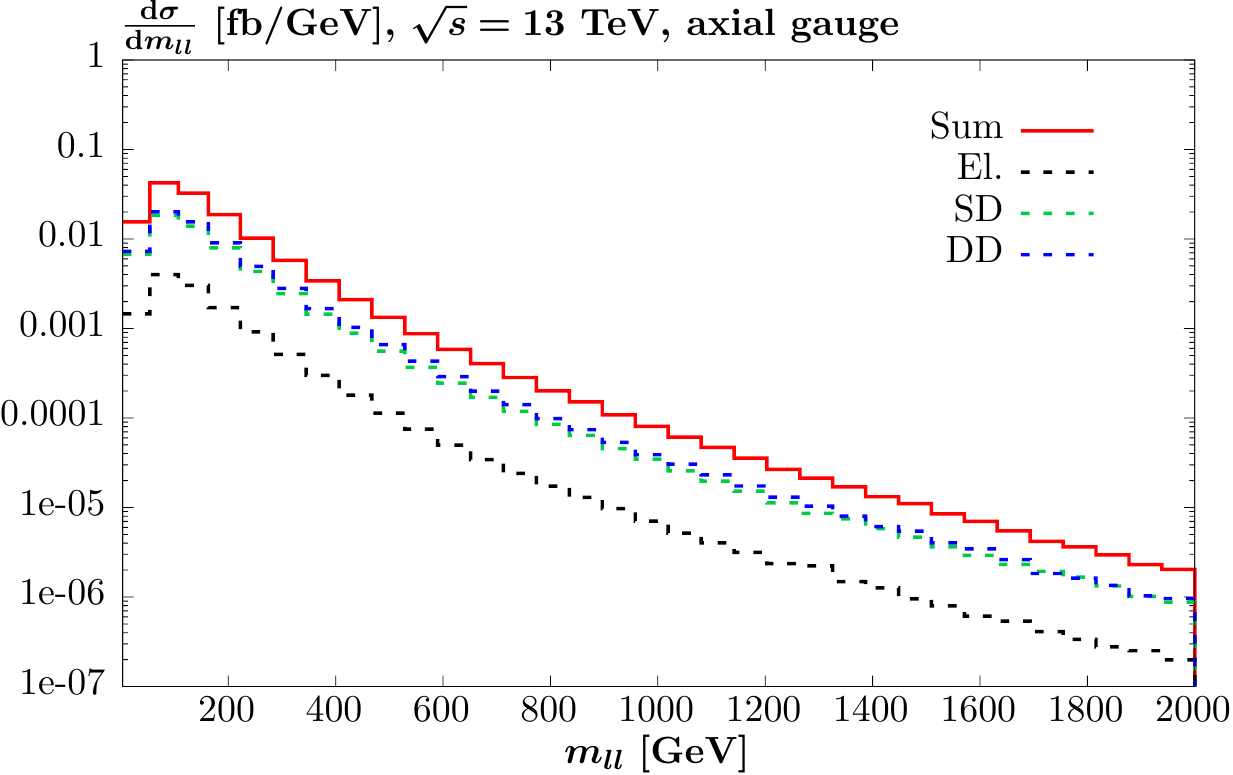}
\includegraphics[scale=0.63]{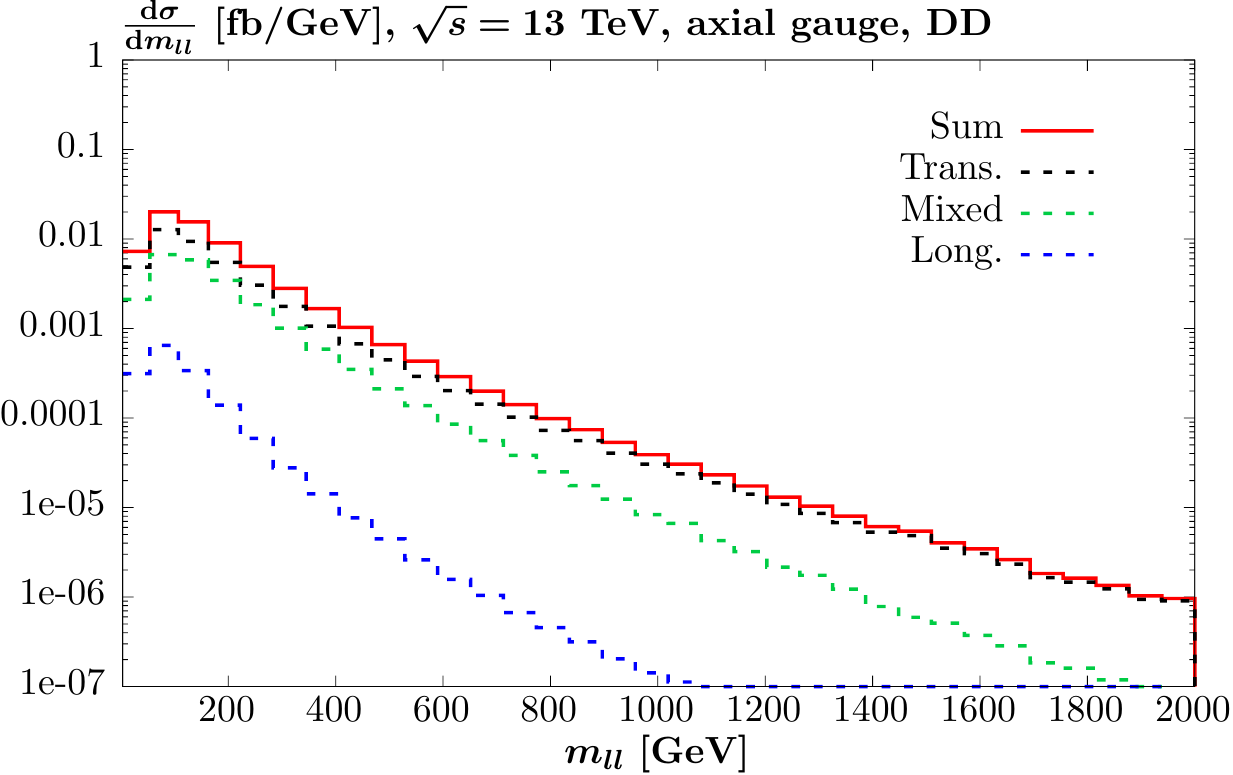}
\caption{As in Fig.~\ref{fig:unitary_nvbf} (top), but for the EW axial gauge.}
\label{fig:axial_nvbf}
\end{center}
\end{figure}

To general solution to the issue of unitarity breaking effects discussed above is rather straightforward. That is, the pure PI diagrams only correspond to a subset of the whole group of gauge invariant diagrams that contribute to $W^+ W^-$ production, shown in Figs.~\ref{fig:wwfig} and~\ref{fig:wwfigsd}, for the DD and SD cases, respectively. It is only once these are included that the issue will be resolved. Nonetheless, it is not immediately obvious that these additional diagrams should provide such a significant contribution. In particular, a straightforward analysis of the propagators that enter in the different cases demonstrates that the pure PI diagrams are the only ones for which the corresponding amplitude contains a $t$--channel, $\sim 1/Q_1^2 Q_2^2$, pole structure. The residue of these poles must therefore be gauge invariant, and indeed is precisely the ingredient that effectively enters the cross section in the on--shell (or equivalent photon) approximation. More broadly, we can expect the additional amplitudes to be suppressed by at least $\sim Q_i^2 / M_Z^2$, or $Q_i^2/M_{WW}^2$, where the former scaling comes from the inclusion of initial--state $Z$ boson in the VBS scattering diagrams. Indeed, in~\cite{Harland-Lang:2021zvr} the impact of the equivalent diagrams in the case of lepton pair production are found to be very small away from the $Z$ peak region, precisely due to these general expectations (see Section~\ref{sec:lep}).

However, as discussed in e.g.~\cite{Borel:2012by}, the $O( Q_i^2 / M_{Z,WW}^2)$ corrections to the pure PI diagrams that enter once one moves away from the on--shell limit are not individually gauge invariant (here and in what follows it is implicit that $M_Z \sim M_W$ for the purpose of such counting). In the unitary gauge, these corrections can receive large $\sim E_W/M_W$ corrections, for longitudinal $W$ bosons, and the power counting breaks down. An interesting possibility, discussed in~\cite{Kunszt:1987tk,Accomando:2006iz,Accomando:2006mc,Borel:2012by}, that avoids this issue is to instead in the EW axial gauge. The basic formalism is described in~\cite{Dams:2004vi,Kunszt:1987tk}, and corresponds to applying the gauge fixing term
\be
\mathcal{L}_{gf} = -\frac{1}{2} \lambda n^\mu A_\mu^a A_\nu^b n^\nu - \frac{1}{2}\lambda (n\cdot B)^2\;,
\ee
to the EW Lagrangian, where $n$ is an in principle arbitrary constant $4$--vector. Here $A_\mu^a$ ($a=1,2,3$) are the SM $SU(2)$ gauge fields and $B_\mu$ is the $U(1)_Y$ gauge field, i.e. defined in the usual way prior to spontaneous symmetry breaking. $\lambda \to \infty$ is then taken in the derivation of the Feynman rules. In comparison to the unitary gauge, this leads to the introduction of intermediate Goldstone bosons $\phi_{W,Z}$, similarly to the $R_\xi$ gauge. We use the realisation of~\cite{Dams:2004vi}, for which the $W,Z$ and $\phi_{Z,W}$ propagators are diagonal, but the interaction vertices (including between the purely `physical' fields $W,Z,\gamma$) are modified. An alternative approach is given in~\cite{Kunszt:1987tk}, for which the interaction vertices are not modified at the expense of introducing mixed propagators between the bosonic fields. 

The full list of Feynman rules are given in~\cite{Dams:2004vi}, and are not repeated here. However we note for concreteness that the vector $n$ enters explicitly in the e.g. the modification to the EW boson propagators:
\be\label{eq:propaxial}
\Delta_{\mu \nu}(k) = -i \frac{g_{\mu\nu} - \frac{n_\mu k_\nu + n_\mu k_\nu}{n\cdot k} + k_\mu k_\nu \frac{n^2}{(n\cdot k)^2}}{k^2 - M^2 + i \epsilon}\;,
\ee
where $M$ is the boson mass, and in the definition of the $W$ (and $Z$) boson polarization vectors. In particular, for $n^2=0$ the longitudinal polarization is simply
\be\label{eq:axiallong}
\epsilon_L^\mu(k) = i \frac{M_W}{k\cdot n} n^\mu\;,
\ee
and similarly for the $Z$, while the transverse polarization vectors satisfy
\be
\epsilon_\pm \cdot k = \epsilon_\pm \cdot n = 0\;,
\ee
as well as the usual orthonormal conditions (note that \eqref{eq:axiallong} does not satisfy the first requirement). We can immediately see that the longitudinal polarization vector no longer scales as $\sim E_W/M_W$ at high energy, and hence no unitarity breaking effects will be expected. Some care is needed over the choice of $n$, as while the full gauge invariant result is independent of it, the gauge dependent pure PI subset is not. In this case,  to avoid instabilities in the result, as in~\cite{Accomando:2006iz} we choose
\be
n=(1,0,0,1)\;,
\ee 
in the lab frame. By choosing $n$ to lie along one of the beam directions, it can in particular be shown that this avoids the case that $n \cdot k = 0$ in the denominator of the $W$ propagators \eqref{eq:propaxial}, which would lead to ill--defined results when all diagrams are not included.

\begin{figure}
\begin{center}
\includegraphics[scale=0.63]{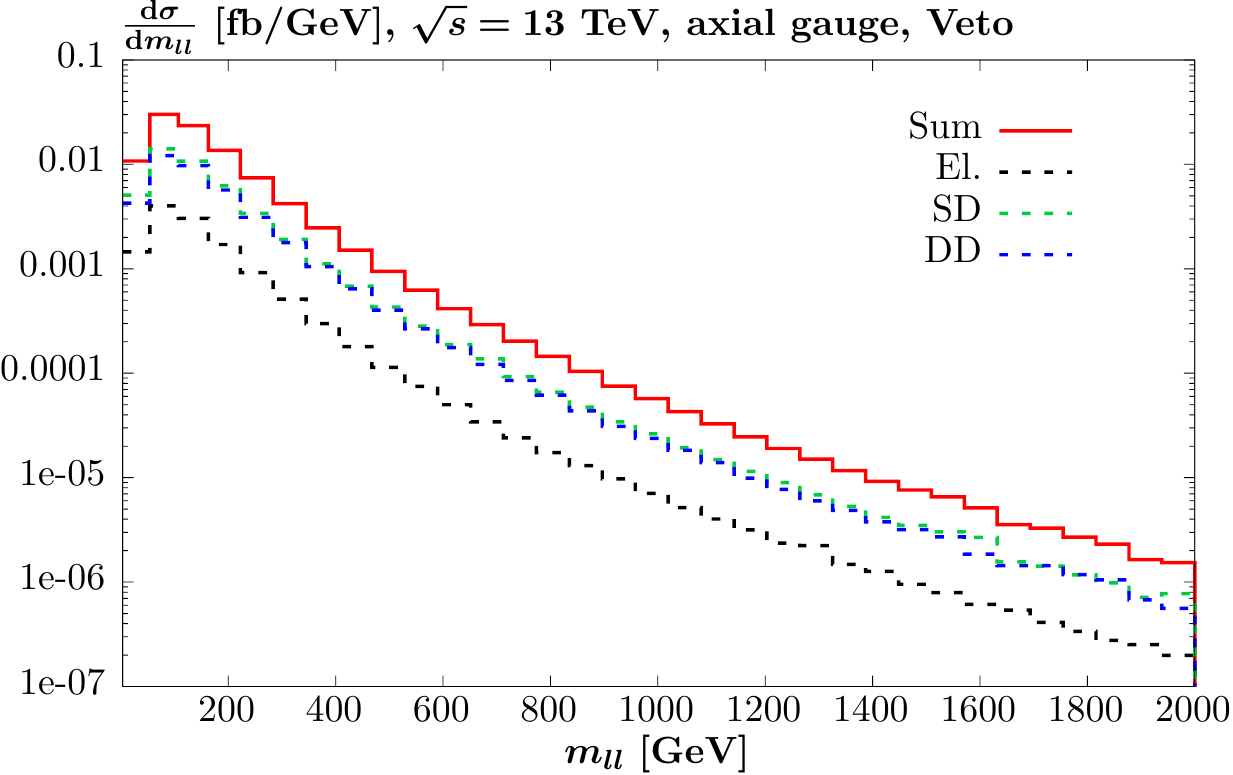}
\includegraphics[scale=0.63]{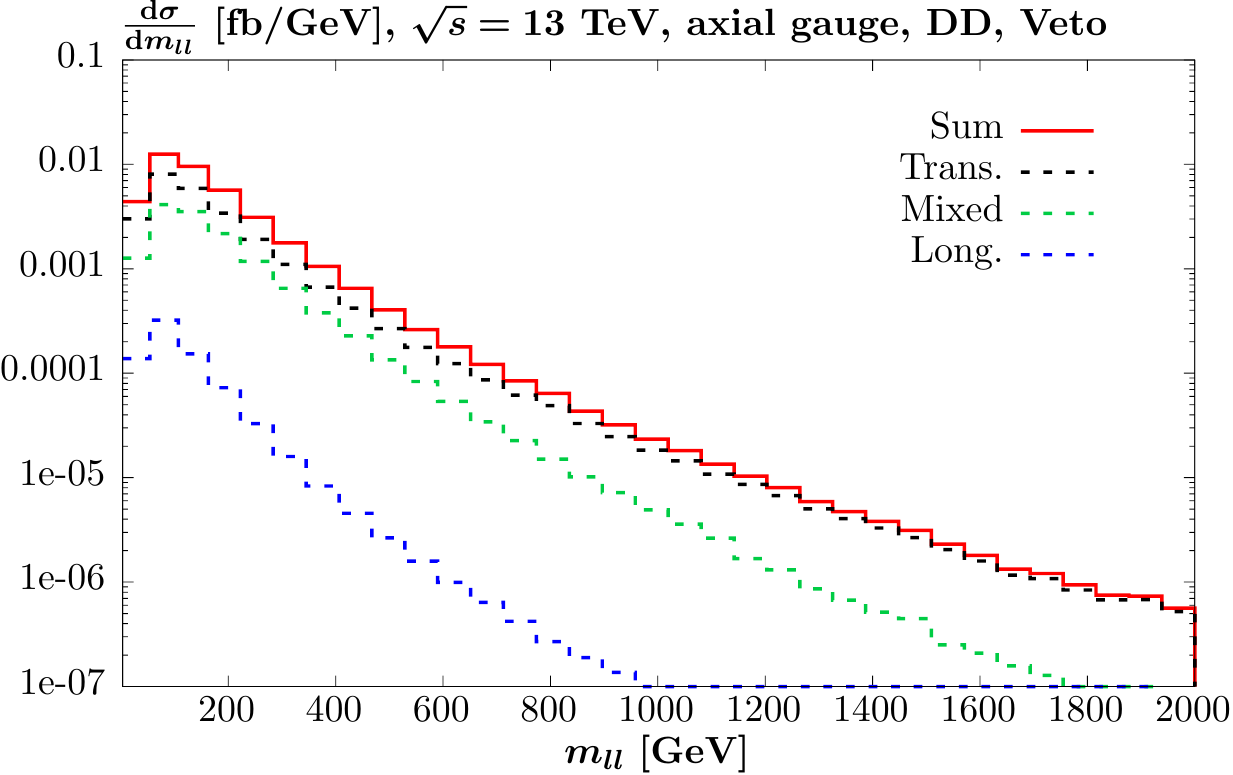}
\caption{As in Fig.~\ref{fig:unitary_vbf2} (top), but for the EW axial gauge.}
\label{fig:axial_vbf2}
\end{center}
\end{figure}

\renewcommand{\arraystretch}{1.5}
\begin{table}
\begin{center}
\begin{tabular}{|c||c|c|c|}
\hline
 $\sigma$ [fb]& \multicolumn{3}{c|}{Axial} \\
 \cline{2-4}
 &EL  & SD& DD\\
\hline 
No veto &0.701& 3.25& 3.64\\
\hline
Veto &0.701  & 2.52& 2.26\\
\hline
\end{tabular}
\end{center}
\caption{Cross sections (in fb) corresponding to Figs.~\ref{fig:axial_nvbf} (left). See figure captions for definitions.}  \label{tab:cs_axons}
\end{table}

The equivalent results to the unitary case in the previous section are shown in Figs.~\ref{fig:axial_nvbf} and~\ref{fig:axial_vbf2}. The unitarity breaking effects are clearly absent, as expected, and indeed the distributions look rather similar to the on--shell case shown in Figs.~\ref{fig:unitary_nvbf} and~\ref{fig:unitary_vbf2}; a more direct comparison will be presented later. The numerical results are shown in Table~\ref{tab:cs_axons} and in fact we can see that these are very close to the  on--shell approximation shown in Table~\ref{tab:cs_unitons}. Thus indeed the $O(Q_i^2 / M_{Z,WW}^2)$ corrections to the pure PI diagram that come from allowing the photons to be off--shell follow the naive scaling we could expect, i.e. are indeed small in this gauge.

However, we are still left with the question of the impact of the non PI diagrams, which might still be non--negligible. To include these clearly requires a more significant modification of the above approach, and we address this in the following section.

\subsection{Hybrid approach: basic idea}\label{sec:hybrid}

\begin{figure}[t]
\begin{center}
\subfigure[]{\includegraphics[scale=0.6]{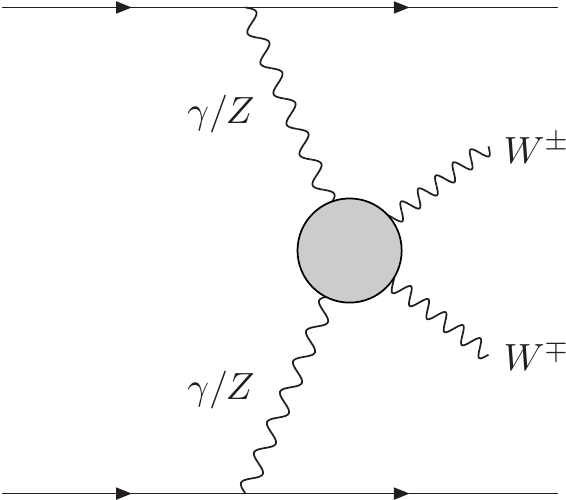}}\quad
\subfigure[]{\includegraphics[scale=0.6]{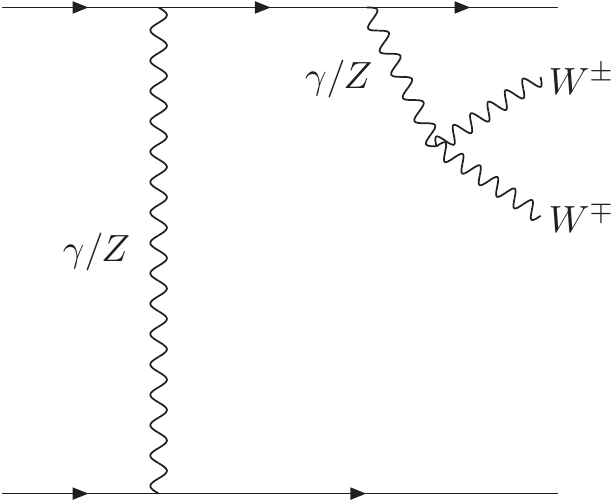}}\quad
\subfigure[]{\includegraphics[scale=0.6]{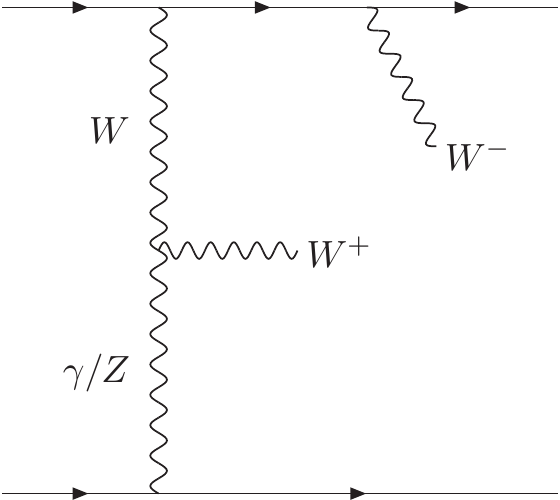}}\quad\\
\subfigure[]{\includegraphics[scale=0.6]{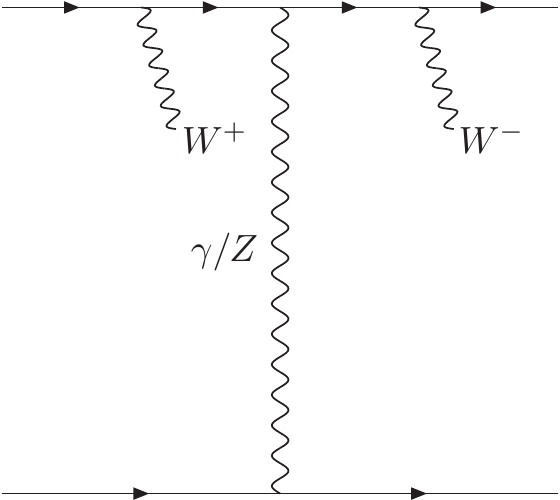}}\quad
\subfigure[]{\includegraphics[scale=0.6]{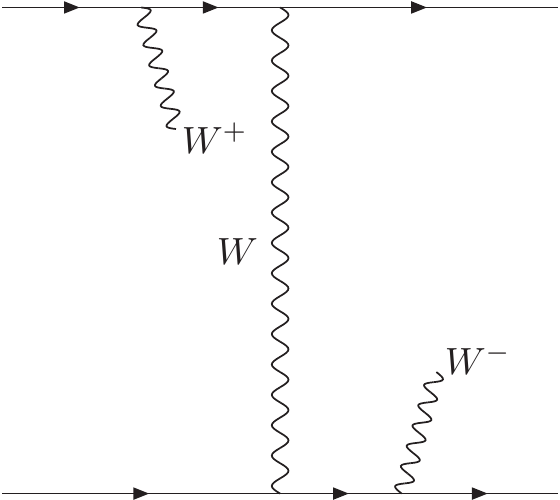}}\quad
\caption{Classes of Feynman diagrams contributing to $W^+ W^-$ DD production at LO in the $qq \to W^+ W^- qq$ process.  The blob in plot (a) denotes the sum of the $t$, $u$--channel and contact diagrams. Diagrams correspond to the case of up--type initiating quarks for concreteness, and with various permutations implied.  }
\label{fig:wwfig}
\end{center}
\end{figure}

\begin{figure}[t]
\begin{center}
\subfigure[]{\includegraphics[scale=0.6]{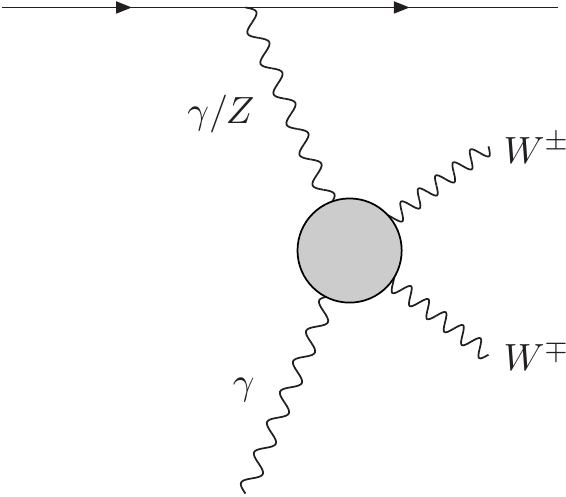}}\quad
\subfigure[]{\includegraphics[scale=0.6]{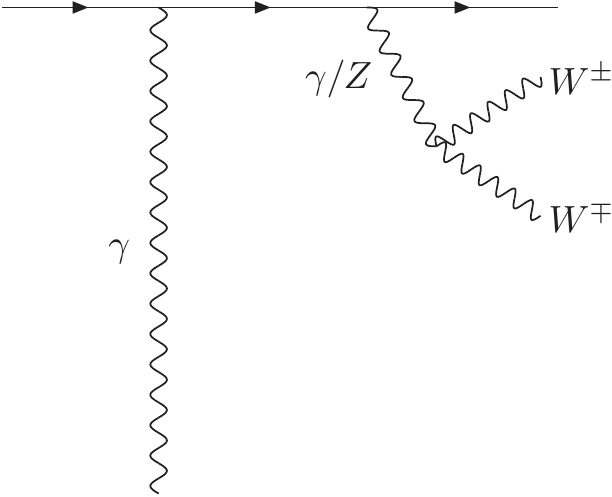}}\quad
\subfigure[]{\includegraphics[scale=0.6]{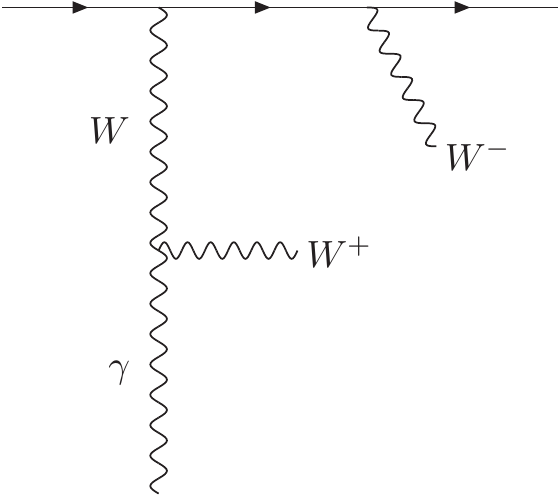}}\quad
\subfigure[]{\includegraphics[scale=0.6]{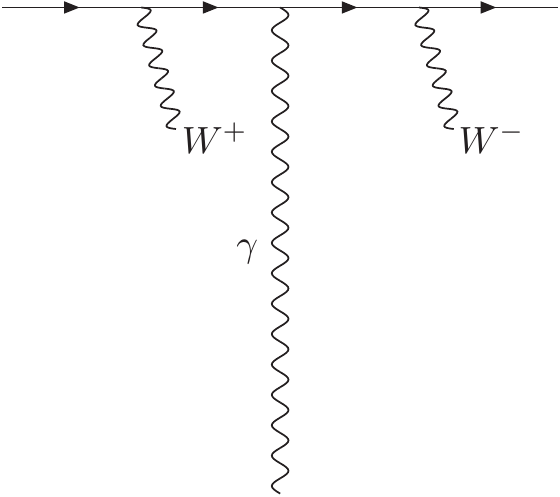}}\quad
\caption{Classes of Feynman diagrams contributing to $W^+ W^-$ SD production at LO in the $q\gamma \to W^+ W^- q$ process.  Diagrams correspond to the case of up--type initiating quarks for concreteness, and with various permutations implied.   Notation as in Fig.~\ref{fig:wwfig}. }
\label{fig:wwfigsd}
\end{center}
\end{figure}

As mentioned in the previous section, the pure PI contributions to $W^+ W^-$ scattering only represent a (gauge dependent) subset of the full set of diagrams that enter into $W^+ W^-$ production. These are shown in  Figs.~\ref{fig:wwfig} and~\ref{fig:wwfigsd}, for the DD and SD cases, respectively. We in particular show the corresponding quark--initiated processes at LO, considering the case of purely up--type quarks for concreteness. We note that we explicitly exclude $s$ and $u$--channel contributions, as defined with respect to the quark momenta, as these will to very good approximation not pass the veto requirement. The PI process corresponds to diagram (a), with the contribution from initial--state $Z$ bosons omitted. While the non--PI diagrams are expected to be kinematically subleading, we have seen that this is only apparent once we work in an appropriate gauge, such as the axial gauge. Moreover, even then the contribution from these additional diagrams may not be negligible. With this in mind we include these in this section.

Now, if we simply calculated the contribution from the diagrams as in Figs.~\ref{fig:wwfig} and~\ref{fig:wwfigsd} at LO, i.e.  with initial--state massless quarks (and photons in the latter case) and using standard collinear factorization,  then these would of course contain  singularities due to the ($Q_i^2 \to 0$) region of collinear $q\to q \gamma$ emission. The textbook approach to deal with this would as usual be to apply appropriate collinear subtractions, as well as to include the corresponding lower order PI diagrams. These latter diagrams would be included via a collinear photon PDF, suitably calculated via the LUXqed approach, e.g.~\cite{Manohar:2016nzj,Bertone:2017bme,Xie:2021equ,Cridge:2021pxm}. This will however introduce a degree of scale variation uncertainty into the result, and moreover has no direct way of dealing with the low $Q^2_i$, $W_i^2$ region (where pQCD is not reliable) differentially, as discussed in~\cite{Harland-Lang:2019eai,Harland-Lang:2021zvr}; the latter point is particularly relevant when it comes to the inclusion of the soft survival factor, as we will discuss later on.

Now, the above points are in many cases inevitable effects of the necessary application of collinear factorization to the problem, which of course provides a robust framework for including successive orders in the calculation within perturbation theory, and hence of reducing the scale variation uncertainty in the result, as well as dealing with e.g. collinear $\gamma \to q\overline{q}$ emission in the initial state, as discussed further in~\cite{Harland-Lang:2021zvr}. However, in the current case the distinct requirement  that comes from imposing a rapidity veto allows us to take a different approach. In particular, while the class of diagrams show in Figs.~\ref{fig:wwfig} (b) and~\ref{fig:wwfigsd} (b) in principle contain a region of collinear $\gamma \to q\overline{q}$ emission, this is removed by the rapidity veto we impose. That is, considering Fig.~~\ref{fig:wwfigsd} (b) for simplicity, the collinear $\gamma \to q\overline{q}$ region only occurs when the outgoing quark on the upper line in the figure is collinear to the initial--state photon, such that the outgoing quark which originates from upper beam is collinear to the lower beam direction. This is in other words an $s$--channel contribution, and is certainly excluded by the rapidity veto. An identical argument applies in the case of Fig.~\ref{fig:wwfig} (b).

We are therefore left with the those due to collinear $q\to q\gamma$ emission, which occurs in the $Q^2_i \to 0$ regime. However, in this case we can more precisely treat the low $Q^2_i$ (collinear) region via a suitable generalization of the SF approach. To see this, we consider first for clarity the SD case shown in Fig.~\ref{fig:wwfigsd}, and in particular the pure PI diagrams shown in (a), i.e. with the $Z$ contribution omitted. Here, the cross section can be (and is in the previous sections) calculated directly within the SF approach, that is by applying \eqref{eq:mtran}, with $\rho_{1,2}$ given in terms of inelastic (elastic) SFs. For the inelastic vertex, there is no collinear singularity present due to $q\to q\gamma$ emission as this is of course absent in the SFs, which are perfectly regular as $Q_i^2 \to 0$, being experimentally parameterised in this region rather than calculated in collinear factorization, where such singularities would appear in the intermediate steps. At high $Q_i^2$ on the other hand, the inelastic SF is calculated using pQCD and hence the SF approach corresponds implicitly to including the PI diagrams at parton level. The SF approach therefore provides a straightforward way to calculate the contribution from these diagrams across the entire kinematic region.

However, the above discussion clearly does not immediately apply to the other diagrams in Fig.~\ref{fig:wwfigsd}, for which \eqref{eq:sighh} cannot be directly used. Nonetheless, it can be straightforwardly generalized to do so. In particular, we apply a hybrid approach, whereby if we have
\\
\be\label{eq:qwcut}
Q_{i}^2 > Q^2_{\rm cut}\;, W^2_i > W^2_{\rm cut}\;,
\ee
\\
with $Q^2_{\rm cut}=1\, {\rm GeV}^2$, $W^2_{\rm cut}= 3.5\, {\rm GeV}^2$ and $i=1$ in the labelling of Fig.~\ref{fig:wwfigsd}, then we directly include the full sum of contributing diagrams at parton level.  If on the other hand \eqref{eq:qwcut} is not satisfied then the contribution from these additional diagrams will be strongly kinematically suppressed, recalling these are $O(Q_i^2/M_{Z,WW}^2)$ and hence should be negligible, given \eqref{eq:qwcut} requires that $Q_i^2 < 1\, {\rm GeV}^2$ for most of the relevant regions of phase space, i.e. other than at very large $x$ values; we will verify this expectation numerically below. We can therefore directly apply the SF calculation  \eqref{eq:sighh}, i.e. purely for the PI diagram (a), here. This will in particular guarantee that the completely smooth (and regular) with respect to the collinear $Q_i^2 \to 0$ region.

This  provides a straightforward way to account for the contribution from all diagrams across the entire kinematic region. We emphasise that is to very good approximation a smooth matching, as above the transition point of \eqref{eq:qwcut} the contribution from the diagrams other than the pure PI component will remain strongly suppressed, even if their (small) contribution is now explicitly included. In particular, the  transition point is chosen to match that in the SF approach, which itself is applied in the approach of~\cite{Manohar:2016nzj} for calculating the photon PDF. Roughly Bbelow this point one does not expect to reliably use pQCD, and hence these represent the minimum cut values, but as we will see below for larger values of the cut the predicted cross section remains almost unchanged. 

We note that when \eqref{eq:qwcut} is satisfied, and the sum over all relevant parton--level diagrams is included, then this will in the limit that the elastic photon $Q_2^2 \to 0$ be gauge invariant. As this is to very good approximation true, any residual gauge dependence will be very small; we will verify this below for the purely elastic case, where it is seen to enter at the sub--percent level. Below this cut, the pure PI contribution is strongly dominant in any gauge, and therefore the SF calculation in this region is large gauge independent. We will again demonstrate this explicitly in later sections. However for concreteness, we note that when the SF approach is applied directly, we apply the axial gauge, guided by the expectation that this is more stable and less sensitive to unitarity breaking effects. For the explicit parton--level calculation we apply the unitary gauge, as is this a default choice available in the \texttt{MadGraph5\_aMC@NLO}~\cite{Alwall:2014hca,Frederix:2018nkq} code we use. However, we emphasise again that the final result is largely independent of these choices.

The situation in the DD case is in practice a little more involved, but in principle follows exactly the same approach. In particular if for both $i={1,2}$ the requirement \eqref{eq:qwcut} is satisfied, then we simply include all diagrams in Fig.~\ref{fig:wwfig} in the usual way, while if both these requirements are not satisfied we can use the SF approach to calculate the PI contribution alone. There is in addition now the possibility of the mixed case, where \eqref{eq:qwcut} is satisfied for $i=1$, but not $i=2$ (or vice versa). Here, we now include all diagrams as in  Fig.~\ref{fig:wwfigsd}, but where the initial--state off--shell photon is coupled to a low scale inelastic SF. In more detail, to do so we simply replace in \eqref{eq:sighh}
\be\label{eq:rhorep}
\rho_{1}^{\mu\mu'}\rho_{2}^{\nu\nu'} M^*_{\mu'\nu'}M_{\mu\nu} \to \frac{Q_1^2}{4\pi \alpha(Q_1^2)} \int  \frac{{\rm d}M_1^2}{Q_1^2} \, \rho_{2}^{\mu\mu'} \sigma_{\mu \mu'}^{1}\;,
\ee
where the integration is as usual performed simultaneously with the other phase space integrals, while for the case that  \eqref{eq:qwcut} is satisfied for $i=2$, but not $i=1$, we simply interchange $1 \leftrightarrow 2$. At LO we have
\be\label{eq:siggq}
\sigma_{\mu \mu'}^{i} = \sum_{j=q,\overline{q}}  f_j(x_{B,i},\mu_F^2)  \left \langle A_{\mu}^i A_{\mu'}^{i*} \right\rangle \;,
\ee
where $A_\mu^i$ is the corresponding $\gamma^* + q \to W^+ W^- + q$ amplitude including all diagrams in Fig.~\ref{fig:wwfigsd}, with a collinear initial--state quark/anti--quark from beam $i$, carrying proton momentum fraction $x_{B,i}$. Further details of the precise implementation of this, and the relation to the standard collinear result are given in Appendix~\ref{app:hybrid}. We note that the same conclusions with respect to the overall gauge independence of the result discussed above in the SD case apply here.

Before concluding this section, a few comments are in order. We first recall that  in the SF approach the $Q^2 > 1$ ${\rm GeV}^2$ continuum component of the SFs, although one could in principle take a direct experimental parameterisation, is more straightforwardly calculated using ZM--VFNS at NNLO in QCD  predictions for the structure functions, in our case using the \texttt{MSHT20qed} NNLO PDF set~\cite{Cridge:2021pxm}. As the PDF set is itself fit to DIS data, most notably from HERA~\cite{Abramowicz:2015mha}, it is important that the order of the PDFs matches the order of the calculation, as this will provide the closest match to the measured SFs, with the NNLO combination being the most precise available. On the other hand, the contribution from the pure PI component of Fig.~\ref{fig:wwfigsd} (a) (and similarly for the DD case) will according to the hybrid approach be calculated using purely LO theory. This therefore will not match the NNLO order of the corresponding PDFs. As will be demonstrated in the following section, out to $Q^2_i \sim 10$ ${\rm GeV}^2$ or more, that is well beyond the transition point \eqref{eq:qwcut}, the pure PI diagrams are strongly dominant and hence in this region we will in effect be including a less precise result for the corresponding prediction. In order to more closely match the experimentally determined SFs in this region, we therefore reweight our parton--level prediction by the NNLO to LO $K$--factor of $F_2(x_{B,i},Q_i^2)$ for the case that the corresponding beam $i$ satisfies \eqref{eq:qwcut}. This will ensure the PI contribution is correctly modelled in the lower $Q_i^2$ region where it dominates. At higher $Q_i^2$ this correction only enters beyond the (leading) order of the calculation and is therefore permissible, though one cannot say whether it provides a more accurate result. 

While we will only calculate the corresponding parton--level diagrams in Figs.~\ref{fig:wwfig} and~\ref{fig:wwfigsd}  at LO in $\alpha_S$ (and $\alpha$), there is nothing preventing a calculation beyond this order being applied. The corresponding process in Fig.~\ref{fig:wwfig} after VBS cuts have been applied (i.e. with the sensitivity to the low $Q_i^2$ removed by these cuts) has indeed been calculated at NLO in QCD in~\cite{Jager:2006zc}. The required results would be the NLO QCD correction to Fig.~\ref{fig:wwfig}, with \eqref{eq:qwcut} applied to both initiating quarks (as well as any further  cuts to e.g. the final--state leptons) and to Fig.~\ref{fig:wwfigsd}, with the same cut applied to the initiating quark. In the latter case the initial--state photon could be treated as on--shell to very good approximation in order to calculate the corresponding $K$--factor straightforwardly. Indeed, all of the above could be calculated using standard off--the--shelf tools, although we leave this to future work. We note that in the lower $Q^2_i$ region where the pure PI contribution is dominant, we already effectively include a calculation up to NNLO QCD precision, via the corresponding SFs as calculated using pQCD. 

We note that for DD $W^+ W^-$ production, we in addition have the contribution from $q q \to q' q' W^+ W^-$ and likewise for antiquark scattering, i.e. due to $W^+ W^- \to W^+ W^-$ scattering, where the flavour of the initial--state and final--state quarks is different. However this has no $1/Q_i^2$ pole structure due to the lack of initial--state photon contributions and hence only the region passing the cut \eqref{eq:qwcut} provides a non--negligible contribution. Hence it can be calculated in the usual manner, at parton--level, and indeed in that ATLAS analysis~\cite{ATLAS:2020iwi} it is accounted for in this way as a background source.

Finally, we note that an alternative approach to the above would be to simply work in standard collinear factorization. At LO we would simply have the on--shell $\gamma \gamma \to W^+ W^-$ process, and therefore the contribution from the additional diagrams in Figs.~\ref{fig:wwfig} and~\ref{fig:wwfigsd} would be absent. Including the impact of these would therefore require going to NLO in $\alpha$ or beyond. Considering the latter DD component, the high $Q^2_i$ contribution, i.e. when \eqref{eq:qwcut} is passed for both $i=1,2$, would effectively be included at the same level of precision as in the hybrid calculation only once one went to NNLO in $\alpha$. The integration down to $Q^2_i \to 0$ would on the other hand result in a collinear singularity due to $q\to \gamma q$ emission from the initial--state quark (or antiquark); as discussed above this is the only form of singularity that occurs at this order once an appropriate rapidity veto is applied. This would be subtracted in the usual way, and matched by the $\overline{\rm MS}$ definition of the corresponding photon PDF that enters the lower order diagrams. The low $Q_i^2$ region would then  effectively be included in the photon PDF, which contains the same experimental inputs for this as in our calculation. However, the combination of subtracted quark--initiated diagram and the on--shell PI diagram is by construction designed in order to match the SF result for the $p \to \gamma X$ vertex to the required level of precision (i.e. NLO wrt to the initial--state photon in this case), as this is precisely how the original LUXqed photon PDF is derived~\cite{Manohar:2016nzj}. Therefore, by applying collinear factorization in the current case, one would effectively be recalculating the full SF result, but at a by definition lower level of precision. Moreover, as we will discuss this would then raise the question of how to treat the survival factor within such an approach.  However, we emphasise again that the above discussion is only intended to apply to the case at hand. For other cases, for example where we do have to deal with initial--state $\gamma\to q\overline{q}$ collinear emission, a calculation within collinear factorization is often in practice to be preferred.

\subsection{$W^+ W^-$ production: results}\label{sec:hybridres}

\begin{figure}[t]
\begin{center}
\includegraphics[scale=0.63]{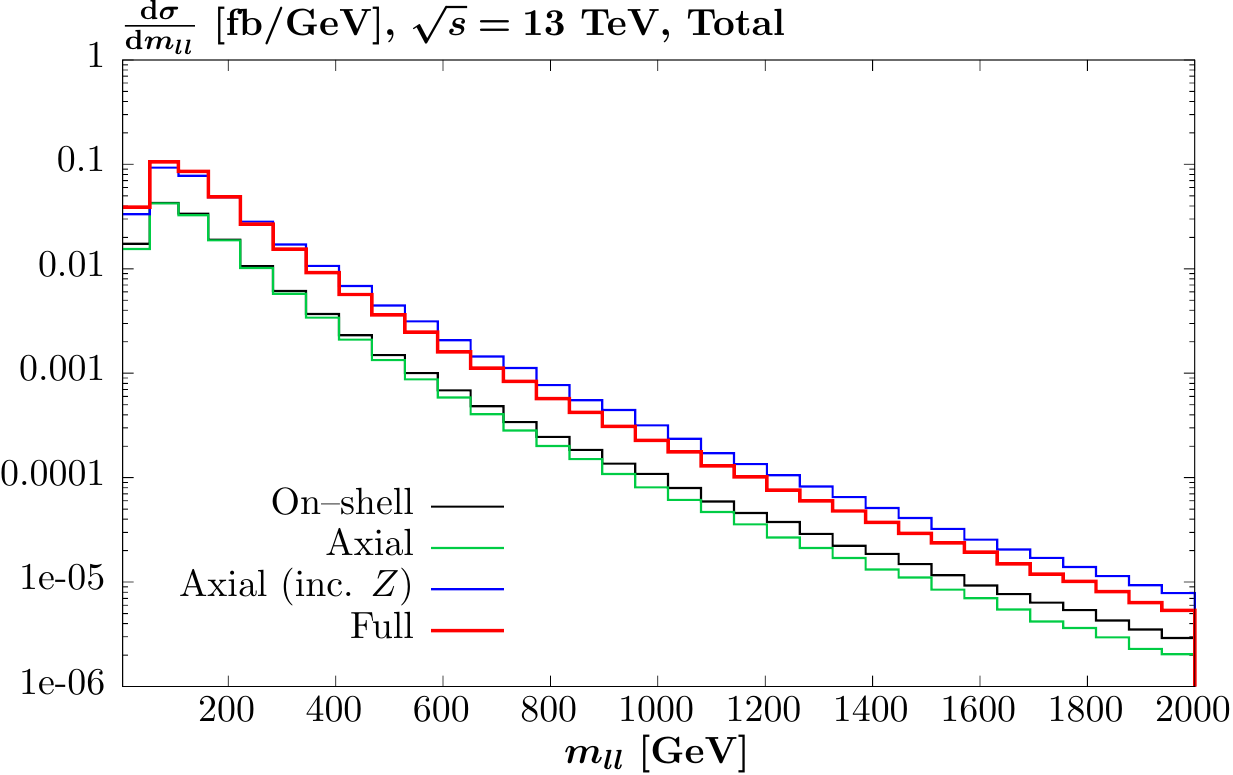}
\includegraphics[scale=0.63]{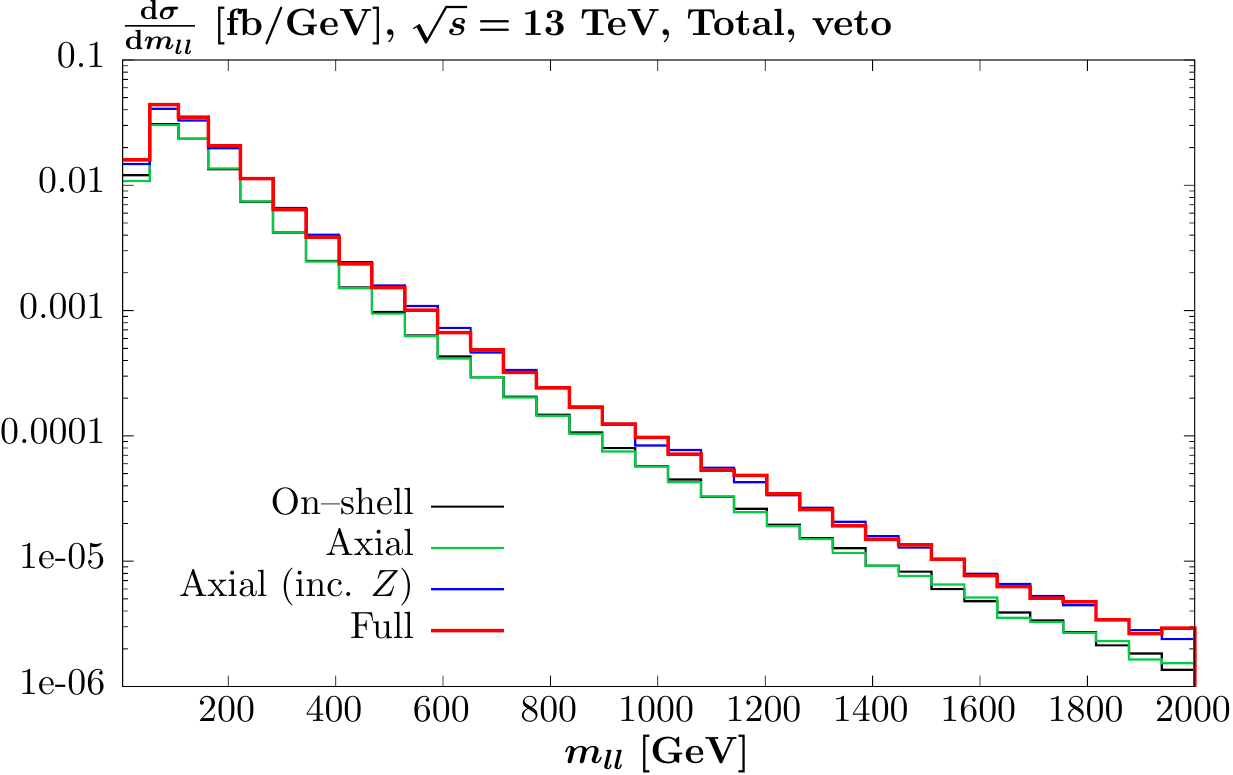}
\includegraphics[scale=0.63]{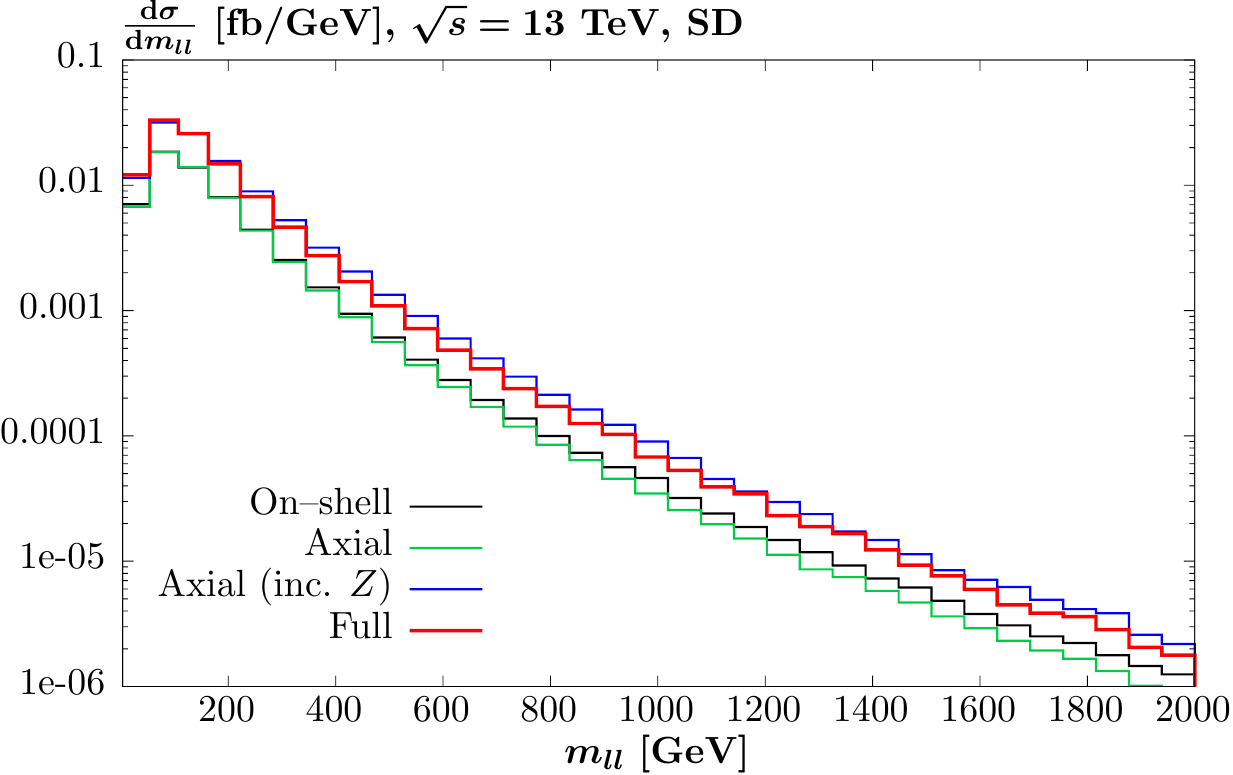}
\includegraphics[scale=0.63]{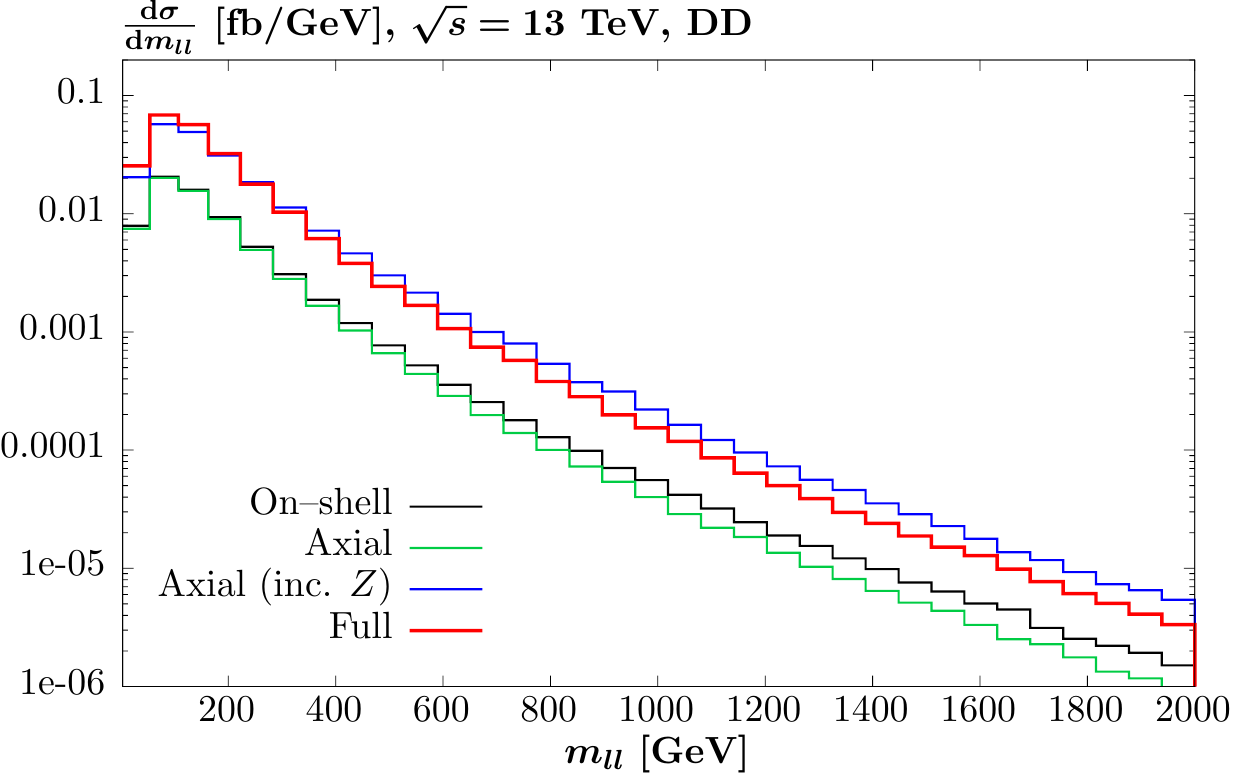}
\caption{Differential cross section with respect to the dilepton invariant mass, $m_{ll}$, for $W^+ W^- \to e^\pm \nu \mu^\mp \nu$ production at the 13 TeV LHC, within the event selection of the ATLAS measurement~\cite{ATLAS:2020iwi}. The top figures show the total cross sections (i.e. the sum of EL., DD and SD), with a veto \eqref{eq:veto} imposed at the parton level (no survival factor included), in the top right figure. The bottom figures show the SD (left) and DD (right) cases, with no veto applied. Results are given for the full (calculated in the hybrid approach), on--shell and SF axial gauge (with and without $Z$--initiated diagrams) cases.}
\label{fig:mll_gcomp}
\end{center}
\end{figure}

In this section we present a selection of results for the hybrid approach described above. We will compare with the on--shell prediction and the SF axial gauge approach for the  illustration; in the latter case we will also now consider the impact of $Z$--initiated production, as calculated in the SF approach. In Fig.~\ref{fig:mll_gcomp} we show the dilepton invariant mass distribution. In the top left plot we consider the case with no veto applied, while the corresponding cross section values are given in Table~\ref{tab:cs_gnvbf}. This is purely shown for illustration, given it is only when a veto is imposed that the VBS--like signal can be effectively isolated. Moreover, as discussed in the previous section we in fact explicitly exclude the contribution from $s$ and $u$--channel diagrams in Fig.~\ref{fig:wwfig}, as these will eventually fail the veto we impose; therefore in the hybrid case we are comparing to the $t$--channel contribution only. This is for directness of comparison but again, for these reasons is only shown for illustration purposes.

As the hybrid calculation includes the contribution from all relevant diagrams, we will often refer to this as the `full' result in what follows and in the figures. We can immediately see that the full result is substantially larger, by over a factor of $2$, than the pure PI contribution in the axial gauge. That is, at the level of the total cross section, with only lepton cuts applied, the kinematic enhancement of the pure PI diagrams (calculated using the axial gauge for the reasons discussed above) is relatively mild. A very similar level of enhancement is observed with respect to the on--shell prediction, which also only includes the PI component; as we would expect, there is good, though not perfect, agreement between the axial gauge result and the on--shell prediction. 

We also show in Table~\ref{tab:cs_gnvbf} the collinear prediction of \eqref{eq:colllo}. The scale variation uncertainties are $\sim 10$ (20) \% for the SD (DD) cases (for elastic production they are below the permille level, and are not shown\footnote{Indeed one can see from \eqref{eq:xfgamma-phys} that the combination of $\alpha(\mu^2) f_\gamma(x,\mu^2)$ is effectively independent of $\mu$ for elastic production. Any  $\mu$ dependence therefore relates to the precise implementation of \eqref{eq:xfgamma-phys} in a PDF fit, and is either numerical in nature or else relates to differences in the  treatment of $\alpha(\mu)$ in the cross section prediction and the order of $P_{\gamma\gamma}$ in the PDF fit. Both effects are very small.}), and are consistent with the on--shell results within these. On the other hand, these systematically undershoot the full result, similarly to the on--shell case, by an amount that is well outside the scale variation band. This is not surprising, given this scale variation effectively only relates directly to the pure PI component show of Fig.~\ref{fig:wwfig}. We note that the difference between the collinear and on--shell predictions in the elastic case is essentially entirely due to the impact of the lepton cuts, that is the fact that only  in the on--shell case are the exact photon kinematics kept track of. For the total cross section without lepton cuts (not shown), the results are extremely close, as they must be by construction. 

The breakdown into SD and DD components is shown in the bottom plots of Fig.~\ref{fig:mll_gcomp}, and qualitatively follows the same trends of the full (i.e. sum of EL. SD and DD) case. As expected, the level of differences between the full and pure PI (axial gauge) results is less in the SD in comparison to the DD (and hence total) case, but we can see that even here it is evident. More precisely, we can see from Table~\ref{tab:cs_gnvbf} that in the SD case the full cross section is just under a factor of $\sim 2$ larger, while for the DD case this is more like a factor of $\sim 3$. 

For elastic production, the results is by construction the same as the axial gauge case, as here we simply apply the latter result. The cross section here is strongly peaked at low photon $Q_i^2$, and hence we expect the result to lie very close to the gauge invariant on--shell case. However, the initial--state photons are not exactly on--shell, and hence some residual gauge dependence will remain. Indeed, we can see from Table~\ref{tab:cs_unitons} that the elastic result in the unitary gauge is $\sim 0.5\%$ higher than in the axial gauge. This difference is small, though not entirely negligible, and would in principle be resolved by including the equivalent of the non--PI diagrams in Fig.~\ref{fig:wwfig} for the elastic case, e.g. due to two--photon contributions with the proton. We leave this rather subtle question to future study, and instead here apply the axial gauge result. This choice is guided by the fact that on general grounds the $Q_i^2/M_{Z,WW}^2$ power counting cannot be applied in the unitary gauge, as discussed in Section~\ref{sec:ax}, and hence this may result in unphysical results when applied to the calculation of the pure PI contribution. In reality, this $\sim 0.5\%$ difference can only be considered as a genuine uncertainty in the result, which in any case enters at a similar level to the experimental uncertainty on the elastic proton form factors. We note that the axial gauge prediction lies $\sim 0.5\%$ above the on--shell result, due to the small but non--negligible impact of non--zero $Q_i^2$ corrections (though again this is of the order of the theoretical uncertainty).

In the top right plot the experimentally more realistic case is shown, that is when the veto \eqref{eq:veto} is imposed at the parton level (with no survival factor included), with the cross section values given in Table~\ref{tab:cs_gvbf2}. This  reduces the level of difference observed between the full and axial gauge results, as we would expect: the impact of the veto is to  reduce the contribution from the higher $Q_i^2$ region, and hence enhance the pure PI component. Nonetheless, we can see that even so this difference is non--negligible. From the table we can see that for  SD (DD) production this is at the level of a factor of $\sim 1.5$ (2), in comparison to $\sim 2$ (3) for the case without a veto. The level of reduction is rather larger for the DD cross section, as we would expect given the impact of the veto is larger there. In particular, for the purely elastic case, and hence for the elastic emission in the SD case, all events pass the parton--level veto.

\renewcommand{\arraystretch}{1.5}
\begin{table}[t]
\begin{center}
\begin{tabular}{|c|c|c|c|c|c|}
\hline
$\sigma$ [fb]  &On--shell  & Collinear & Axial &Axial (inc. $Z$) & Full \\
\hline 
EL &0.696& 0.713& 0.701&0.701& 0.701\\
\hline
SD &3.31  & 3.73${}^{+0.40}_{-0.41}$& 3.25&6.11& 6.00\\
\hline
DD &3.81  & 4.71${}^{+1.07}_{-0.95}$& 3.64&11.9& 13.1\\
\hline
Total &7.82  & 9.15${}^{+1.47}_{-1.36}$& 7.59&18.7& 19.8\\
\hline
\end{tabular}
\end{center}
\caption{Cross sections (in fb) for $W^+ W^-$ production at $\sqrt{s}=13$ TeV, as described in the text. Lepton cuts \eqref{eq:atcuts} applied. Scale variation uncertainty given for collinear SD and DD predictions (for the EL case these are below the quote level of precision so are omitted), but otherwise central values shown only. `Total' corresponds to sum of EL, SD and DD. On--shell and axial gauge numbers are as in Tables~\ref{tab:cs_unitons} and~\ref{tab:cs_axons}, respectively, and are repeated for comparison. The EL cross section for the axial (inc. $Z$) case is by construction the same as the pure axial gauge result.}  \label{tab:cs_gnvbf}
\end{table}

\renewcommand{\arraystretch}{1.5}
\begin{table}[t]
\begin{center}
\begin{tabular}{|c|c|c|c|c|}
\hline
$\sigma$ [fb]   &On--shell  &Axial &Axial (inc. $Z$) & Full \\
\hline 
EL &0.696&  0.701&0.701& 0.701\\
\hline
SD &2.53  &  2.52&3.27& 3.39\\
\hline
DD &2.30  & 2.26&3.80& 4.04\\
\hline
Total &5.53  & 5.48&7.77& 8.13\\
\hline
\end{tabular}
\end{center}
\caption{As in Table~\ref{tab:cs_gnvbf}, but with veto \eqref{eq:veto} imposed at the parton level (no survival factor included).  On--shell and axial gauge numbers are as in Tables~\ref{tab:cs_unitons} and~\ref{tab:cs_axons}, respectively, and are repeated for comparison.}  \label{tab:cs_gvbf2}
\end{table}

The impact of non--PI production diagrams is therefore clearly significant, despite the $\sim 1/Q_i^2$ enhancement in the PI case. To analyse this question in further detail, it  is interesting to  consider the axial gauge result, but now also including the $Z$--initiated contributions as in Figs.~\ref{fig:wwfig} and~\ref{fig:wwfigsd} (a) to the pure VBS diagrams  (i.e. $\gamma/Z \gamma/Z \to W^+ W^-$ with off--shell bosons in the initial--state), including the $s$--channel contribution from the Higgs boson. These are suppressed by (powers of) $\sim Q_i^2/M_Z^2$ with respect to the PI case at the integrand level, the impact of which is logarithmic after the phase space integration is performed; numerically this leads to roughly up to an order of magnitude suppression for each beam $i=1,2$, with the precise amount depending on the kinematics and cuts applied (the suppression is in particular rather less in the absence of the parton--level veto). On the other hand, the $Z WW$ coupling is $g_W \cos\theta_W$, which is  enhanced in comparison to the $\gamma WW$ case by a factor of $\cos\theta_W /\sin \theta_W \sim 2$ at the amplitude level. Moreover, we can see that the additional diagrams in Figs.~\ref{fig:wwfig} and~\ref{fig:wwfigsd} (b) onwards are expected to be more strongly suppressed kinematically, i.e by powers of $\sim Q_i^2/M_{WW}^2$. We may therefore expect these $Z$--initiated contribution to provide the dominant non--PI contribution. Such a comparison can as always only be performed in the axial gauge, due to the unitarity breaking effects that are present in the unitary gauge which will spoil the power counting arguments above.

Results for  $\gamma/Z \gamma/Z \to W^+ W^-$  production in the axial gauge are also given in Fig.~\ref{fig:mll_gcomp} and Tables~\ref{tab:cs_gnvbf} and~\ref{tab:cs_gvbf2}. We can see that indeed the matching with the full calculation is greatly improved, with agreement reached at the $10\%$ level or less in the tabulated cross sections. Therefore, once we work in an appropriate gauge, we do indeed find that in all cases the dominant contribution comes from pure  $\gamma/Z \gamma/Z \to W^+ W^-$  production. On the other hand, the agreement with the full result is not perfect, and indeed in some kinematic regions (e.g. at larger $Q_i^2$) will be expected to deteriorate further. This is clear from Fig.~\ref{fig:mll_gcomp} for the DD case when no veto is imposed, where the difference is larger as $m_{ll}$ increases. On the other hand, once the phenomenologically relevant case with a veto impose is considered, we can see that the agreement is very good at the level of the total (sum of EL, SD and DD) cross sections across the entire $m_{ll}$ region. Even so, we consider the hybrid calculation as the more complete one, and so take this in our MC implementation.

 We note that the $\sim 10\%$ difference between the axial gauge ($\gamma + Z$) and the hybrid results cannot be completely associated to the contribution from the additional diagrams in Figs.~\ref{fig:wwfig} and~\ref{fig:wwfigsd}, as the contribution from diagrams (a) are calculated at LO parton--level in the hybrid result, but at NNLO level in the SF calculation. The effect of this is considered further in Section~\ref{sec:theorunc}, and may lead to $\sim 2 (5) \%$ level differences in the SD (DD) cases. Nonetheless, the dominant impact of this is accounted for following the procedure discussed in Section~\ref{sec:hybrid}, and hence the real size of the effect will be smaller than this. We note that in~\cite{Accomando:2006mc},  the  $W^+ W^- \to W^+ W^-$ (and related contribution not due to pure VBS) case was considered, and here the axial gauge result lies somewhat above the full calculation. The underlying $W$--initiated process in this case is rather distinct, given it features no $1/Q_i^2$ pole structure in the VBS diagrams.

\begin{figure}[t]
\begin{center}
\includegraphics[scale=0.63]{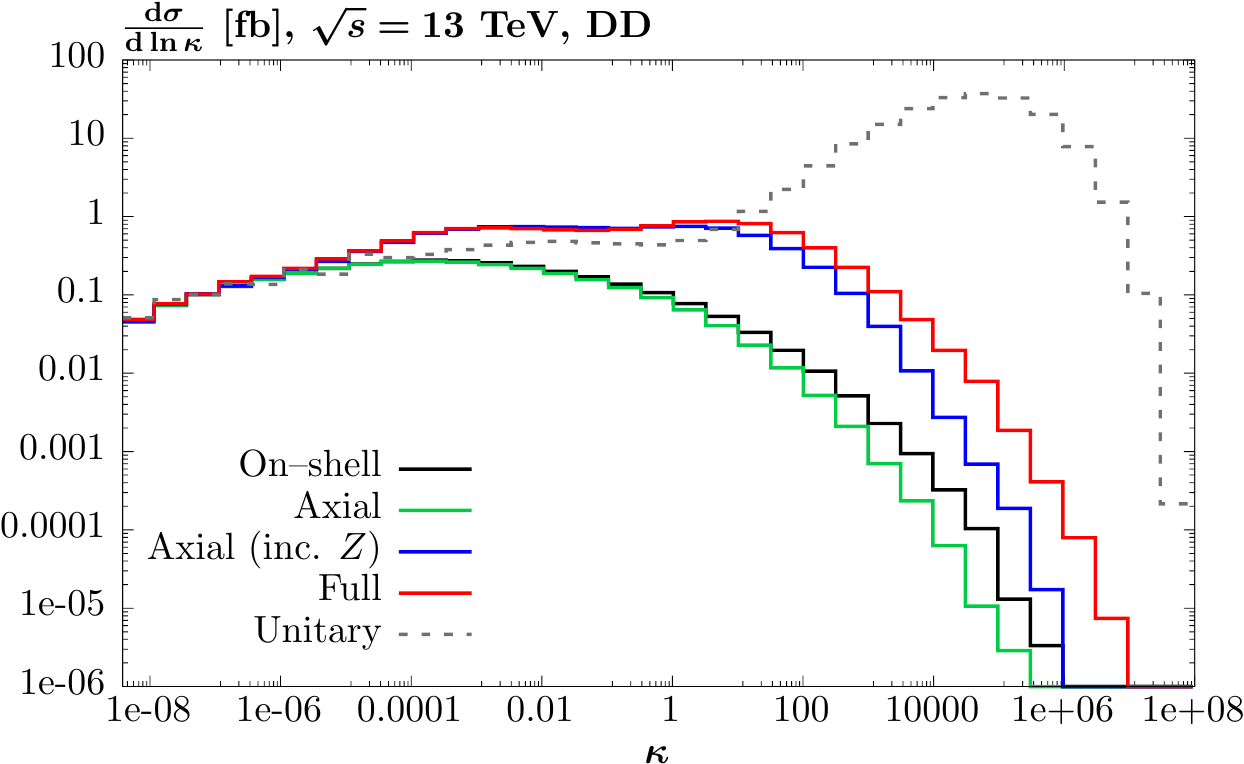}
\includegraphics[scale=0.63]{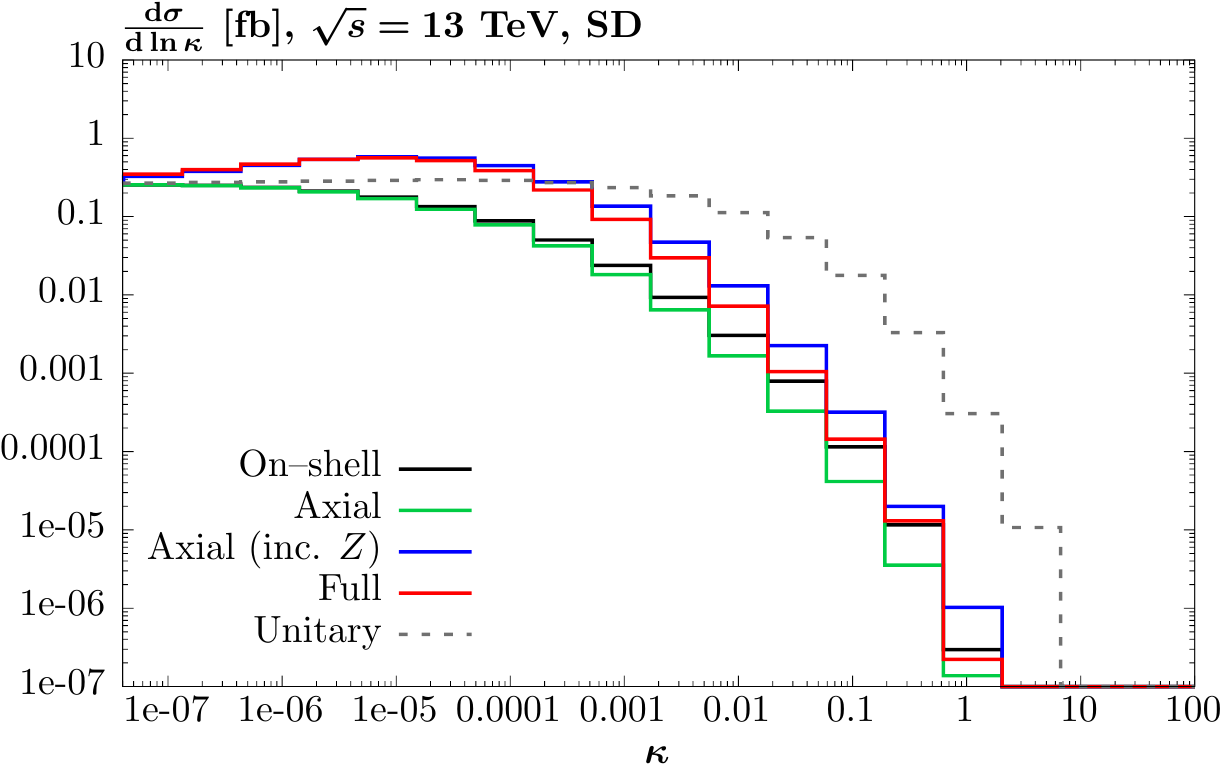}
\includegraphics[scale=0.63]{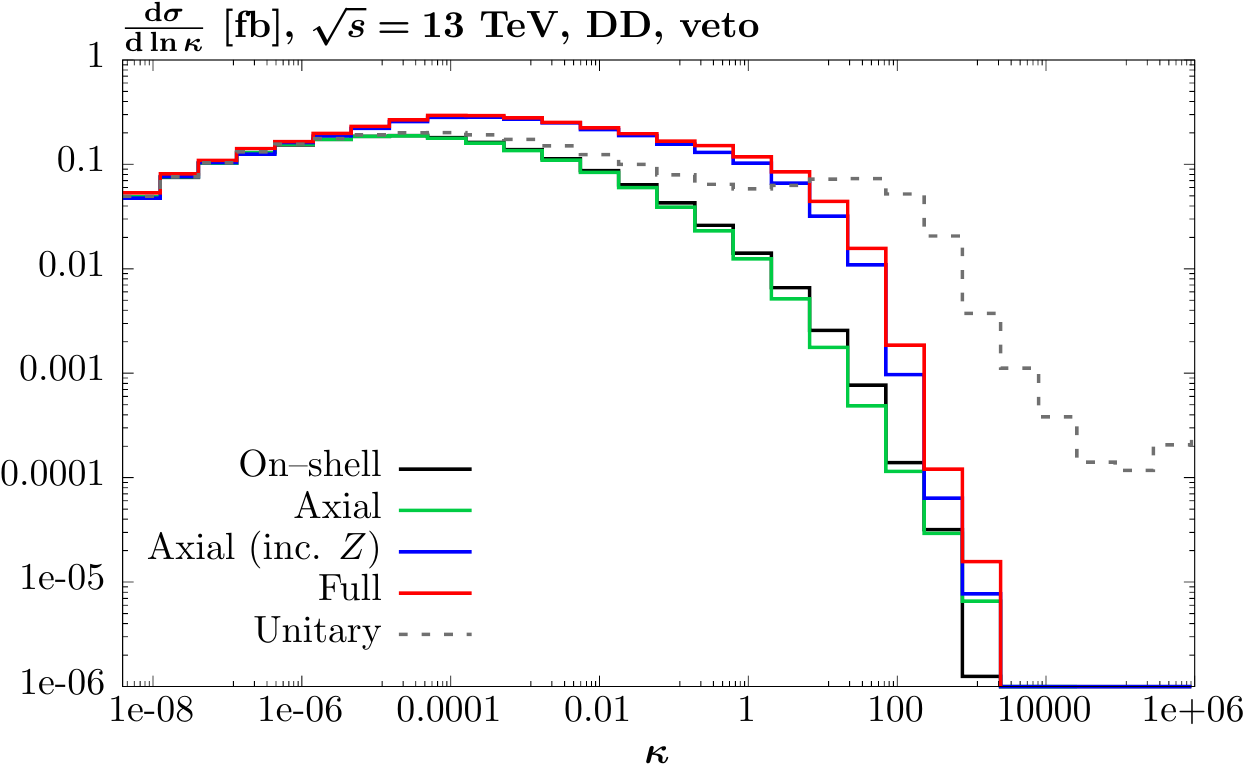}
\includegraphics[scale=0.63]{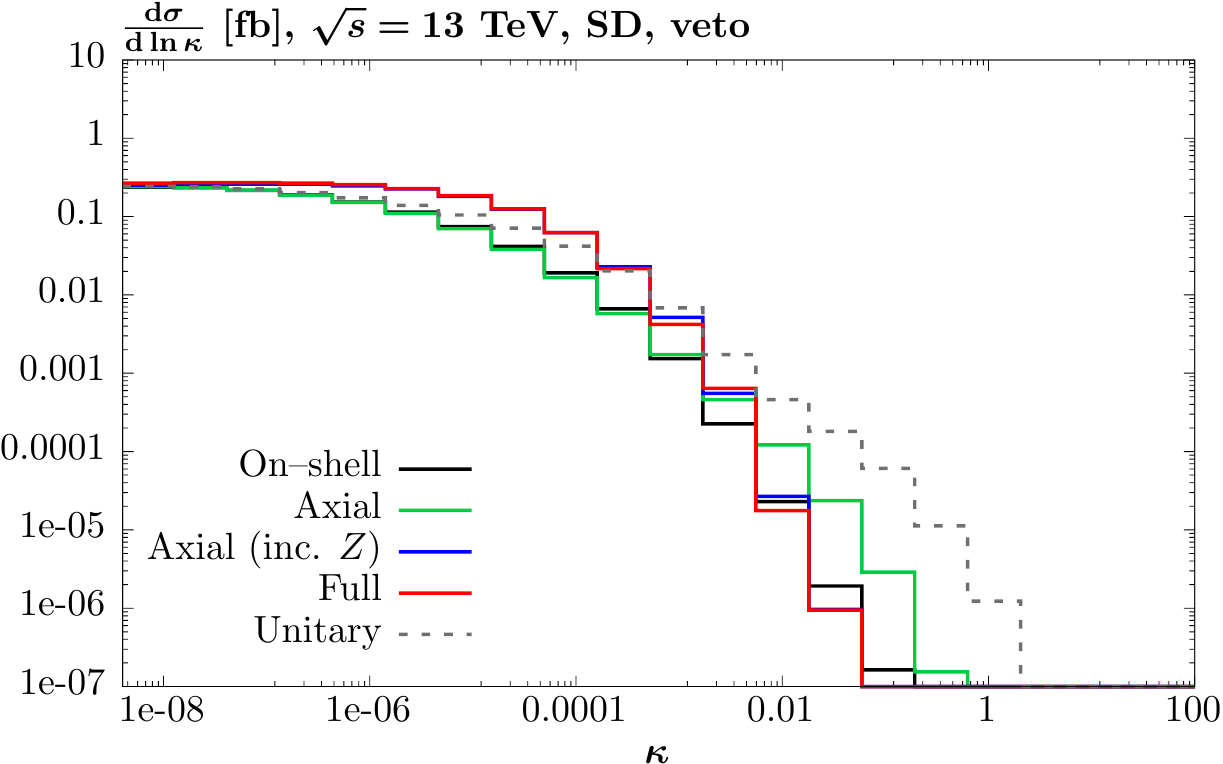}
\caption{Differential cross section with respect to $\ln(\kappa) = \ln(Q_1^2 Q_2^2/M_Z^4)$, for $W^+ W^- \to e^\pm \nu \mu^\mp \nu$ production at the 13 TeV LHC, within the event selection of the ATLAS measurement~\cite{ATLAS:2020iwi}. The left (right) figures show the case of DD (SD) production, while the top (bottom) cases are with (without) a veto \eqref{eq:veto} imposed at the parton level (no survival factor included).}
\label{fig:lnk_gcomp}
\end{center}
\end{figure}

To understand the above results better, in Fig.~\ref{fig:lnk_gcomp} we show distributions with respect to 
\be
\kappa = \frac{Q_1^2 Q_2^2}{M_Z^4}\;.
\ee
This is clearly not an observable quantity, but is illustrative from the point of demonstrating the impact of the non--PI diagrams with the photon $Q_i^2$. The normalization with respect to $M_Z$ is applied in order to define a dimensionless quantity, and because from the discussion above we know that the factor of $Q_i^2/M_Z^2$ is a relevant ratio when defining the impact of corrections away from the on--shell limit. We could alternatively have normalized with respect to $M_{WW}$, but find this lead to somewhat less transparent results, as it is less straightforward to identity the regions of low and high $Q_i^2$ unambiguously. We plot ${\rm d}\sigma/{\rm d}\ln \kappa$, such that the contribution to the cross section in each decile is the same.

The DD (SD) case is shown in the left (right) plots and without (with) the usual parton--level veto applied in the top (bottom) plots. Considering the DD case first, we can see that out to $\kappa \lesssim 10^{-6}$, there is very close agreement between the SF results in both axial and unitary gauges, as well as the on--shell and full results. In this region, the photon $Q_i^2$ is therefore sufficiently small that the contribution from non--PI diagrams is strongly kinematically suppressed, even in the unitary gauge. This results applies irrespective of whether a parton--level veto is applied or not, although the agreement is  pushed to slightly higher values of $\kappa$ when the veto is applied. If we assume that $Q_1^2=Q_2^2$, then this corresponds to $Q_i^2 \lesssim 10$ ${\rm GeV}^2$, while for unequal values of $Q_i^2$, the upper limit on the larger value will be above $\sim 10$ ${\rm GeV}^2$. This is therefore  indeed well beyond the transition point applied in \eqref{eq:qwcut}, and hence as we argued above we expect the corresponding transition as this point between the SF and full result to be smooth. This is in addition demonstrates that in the low $Q_i^2$ (i.e. $\kappa$) region the result is largely gauge independent.

In further detail, in the DD case we can see that SF axial gauge and on--shell results have a very similar scaling with $\kappa$, out to $\kappa \sim 1$ (i.e. $Q_i^2 \sim M_Z^2$), where some deviation is observed, as we might expect. The full result on the other hand, begins to deviate from these beyond 
$\kappa \lesssim 10^{-6}$, and is roughly an order of magnitude higher by $\kappa \sim 1$. This is again roughly independent of whether a parton--level veto is applied or not, although we can see that the contribution from the cross section in the $\kappa \gtrsim 1$ region is suppressed by this veto.  This is entirely as expected, given in this regime there is no kinematic enhancement of the pure PI diagrams and hence no justification for omitting the other contributions. On the other hand, once $Z$--initiated production is included in the axial gauge,  the trend observed in the full case is  rather closely followed out to $\kappa \sim 1-10$. This again highlights the fact that once these diagrams are included the dominant non--PI contribution is accounted for, with the remaining difference entering at higher $\kappa$, where the kinematic suppression in the contribution from Figs.~\ref{fig:wwfig} (b) onwards is no longer present. The unitary gauge result is also shown for comparison, and the strongly unphysical behaviour at large $\kappa$ is clear.

For the SD case, shown in the right hand plots, the basic qualitative trends described above remain. More precisely, we can see that the distributions are peaked at lower values of $\kappa$, as we would expect given this occurs with an initiating elastic photon of rather low $Q_i^2$. The full result closely follows that of the axial or on--shell case out to $\kappa \lesssim 10^{-7}$, i.e. an inelastic $Q_i^2 \lesssim 10$ ${\rm GeV}^2$ (assuming the elastic photon has $Q_i^2 \lesssim 0.5 \,{\rm GeV}^2$). Therefore, this again is well beyond the transition point applied in \eqref{eq:qwcut}. Once again, we see a significantly improved matching once $Z$--initiated production is included in the axial gauge. The behaviour of the unitary gauge PI prediction, again shown for illustration, is less marked than in the DD case but  nonetheless displays some level of unphysical enhancement, in particular in the absence of the parton--level veto.

\subsection{VBS cuts: a comparison}\label{sec:vbs}

\begin{figure}
\begin{center}
\includegraphics[scale=0.63]{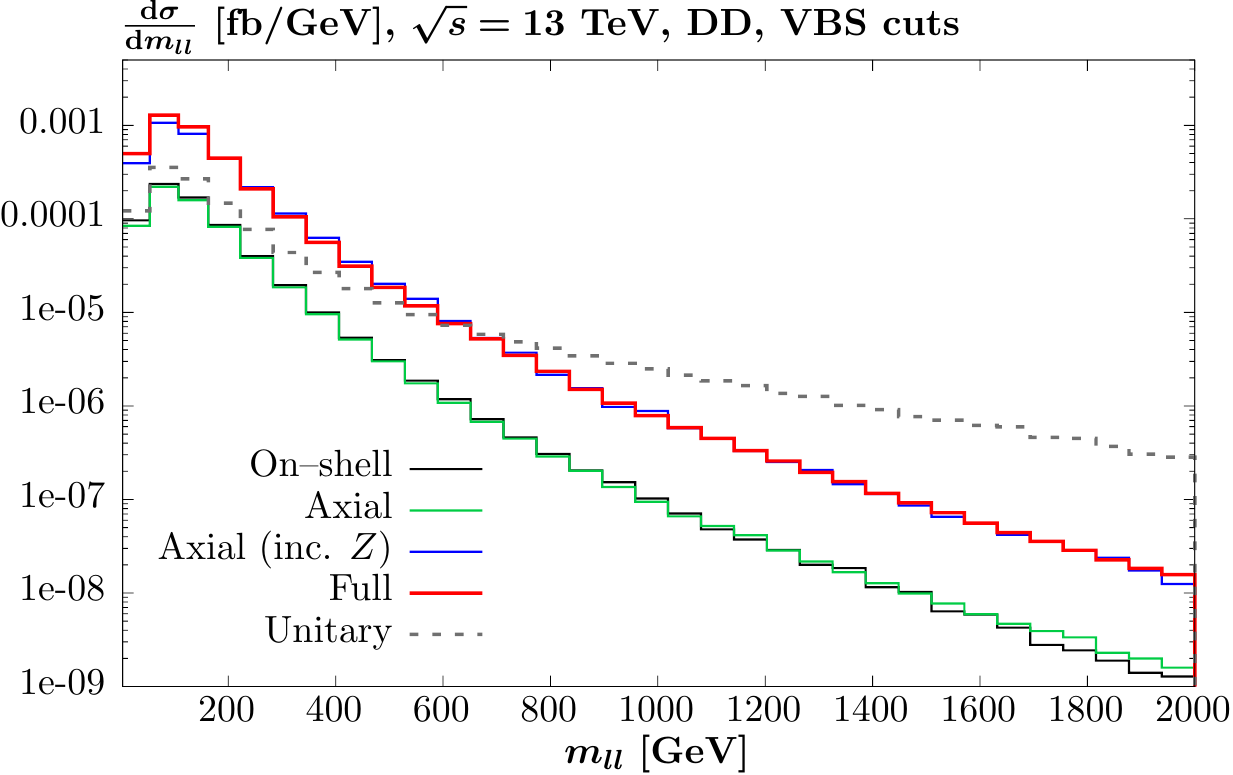}
\caption{Differential cross section with respect to the dilepton invariant mass, $m_{ll}$, for $W^+ W^- \to e^\pm \nu \mu^\mp \nu$ production at the 13 TeV LHC, with VBS cuts applied as described in the text.}
\label{fig:mll_gcomp_vbf1}
\end{center}
\end{figure}

\renewcommand{\arraystretch}{1.5}
\begin{table}
\begin{center}
\begin{tabular}{|c|c|c|c|c|c|}
\hline
 &On--shell  &Axial &Axial (inc. $Z$) &Axial (inc. $Z$), LO SF & Full \\
\hline 
$\sigma$ [fb] &0.037  & 0.035&0.179& 0.195 & 0.205\\
\hline
\end{tabular}
\end{center}
\caption{As in Table~\ref{tab:cs_gnvbf}, but with VBS cuts, described in text, applied. Note in this case only the DD contribution is non--zero, hence this is shown only.}  \label{tab:cs_gvbf1}
\end{table}

As a brief aside, it is interesting to consider the implications of the above discussion for the case when VBS cuts are applied, that is when two jets sufficiently separated in rapidity are required to be present in the detector. To be precise, we apply the cuts described in~\cite{Jager:2006zc} at parton--level, namely we require the two tagging jets (which at our LO level are just the outgoing quark/antiquarks) to have
\begin{align}
p_{j_\perp} &\geq 20\,{\rm GeV}\;,\qquad |y_j| \leq 4.5\;,\\
\Delta y_{jj} &> 4\;,\qquad y_{j_1} \cdot y_{j_2} < 0\;,\qquad M_{jj} > 600\,{\rm GeV}\;.
\end{align}
The same lepton cuts as in \eqref{eq:atcuts} are applied, but we in addition require that
\be
\Delta R_{jl} \geq 0.4\;,\qquad y_{j,{\rm min}} < \eta_l < y_{j,{\rm max}}\;. 
\ee
Results are shown in Fig.~\ref{fig:mll_gcomp_vbf1} and Table~\ref{tab:cs_gvbf1}. We note that the VBS cuts now imply that only the DD contribution is present, and  there are no subtleties related to the treatment of the low $Q_i^2$ region as in the previous case. We can see that very similar trends are observed to those discussed above. Namely, the unitary gauge SF result shows as expected an unphysical growth with invariant mass. This is tamed by working in the axial gauge (or on--shell calculation), but here the result lies significantly below the full calculation. This is as we would expect, given that the VBS cuts  require somewhat larger $Q_i^2$ values in order for the tagging jets to be registered (although the $m_{jj}$ and $\Delta y_{jj}$ requirements impose upper limits on these). Interestingly, once the $Z$--initiated contributions are included, the axial gauge SF result again lies rather close to the full calculation. Moreover, if we instead use purely LO SFs, i.e. to   match the treatment of the full result (which we recall is at LO parton level), then the agreement is improved further. The remaining difference is then purely due to the impact of the additional diagrams, other than the PI and $Z$--initiated. Clearly, for phenomenological applications one can and should apply the full calculation. However, in principle this might provide some guidance as to the potential impact of higher order (NNLO...) corrections in the full case, given these are particularly simple for the SF calculation.

\section{Lepton pair production revisited}\label{sec:lep}

\begin{figure}
\begin{center}
\subfigure[]{\includegraphics[scale=0.6]{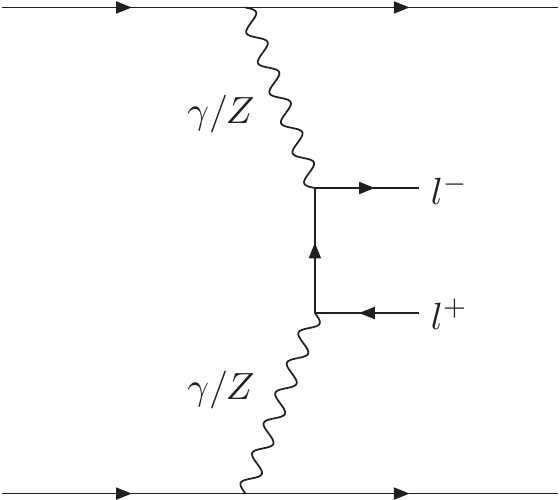}}\quad
\subfigure[]{\includegraphics[scale=0.6]{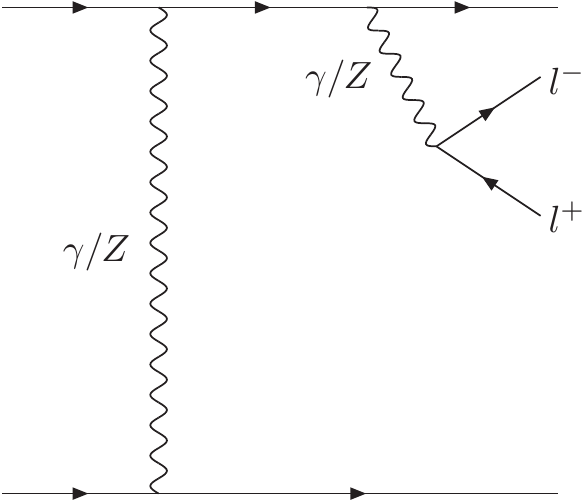}}\quad
\caption{Classes of Feynman diagrams contributing to $l^+ l^-$ DD production at LO in the $qq \to l^+ l^- qq$ process, with various permutations implied.  }
\label{fig:llfig}
\end{center}
\end{figure}

Given the issues raised above, it is worth revisiting the predictions of~\cite{Harland-Lang:2020veo} for lepton pair production. In this case, it has been explicitly demonstrated in~\cite{Harland-Lang:2021zvr} that the pure PI contribution, as calculated within the SF approach, provides the strongly dominant contribution away from the $Z$ peak region (see Fig. [2] of that paper). This is in particular true of the PI contribution to inclusive lepton pair production, but as discussed above once we impose a rapidity veto the $s$--channel DY topology will be strongly suppressed even on the $Z$ peak, and hence we can expect the PI contribution to dominate across the entire phase space.

Nonetheless, there are additional contributions as well as the pure PI one shown in Fig.~\ref{fig:llfig} (a), i.e. due to $Z$--initiated production and direct emission from the quark lines. These are included in the \texttt{SFGen} MC, described in~\cite{Harland-Lang:2021zvr}, following an approach that is similar though not identical to the hybrid calculation described above; due to simpler set of additional diagrams that contribute here, these can be included more straightforwardly. In Table~\ref{tab:lpair} we therefore use this to present predictions in a similar kinematic regime to the $W^+ W^-$ case, with in particular
\be\label{eq:lcuts}
|\eta_l| < 2.5\;,\quad p_{l_\perp} > 20\, {\rm GeV}\;, \quad m_{ll} > 2 \,m_{WW}\;.
\ee
This will in particular be relevant when comparing to the ATLAS 13 TeV data on semi--exclusive $W^+ W^-$ production~\cite{ATLAS:2020iwi}, which we will present in Section~\ref{sec:ATLAS}. We note that the precise $p_{l_\perp}$ cut  differs somewhat from the $W^+ W^-$ case, and is chosen so as to match the fiducial region given in~\cite{Aaboud:2016dkv}. However, the results do not depend sensitively on this specific choice.

For comparison, we show results with and without the veto~\eqref{eq:veto} imposed at parton--level, i.e. with no survival factor included. The PI results are calculated using the SF approach~\eqref{eq:sighh}, while the non--PI contributions are included via the calculation of~\cite{Harland-Lang:2021zvr}, that is at LO parton level. We note that in this case the pure PI component is individually gauge invariant, a fact that follows upon considering the distinct scaling of the PI contribution with the fractional charge of the corresponding quark lines. Without the veto imposed, the inclusion of non--PI diagrams leads to a $\sim 5\%$ (8\%) increase in the SD (DD) cases. However, this is significantly reduced in the  phenomenologically relevant case, with a veto, to $\sim 1\%$ (2\%). This is qualitatively as we would expect: by imposing a rapidity veto we reduce the impact from the larger $Q_i^2$ region, where these non--PI diagrams are more significant. For comparison, we also show the same results but with the rather high cut of $m_{ll} > 1$ TeV imposed, in order to evaluate the $m_{ll}$ dependence of this result. At these higher masses, the contribution from the diagrams of type Fig.~\ref{fig:llfig} (b) is  negligible, and so the enhancement is essentially entirely due to the inclusion of $Z$--initiated diagrams a in Fig.~\ref{fig:llfig} (a); at lower masses both play a role.

\renewcommand{\arraystretch}{1.5}
\begin{table}
\begin{center}
\begin{tabular}{|c|c|c|c|c|}
\hline
$\sigma_{\gamma\gamma}/\sigma_{\rm tot}$& \multicolumn{2}{c|}{No Veto}& \multicolumn{2}{c|}{Veto} \\
\cline{2-5}
 &SD  &DD & SD & DD  \\
\hline 
$m_{ll} > 2 \,m_{WW}$&1.052  & 1.083&1.008 &1.018\\
\hline
$m_{ll} > 1$ TeV &1.089  & 1.165&1.019 & 1.040\\
\hline
\end{tabular}
\end{center}
\caption{Ratio of the total (including $Z$--initiated production and $\gamma/Z \to l^+ l^-$ emission from the quark lines) to PI cross section at 13 TeV. Results are shown for the lepton cuts \eqref{eq:lcuts} applied, and with two different cuts on the dilepton invariant mass. Results with and without a veto \eqref{eq:veto} imposed at the parton level (no survival factor included) are shown.}  \label{tab:lpair}
\end{table}

Therefore, we can expect that once an experimentally realistic rapidity veto is imposed, the contribution from non--PI production will be at the percent level or less. This is to be contrasted with the results of Table~\ref{tab:cs_gvbf2} for $W^+ W^-$ production, where the  appropriate comparison is between the axial (or on--shell)  and the `full' results. There, the inclusion of non--PI diagrams lead to a $\sim 40\%$ (80\%) increase in the SD (DD) cases. This is over an order of magnitude larger relative increase in comparison to the lepton case. This will be  due in part to the  different class of additional diagrams, as per Fig.~\ref{fig:wwfig}, that enter in addition to the PI case, but also to a significant extent due to the relative impact of $Z$--initiated production in the two cases. In particular, while as discussed above  the $ZWW$ vertex  is enhanced by a factor $\cos\theta_W /\sin \theta_W \sim 2$ relative to the $\gamma WW$ case, for the $Zll$ vertex the coupling is instead $\sim g_W a_l/(2 \cos\theta_W)$ (recalling that $ v_l \ll a_l$), and hence the corresponding factor is $\sim a_l/(2\cos\theta_W \sin \theta_W)\sim 0.6$. It is therefore natural to expect the impact of $Z$--initiated production to be significantly larger in the case of $W^+ W^-$ production, and this is precisely what we observe.

The above results are however only produced with an approximate parton--level veto imposed, and without the survival factor included. We will  address these points in the following sections. 

\section{Soft survival effects}\label{sec:S2}

In the previous sections we have at various points considered the impact of a veto \eqref{eq:veto} on any additional charged tracks above a (low) $p_\perp$ threshold within the detector acceptance. As discussed in the introduction, this is a possible way (alternative to standard VBS cuts) to isolate the VBS signal, and suppress the  $s$--channel component. More precisely, one may hope in this way to enhance the pure PI signal. Indeed in the ATLAS measurement of semi--exclusive $W^+ W^-$ production~\cite{ATLAS:2020iwi} precisely such a veto is applied. 
 
However, we have so far only considered the impact of this veto effectively on the proton dissociation system; this was applied in an approximate way above, although we will discuss how this can be done more precisely at the MC level in the following section. However, before doing that we must deal with a separate important issue. Namely, that as well as being produced directly due to  proton dissociation (i.e. the outgoing quarks in the LO parton--level picture), we must also account for the fact that additional particles can be produced by  soft proton--proton interactions, independent of the $W^+ W^-$ production process considered so far. Such MPI activity can then lead to particle production in the veto region, and must be taken account of if a reliable evaluation of the cross section in the presence of such a veto is to be achieved.

We note that in the absence of MPI there is no colour flow between the colliding protons for the VBS $W^+ W^-$ production process  we consider. We are in particular only interested in this scenario, in order to satisfy the experimental rapidity veto that is imposed\footnote{We recall that we explicitly only include the $t$--channel diagrams in Fig.~\ref{fig:wwfig} for which this is the case, precisely in order to pass the experimental veto.}. As discussed in e.g.~\cite{Khoze:2010by}, once we allow MPI activity and therefore colour flow between the proton beams, the probability of producing a rapidity gap of sufficient size, and with sufficiently low $p_\perp$ threshold, due to fluctuations in the underlying event activity alone are extremely small. Hence to very good approximation we are simply interested in including the probability that no additional particles are produced by accompanying soft proton--proton interactions. This is known as the soft `survival factor', see e.g.~\cite{Harland-Lang:2014lxa} for further discussion and references. Further discussion of this point can be found in~\cite{Harland-Lang:2020veo}.

To account for this, we follow the approach described in~\cite{Harland-Lang:2016apc,Harland-Lang:2020veo}, to which we refer the reader for further details.  The key point is that the survival factor is not a universal, multiplicative constant. In other words, it is not fully factorized from the underlying production process, but rather depends sensitively on it. Broadly speaking, the survival factor is sensitive to the impact parameter of the colliding protons: as the average proton--proton impact parameter is increased, we should expect the probability for additional particle production to be lower, and so for the survival factor to be closer to unity. More specifically, the proton--proton impact parameter $b_{\perp}$ is related to the Fourier conjugate of the momenta $q_{i_\perp}$ as defined in \eqref{eq:sighhf}, which in the PI case corresponds to the photon transverse momenta and more generally are associated with the momentum transfer to the protons. Hence we expect the survival factor to depend on the differential distribution with respect to these variables. This leads to the well known effect (see e.g.~\cite{Harland-Lang:2016apc,Harland-Lang:2020veo,Harland-Lang:2021ysd}) that for elastic PI production, the survival factor is very close to unity, due to the strongly peaked nature of the elastic proton form factors towards small $Q_i^2$, and hence $q_{i_\perp}^2$, i.e. large $b_{i_\perp}$. Essentially, due to the long--range nature of the elastic photon--proton interaction, the collision is dominantly outside the range of QCD and hence additional MPI effects. As discussed in~\cite{Harland-Lang:2020veo} this persists to a large extent for SD production, but is less apparent in the DD case.

In more detail, the average survival factor in impact parameter space is given by\footnote{For simplicity we consider here a simplified `one--channel' model, which ignores any internal structure of the proton; see~\cite{Khoze:2013dha} for discussion of how this can be generalised to the more realistic `mutli--channel' case, which we apply in all calculations.}
\begin{equation}\label{S2}
\langle S^2 \rangle=\frac{\int {\rm d}^2 b_{1_\perp}\,{\rm d}^2 b_{2_\perp}\, |T(b_{1_\perp},b_{2_\perp})|^2\,{\rm exp}(-\Omega(s,b_\perp))}{\int {\rm d}^2\, b_{1_\perp}{\rm d}^2 b_{2_\perp}\, |T(b_{1_\perp}, b_{2_\perp})|^2}\;,
\end{equation}
where $b_{i_\perp}$ is the impact parameter vector of proton $i$, so that $b_\perp=b_{1_\perp}+ b_{2_\perp}$ corresponds to the transverse separation between the colliding protons.   $\Omega(s,b_\perp)$ is the proton opacity, which can be extracted from such hadronic observables as the elastic and total cross sections as well as, combined with some additional physical assumption about the composition of the proton, the single and double diffractive cross sections. Physically $\exp(-\Omega(s,b_\perp))$ represents the probability that no inelastic scattering occurs at impact parameter $b_\perp$.

In the above expression $T(b_{1_\perp},b_{2_\perp})$ is the Fourier transform of the hadron--level production amplitude:
  \begin{equation}\label{Mfor}
T(q_{1_\perp},q_{2_\perp})=\int {\rm d}^2 b_{1_\perp}\,{\rm d}^2 b_{2_\perp}\,e^{-i (q_{1_\perp} b_{1_\perp})}e^{+i (q_{2_\perp}  b_{2_\perp})}T(b_{1_\perp}, b_{2_\perp})\;.
\end{equation}
In particular, when the SF calculation applied, $T(q_{1_\perp},q_{2_\perp})$ corresponds to the hadron--level production amplitude that enters the cross section as in \eqref{eq:sighhf}; we have omitted the additional arguments of these amplitudes (on $s$, $x_{B,i}$ etc) for brevity. With this, it is straightforward to show that \eqref{S2} can be written as
\begin{equation}\label{seikav}
\langle S^2\rangle= \frac{\int {\rm d}^2 q_{1_\perp}\,{\rm d}^2 q_{2_\perp}\,|T(q_{1_\perp},q_{2_\perp})+T^{\rm res}(q_{1_\perp},q_{2_\perp})|^2}{\int {\rm d}^2q_{1_\perp}\,{\rm d}^2q_{2_\perp}\,|T(q_{1_\perp},q_{2_\perp})|^2}\;,
\end{equation}
where $T^{\rm res}$ includes the `rescattering' effect of potential proton--proton interactions, and is given in terms of the original amplitude $T$ and the elastic proton--proton scattering amplitude, with
\begin{equation}\label{skt}
T^{\rm res}({q}_{1_\perp},{q}_{2_\perp}) = \frac{i}{s} \int\frac{{\rm d}^2  {k}_\perp}{8\pi^2} \;T_{\rm el}(k_\perp^2) \;T({q}_{1_\perp}',{q}_{2_\perp}')\;,
\end{equation}
where $q_{1_\perp}=q_{1_\perp}'+k_\perp$ and $q_{2_\perp}'=q_{1_\perp}-k_\perp$. The elastic amplitude itself can be written in impact parameter space in terms of the probability $\exp(-\Omega(s,b_\perp))$ given above, such that \eqref{skt} is indeed equivalent to \eqref{S2}.

Now, as mentioned above  $T(q_{1_\perp},q_{2_\perp})$ corresponds to the hadron--level production amplitude that enters the cross section as in \eqref{eq:sighhf}. To make that connection more precise we can decompose the  momenta $q_i$ as in \eqref{eq:qdecomp}, i.e.
 \be\label{eq:qdecomps2}
 q_1= \xi_1 p_1 + \tilde{x}_1 p_2 + q_{1_\perp}\;,\qquad q_2 = \tilde{x}_2 p_1 + \xi_2 p_2 + q_{2_\perp}\;.
 \ee
 One can show that in the high energy limit we have
 \be
 \tilde{x}_i = -\frac{Q_i^2}{s}\frac{1}{x_{B,i}}\;.
 \ee
 Once we impose a rapidity veto this will tend to suppress larger values of $M_i$, for which $x_{B,i} \ll 1$, and hence we safely assume $\tilde{x}_i \approx 0$ for the purpose of this discussion, although we do not make such an approximation in actual calculations. Using this, and the gauge invariance of the PI amplitude to drop the terms $\sim q_i$ in \eqref{eq:rho}, gives 
 \be\label{eq:rhosub}
  \rho_i^{\alpha\beta}\approx 2\int \frac{{\rm d}M_i^2}{Q_i^2}  \bigg[-g^{\alpha\beta} F_1(x_{B,i},Q_i^2)+ 2\frac{q_{i_\perp}^\alpha q_{i_\perp}^\beta}{Q_i^2}\frac{ x_{B,i} }{\xi_i^2}F_2(x_{B,i},Q_i^2)\bigg]\;.
 \ee
The first term $\sim F_1$ is therefore suppressed by  $\xi_i^2/x_{B,i}^2$; again once a veto is applied we can safely assume that the factor of $1/x_{B,i}^2$ will not compensate for the $x_i^2$ suppression in the inelastic case. We can therefore drop this, which allows to write \eqref{eq:sighhf} as 
 \be\label{eq:sigpps2a}
 \sigma_{pp} = \frac{1}{8 \pi^2 s}  \int  {\rm d}x_1 {\rm d}x_2\,{\rm d}^2 q_{1_\perp}{\rm d}^2 q_{2_\perp
} {\rm d \Gamma}  \frac{{\rm d}M_1^2}{Q_1^2}  \, \frac{{\rm d}M_2^2}{Q_2^2} \, \frac{1}{\tilde{\beta}} \,|\mathcal{T}(q_{1_\perp},q_{2_\perp};\{k_{\rm var}^j\})|^2 \delta^{(4)}(q_1+q_2 - p_X)\;,
 \ee
 where
 \be\label{eq:tqq}
 \mathcal{T}(q_{1_\perp},q_{2_\perp};\{k_{\rm var}^j\})= \left[\frac{8\pi (\alpha(Q_1^2)\alpha(Q_2^2))^{1/2}}{\xi_1 \xi_2 } \left(x_{B,1} F_2(x_{B,1},Q_1^2)x_{B,2} F_2(x_{B,2},Q_2^2)\right)^{1/2}\right]\, \frac{q_{1_\perp}^\mu q_{2_\perp}^\nu}{Q_1^2 Q_2^2} M_{\mu\nu}\;.
 \ee
 Here $\{k_{\rm var}^j\}$ indicates the various kinematic arguments of $\mathcal{T}$, i.e. $x_{B,i}$ and so on. We have introduced these explicitly for clarity, but will suppress these as in e.g. \eqref{Mfor} for brevity from now on. We note also that the above expression could  in principle be written in terms of $x_i$ rather than $\xi_i$, as for  $\tilde{x}_i \approx 0$ in \eqref{eq:qdecomps2} we have $x_i \approx \xi_i$.
 
 At this stage, this simply corresponds to a  rewriting of the full cross section  \eqref{eq:sighhf} under the approximations discussed above, i.e. \eqref{eq:sigpps2a} combined with \eqref{eq:tqq} is simply by construction equal to \eqref{eq:sighhf} once these approximations are made. However, we can see the somewhat unusual factors of $F_2(x_{B,i},Q_i^2)^{1/2}$ in \eqref{eq:tqq}, which should certainly make us reluctant to associate $\mathcal{T}$ directly with the corresponding  amplitude we need. 
 To clarify this point, we can consider the purely elastic case, for which we have
 \be
 F_2(x_{B,i},Q_i^2) = F_E(Q^2_i)\, \delta(1-x_{B,i})\;,
 \ee
  where the $\delta(1-x_{B,i})$ is applied  directly in \eqref{eq:sigpps2a} to eliminate the $\int {\rm d} M_i^2/Q_i^2$, and 
  \begin{equation}\label{eq:fe}
F_E(Q^2_i)=\frac{G_E^2(Q_i^2)+\tau_i G_M^2(Q_i^2)}{1+\tau_i}\;,
\end{equation}
where $\tau_i = Q_i^2/4 m_p^2$ and $G_E$ and $G_M$ are the electric and magnetic Sachs form factors, respectively. These form factors are given in terms of the Dirac and Pauli form factors $F_{1,2}$ via
\be
G_E(Q_i^2) = F_1(Q_i^2) - \tau_i F_2(Q_i^2)\;,\qquad G_M(Q_i^2) = F_1(Q_i^2) + F_2(Q_i^2)\;,
\ee
such that
\be\label{eq:fet}
F_E(Q^2_i)=F_1(Q_i^2)^2 + \tau_i F_2(Q_i^2)^2\;.
\ee
The notation here is conventional, and in particular $F_{1,2}(Q_i^2)$ are not to be confused with the inelastic SFs; the lack of $x_{B,i}$ argument makes this clear. The key point is that the $p \to \gamma p$ vertex can be decomposed at the amplitude level (linearly) in terms of $F_{1,2}$. In particular, we can see that at the small $Q_i^2$ value relevant for elastic scattering the factor of $\tau_i \ll 1$ in \eqref{eq:fet}, and indeed numerically it turns out that $F_2$ is rather subleading with respect to $F_1$ even at larger $Q_i^2$. Therefore we can safely  drop the second term in \eqref{eq:fet}, and can then see that the factor of $F_2(x_{B,i},Q_i^2)^{1/2}$  in \eqref{eq:tqq} does indeed become $F_1(Q_i^2)$. That is, $\mathcal{T}$ can be (correctly) associated with the amplitude one would arrive at by starting directly with the amplitude--level decomposition. More precisely, $F_1$ and $F_2$ are associated with the amplitudes where the proton helicity is conserved and flipped, respectively. Therefore, one should account for these amplitudes independently, with the incoherent squared sum giving precisely the sum \eqref{eq:fet}. However, the effect of doing this on the survival factor is numerically very close (at the $\lesssim 1\%$ level) to simply working with $F_E(Q^2_i)^{1/2}$ directly at the amplitude level, and hence in practice we can do this. 

For the elastic case, we can therefore safely associate \eqref{eq:tqq} with the amplitude in \eqref{Mfor}. 
That is, we can account for survival effects by simply replacing $T$ in \eqref{eq:sigpps2a} with the corresponding screened amplitude $T^{\rm res}$ in \eqref{skt}. This will give a correct account of the $q_{i_\perp}$ dependence of the amplitude in the survival factor calculation, which induces the appropriate process dependence. This occurs both in terms of the dependence on the (elastic or inelastic) SFs in \eqref{eq:tqq}, but also the object being produced, where as discussed in e.g.~\cite{Harland-Lang:2015cta} the specific form of the PI helicity amplitudes modifies the $q_{i_\perp}$ dependence of $T$ and hence the survival factor. In the full calculation, we in addition effectively account for the subleading terms $\propto F_1(x_{B,i},Q_i^2) \sim G_M^2(Q_i^2)$ in \eqref{eq:rhosub}, by adding these incoherently to the amplitude, see~\cite{Harland-Lang:2015cta} for further details\footnote{To be precise, in the calculation of the survival factor we assume that $\tilde{x}_i =0$ in order to apply \eqref{eq:rhosub} in both the numerator and denominator of \eqref{seikav} for each phase space point. We then weight the full result, without any assumptions on $\tilde{x}_i$, by this.}.

Considering now the inelastic case, the situation is clearly more complicated. In particular, the inelastic structure functions in \eqref{eq:rhosub} are defined only at the cross section, and not the amplitude level,  by suitably summing over the final state from the proton dissociation. Therefore we are now left with the rather unphysical factors of $F_2(x_{B,i},Q_i^2)^{1/2}$ in \eqref{eq:tqq} if we wanted to apply this directly at the amplitude level. To resolve this, we take a somewhat phenomenological approach. In particular, provided the photon $Q_i^2$ is sufficiently low, i.e. broadly in the $Q^2_i \lesssim Q_0^2 = 1 \, {\rm GeV^2}$ regime, we continue to apply the approach of the elastic case, but with the elastic form factors suitably replaced by the inelastic SFs. On physical grounds, we can expect a relatively smooth transition between the elastic and inelastic cases in the low $Q_i^2$, and by doing so we account for the physically relevant process and $q_{i_\perp}$ dependence of the survival factor.

On the other hand, for higher $Q_i^2$ we might expect this approach to break down. However, in the $Q^2 \gtrsim Q_0^2$ regime in  \eqref{skt} we have $q_{i_\perp}' \approx q_{i_\perp}$ and the amplitude $T$ factorizes from the integral; as discussed in~\cite{Harland-Lang:2016apc} the average $k_t^2$ in $T_{\rm el}$ is $\sim 0.1 \,{\rm GeV}^2 \ll Q_0^2$. In this case, we only ever deal with $|T|^2$ directly in \eqref{seikav}, and hence can work at the correct cross section level appropriate for the corresponding inelastic SFs. In this case, as the survival factor is factorized from $|T|^2$, the process dependence no longer remains; this is physically in line with our expectations that for larger $Q^2$ we have a relatively short distance production process that takes place independently from the MPI that occurs. However, there will remain a kinematic dependence on photon $x_i$.  More precisely, one can model the $k_\perp$ dependence entering the integral in \eqref{skt} from the inelastic cross section with reference to the dependence in the case of the `generalized' PDFs~\cite{Diehl:2003ny,Belitsky:2005qn}, which allowed for such a non--zero momentum transfer, in terms of the proton Dirac form factor $F_1$. This complete factorization only applies for the case that both emitted photons are emitted inelastically with $Q^2 \gtrsim Q_0^2$, but in the mixed case where one photon is emitted elastically or inelastically but at low $Q^2$, a similar procedure can be performed. Further details of this are given in~\cite{Harland-Lang:2016apc}.

We note that the above discussion has so far only directly considered the pure PI case. However, the extension beyond this is now clear. In particular, the non--PI contributions are indeed only relevant well beyond the $Q^2_i \sim Q_0^2$ region, and hence here we have the same factorization of the production process and survival factor described above. Therefore, we can straightforwardly generalise the approach to include the survival effects in this region, without making any assumption about the form of the underlying production process.

Finally, it is worth recalling the qualitative predictions of the above results, and their impact on the relative EL, SD and DD components for semi--exclusive production. This is discussed in detail in~\cite{Harland-Lang:2016apc,Harland-Lang:2020veo}, and we only recall the key issues here. In particular, due to the peripheral nature of the interaction discussed above, for purely elastic production, the predicted average survival factor is $\sim 0.8-1$, depending on the precise kinematics and process, and the impact of this is therefore rather small. This is also true for SD production, due to the elastic photon emission on one beam side, such that the interaction itself remains rather peripheral; more precisely, the expected survival factor is somewhat lower due to the inelastic photon emission, being $\sim 0.5-0.8$, again  depending on the precise kinematics and process. For DD production on the other hand, a significant fraction of the interaction happens at rather small proton--proton impact parameters, where the MPI probability is high. The expected survival factor is in this case $\sim 0.1$, and we will therefore expect any rapidity veto to significantly reduce the relative DD fraction, due to this. We note that if instead of the above calculation one naively applied the default MPI treatment of a general purpose MC, then this would by default not distinguish between the three (EL, SD and DD) cases, and would predict a survival factor of the same order as the DD case, i.e. $\sim 0.1$, with the precise value depending on the specific model of MPI. This is clearly significantly different from the results above, as it misses some of the key physics involved. 

\section{MC implementation}\label{sec:mc}

With the above ingredients, we have implemented $W^+ W^-$ production in the \texttt{SuperChic 4.1} MC generator. The procedure for doing this very closely follows that described in~\cite{Harland-Lang:2020veo}, and we repeat the key elements here for clarity. 

Using the formalism described above, we can generate a fully differential final state in terms of not just the centrally produced system, but the squared photon virtualities $Q^2_i$ and the invariant masses $M_{i}$ of the proton dissociation systems, for the case of inelastic emission, while for elastic emission the corresponding structure functions are simply $\propto \delta(x_{B,i}-1)$, implying $M_i=m_p$ as expected. In this way we can then generate appropriately formatted unweighted Les Houches events (LHE) that can be passed to \texttt{PYTHIA 8.2} (other general purpose MCs could in principle be used, although this has not been investigated by us so far), with a suitable choice of run parameters. We in particular note that in the inelastic case, the amount of particle production due to the proton dissociation should be driven by the $Q_i^2$ transfer to the proton and the invariant mass $M_i$ of the dissociation system, and should occur essentially independently of any dissociation on the other proton side, being colour disconnected from it. As this general purpose MC is set up to read in parton--level events, with collinear initiating partons, we simply map the kinematics of the $p \to \gamma + X$  process onto a suitable parton--level one. That is, we assume a LO quark--initiated vertex as in Fig.~\ref{fig:wwfig}, with the initial--state (collinear) and final--state quark momenta set according to the 4--momenta $q_i$ generated by the MC. For the PI case these $q_i$, which at the event level corresponds to the 4--momentum transfer to the beam $i$, are associated with the photon momenta, but for a general diagram there is no need to make this association in order to correctly assign the LO kinematics to the quarks. For concreteness we assign the collinear initiator to be an up quark, but the final result should not be sensitive to this choice.

The generated Les Houches events (LHE) may then be passed to \texttt{PYTHIA} for showering and hadronization. For the  \texttt{PYTHIA} settings, we first of all set \texttt{PartonLevel:MPI=off}, as we only consider those events with no addition MPI, as accounted for via our calculation of the survival factor. We also use the dipole recoil scheme discussed in~\cite{Cabouat:2017rzi}, which is specifically designed for cases where there is no colour flow between the two initiating protons (or in the parton--level LHE, quarks), as is the case here; that is, we set \texttt{SpaceShower:dipoleRecoil = on}. Taking the default global recoil scheme leads to a significant overproduction of particles in the central region. As recommended in the \texttt{PYTHIA} user manual, we take \texttt{SpaceShower:pTmaxMatch = 2 }, in order to fill the whole phase space with the parton shower, but we set \texttt{SpaceShower:pTdampMatch=1} to damp emission when it is above the scale \texttt{SCALUP} in the LHE, which we set to the maximum of the two photon $\sqrt{ Q_{i}^2}$; in practice, this latter option is found to have little effect on the results. We in addition set \texttt{BeamRemnants:primordialKT = off}, as we wish to keep the initiating quark completely collinear to fully match the kinematics from the structure function calculation. 

For an elastic proton vertex, we must include the initiating photon in the event in order for \texttt{Pythia} to process it correctly. For the case of SD, this requires the kinematics to be modified in order to keep the elastic photon collinear and on--shell. This is achieved by setting the photon transverse momentum to zero in the event (but not in the cross section calculation), and keeping the momentum fraction fixed.  We note this is only a technical necessity in order for \texttt{Pythia} to correctly handle the event, for the specific case of SD production, which features one elastic and one inelastic vertex. In particular we set \texttt{SpaceShower:QEDshowerByQ = off}, such that there is no back evolution from the photon, consistent with this being an elastic emission. Treating the initiating photon as on--shell in the event kinematics is of course an approximation to the true result, but for most purposes is a very good one. 

Finally, for the calculation of the LO quark initiated diagrams in Figs.~\ref{fig:wwfig} and~\ref{fig:wwfigsd} we make us of \texttt{MadGraph5\_aMC@NLO}~\cite{Alwall:2014hca,Frederix:2018nkq} to generate the corresponding amplitudes. In particular, the stand--alone output of the code has been suitably implemented within the MC code, having been cross--checked against our own implementation. We have further reweighted the \texttt{MadGraph5\_aMC@NLO} results  to include a running value of $\alpha$, in order to match the SF result at low $Q_i^2$. We in addition include full spin correlations in the leptonic decay of the $W$ bosons.

 \section{Theoretical Uncertainties}\label{sec:theorunc}

In this section, we consider the corresponding theoretical uncertainties on our calculation. When quoting uncertainties we will for concreteness consider the cross section predictions relevant to the  ATLAS 13 TeV data on semi--exclusive $W^+ W^-$ production~\cite{ATLAS:2020iwi}, that is Table~\ref{tab:wwATLAS}, which we present in the following section. 

First we have the experimental uncertainty due the SF inputs at lower $Q_i^2$ and quark PDFs at higher $Q_i^2$. For the structure functions, we include the corresponding elastic and inelastic contributions  and include an uncertainty due to the experimental inputs on these, as described in~\cite{Harland-Lang:2019eai}, with the exception that we now use the updated \texttt{MSHT20qed} NNLO PDF set~\cite{Cridge:2021pxm} for the high $Q^2$ region. These are evaluated following the procedure discussed in~\cite{Harland-Lang:2019pla}, which is closely based on that described in~\cite{Manohar:2016nzj,Manohar:2017eqh}. We refer the reader to these references for further details, but in summary we include: an uncertainty on the A1 collaboration~\cite{Bernauer:2013tpr} fit to the elastic proton form factors, based on adding in quadrature the experimental uncertainty on the polarized extraction and the difference between the unpolarized and polarized; a $\pm 50\%$ variation on the ratio $R_{L/T}$, relevant to the low $Q^2$ continuum inelastic region; a variation of $W^2_{\rm cut}$, the scale below which we use the CLAS~\cite{Osipenko:2003bu} fit to the resonant region, and above which we use the HERMES~\cite{Airapetian:2011nu} fit/pQCD calculation (for $Q^2$ below/above $1\,{\rm GeV}^2$), between $3$--$4$ ${\rm GeV}^2$; the symmetrised difference between the default CLAS and Cristy--Bosted~\cite{Christy:2007ve} fits to the resonant region; the standard PDF uncertainty on the \texttt{MSHT20qed\_nnlo} quark and gluon partons in the $Q^2_i > 1 \,{\rm GeV}^2$ continuum region, as calculated via NNLO in QCD  predictions for the structure functions in the ZM--VFNS,  implemented in~\texttt{APFEL}~\cite{Bertone:2013vaa}. In the latter case, the same standard PDF uncertainty is also explicitly present in the LO calculation of the quark--initiated diagrams of Figs.~\ref{fig:wwfig} and~\ref{fig:wwfigsd}, and is included. 

Using the above procedure, we evaluate the uncertainty on the EL, SD and DD components to be $\sim 1-1.5\%$, with the larger error being in the DD case. We note that there is some degree of correlation in these uncertainties between the EL and SD cases, as these depend on the same elastic form factors, and the SD and DD cases, as these depend on the same inelastic SFs and quark/anti--quark PDFs in similar kinematic regions. A full analysis of this is beyond the currently required precision, but can certainly in principle be evaluated.  We also note that these uncertainties are to a large extent correlated between the $W^+ W^-$ and lepton pair ($m_{ll} > 2 m_{WW}$) cases considered in Tables~\ref{tab:wwATLAS} and~\ref{tab:llATLAS} in the following section.

We next consider the impact from higher order corrections, in particular to the LO parton--level diagrams in Figs.~\ref{fig:wwfig} and~\ref{fig:wwfigsd}. The factorization and renormalization scales are both set to $\sqrt{Q_i^2}$ by default, guided by the fact that this is the appropriate scale for the pure PI diagrams, as per the analysis in the SF approach.
Varying these by the usual factor of 2, gives a $\sim 1\%$ (2-3\%) variation in the SD (DD) cross sections; no equivalent uncertainty is present in the EL case of course. Taking instead $\mu_{F,R} = M_W$ as in~\cite{Jager:2006zc} gives results that are consistent within these variation bands. Thus according to this measure the uncertainty from this is very small. It is indeed true that at least for the VBS case as in~\cite{Jager:2006zc}, the difference between the LO and NLO cases for the central scale choice is very small, however clearly this deals with a rather different set of kinematics to those considered here. To be conservative, we also try removing the reweighting applied in order to introduce a running $\alpha$, i.e. reverting to the default fixed value used in \texttt{MadGraph5\_aMC@NLO}. This leads to a similar level of variation to that found by simply varying the factorization/renormalization scales. We in addition investigate the impact of removing the reweighting by the NNLO to LO $K$--factor of $F_2(x_{B,i},Q_i^2)$ discussed in Section~\ref{sec:hybrid}. This leads to a $\sim 2\%$ (5\%) change in the SD (DD) cross section, and is a particularly conservative variation, as we expect the default choice to be more accurate. Combining the above we arrive at a rather small $\sim 2\%$ (5\%) in the SD (DD) cross sections, although the precise values are of course only estimates, which can be checked by suitably calculating the NLO QCD/EW corrections to the LO diagrams we include. We can to some extent expect the impact of QCD corrections at least to be largely correlated between the SD and DD cases, especially given at NLO these are entirely factorizable, i.e. do not connect to the two quark lines in the DD case. For lepton pair production, which is dominated by the PI contribution, there is no explicit factorization scale, and as discussed in~\cite{Harland-Lang:2019eai} the corresponding uncertainty due to non--factorizeable corrections (which are not included) is expected to be very small.

Finally, we consider the uncertainty due to the survival factor. For the purely elastic case, as discussed in detail in~\cite{Harland-Lang:2021ysd} the theoretical uncertainty due to this is expected to be very small, at the $\sim 1\%$ level. This is due to the peripheral nature of the interaction; as this is largely outside the range of QCD no amount of model variation in these soft QCD interactions can lead to any significant change in the result. To some extent we can expect a similar argument to apply in the SD case, whereas for DD production this is far from clear. To evaluate the uncertainty in these latter cases we follow the approach of~\cite{Harland-Lang:2016apc}. That is, in evaluating the short--distance component of the survival factor discussed at the end of Section~\ref{sec:S2}, we apply the two--channel eikonal model of~\cite{Khoze:2013dha}, in which the incoming proton is considered to be a coherent superposition of two diffractive `Good--Walker' (GW) eigenstates~\cite{Good:1960ba}, each of which may scatter elastically. There is then some freedom in the modelling of how the $W^+ W^-$ production process couples to these individual eigenstates, i.e. in the pure PI case in the coupling $\gamma_i$ between the photon (generalized) PDF and the eigenstate $i$ ($=1,2$ in the two--channel case). By default, we assume a universal coupling ($\gamma_1 = \gamma_2$), however another possibility is to assume the same coupling as that of the Pomeron in the model we apply. This results in a survival factor that is $\sim 40\%$ lower in the DD case. For the SD cross section the reduction is instead $\sim 10\%$; the impact is rather less due to the smaller contribution from the short distance regime. Although this particular model choice leads to a reduction, it is in principle possible that a different choice may act in the opposite direction, at least for the DD component. Therefore to be conservative, we consider there to be a $\pm 50\%$ uncertainty in the DD case, and a $\pm 10\%$ uncertainty in the SD case. As discussed above, the uncertainty in the EL case (or to be precise the change in the survival factor one can arrive at by reasonable model variations in this approach) is rather smaller; we take this to be $\pm 1\%$, which as discussed in~\cite{Harland-Lang:2021ysd} is a reasonable estimate.  We expect this uncertainty to be largely fully correlated between the different components, as we see for the explicit model variation considered above in the SD and DD cases. Finally, for lepton pair production as in Table~\ref{tab:llATLAS} we find a very similar level of difference in the SD and DD survival factors when considering the model variation described above, which is again as we would expect, given this impacts on the short distance and relatively process independent component of the survival factor. That is, the uncertainty between the lepton and $W$ pair cases will be rather strongly correlated, although not entirely due to the long distance component of the survival factor, which is more process dependent.
 
 \section{Comparison to ATLAS 13 TeV data}\label{sec:ATLAS}
 
 In this section we present comparisons to the ATLAS 13 TeV data on semi--exclusive $W^+ W^-$ production~\cite{ATLAS:2020iwi}. These are calculated using the \texttt{SuperChic 4.1} MC implementation described in the previous sections.

\begin{figure}[t]
\begin{center}
\includegraphics[scale=0.45]{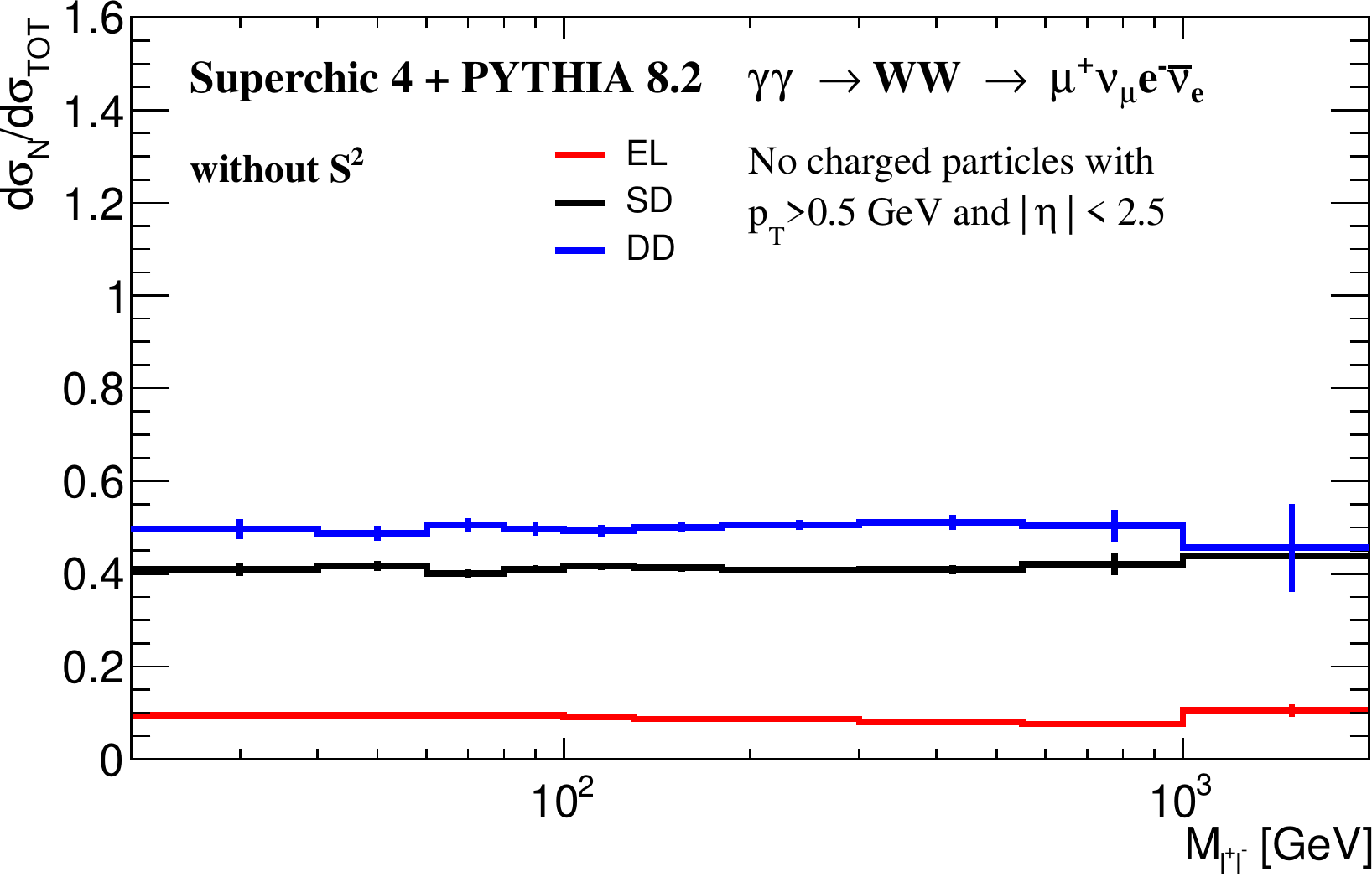}
\includegraphics[scale=0.45]{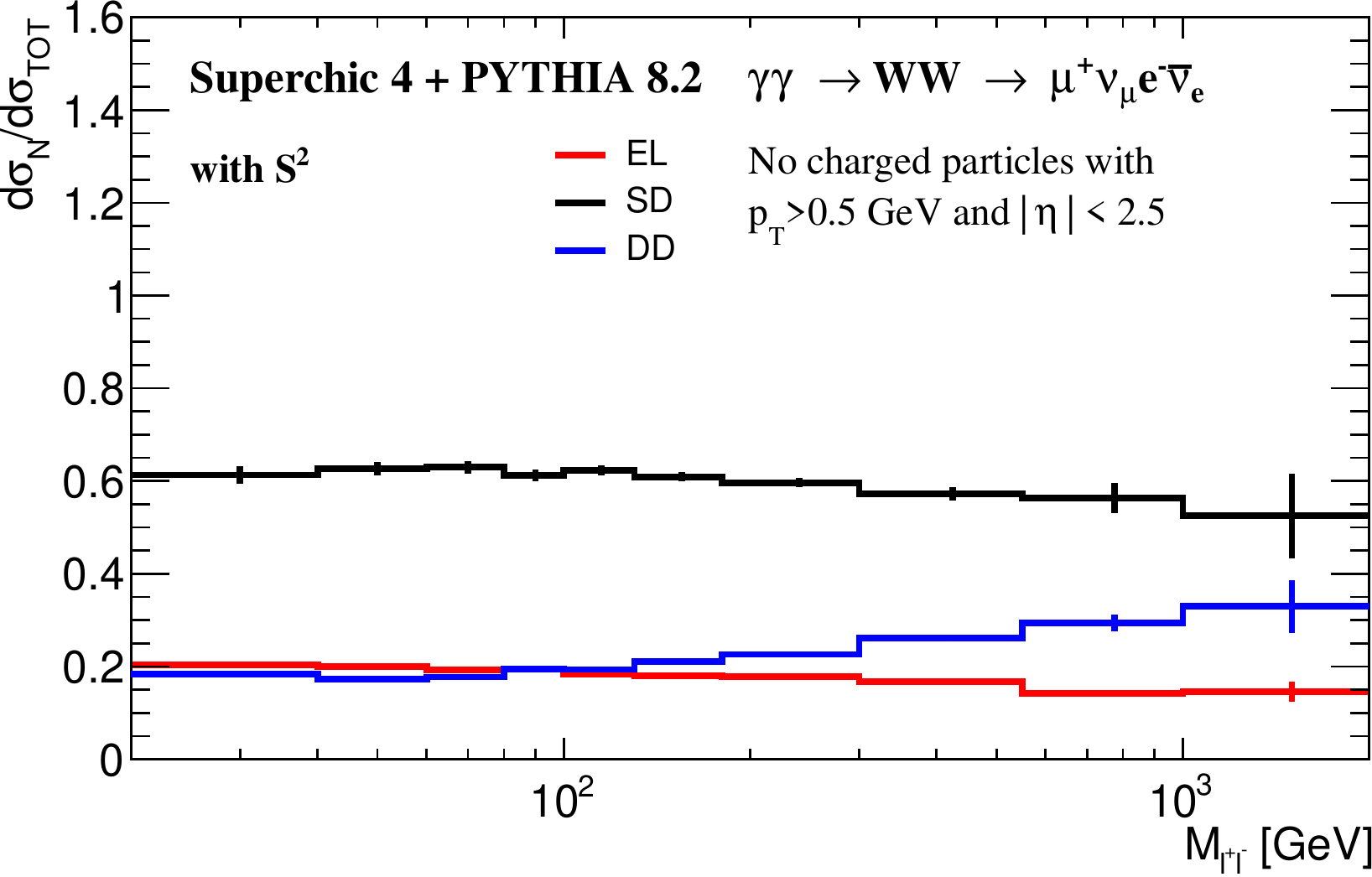}
\caption{Fractional contributions from EL, SD and DD $W^+ W^-$ production at $\sqrt{s}=13$ TeV, within the  ATLAS~\cite{ATLAS:2020iwi} event selection. The rapidity veto~\eqref{eq:veto} is applied at the hadron level, and the survival factor is excluded (included) in the left (right) plots.}
\label{fig:wwATLAS}
\end{center}
\end{figure}

In Table~\ref{tab:wwATLAS} we give the cross section values within the ATLAS event selection. We show results with and without the rapidity veto applied~\eqref{eq:veto} at the hadron--level, and with and without the survival factor included; as described in Section~\ref{sec:S2} this is  accounted for directly in \texttt{SuperChic}, rather than via the MPI implementation in \texttt{PYTHIA 8.2}, which we therefore disable. The case with the veto and survival factor applied is therefore to be compared with the ATLAS data directly, while the other results are only shown for illustration. We also give the breakdown between EL, SD and DD production. The corresponding fractional contributions as a function of the dilepton invariant mass are shown in Fig.~\ref{fig:wwATLAS}, with the rapidity veto applied and without (with) the survival factor included in the left (right) plots.

\begin{figure}[t]
\begin{center}
\includegraphics[scale=0.45]{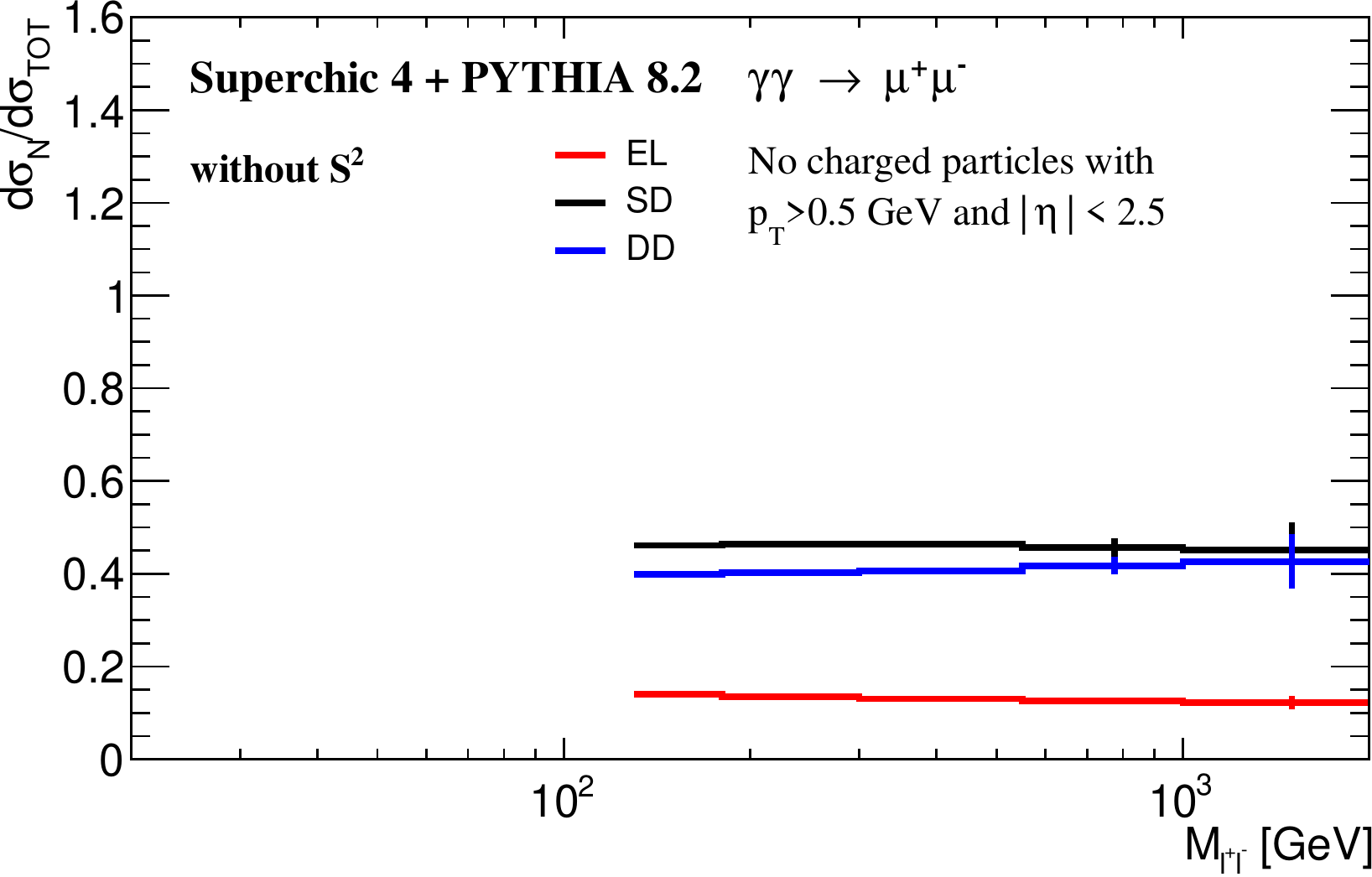}
\includegraphics[scale=0.45]{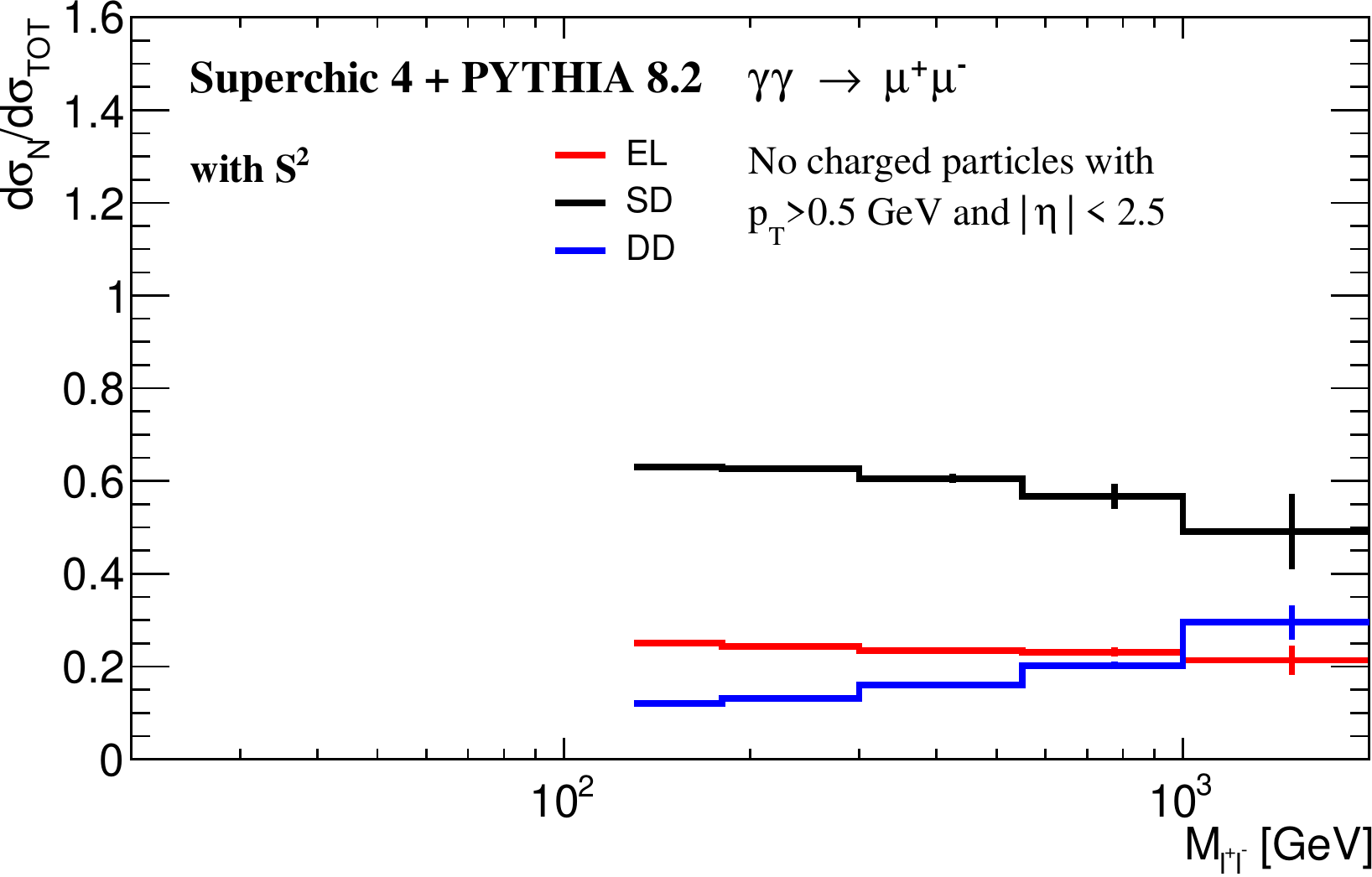}
\caption{As in Fig.~\ref{fig:llATLAS}, but for lepton pair production. Lepton cuts \eqref{eq:lcuts} are applied, rather than \eqref{eq:atcuts}, and we require that $m_{ll} > 2 \,m_{WW}$.}
\label{fig:llATLAS}
\end{center}
\end{figure}

Starting with Table~\ref{tab:wwATLAS}, when no veto is applied we can see that the DD component is largest, and the EL component very small. This is as we would expect, and we note that these values are by construction the same as the corresponding ones in Table~\ref{tab:cs_gnvbf}. We now apply the rapidity veto at the hadron level, after passing to  \texttt{PYTHIA 8.2}. The EL component is by construction unchanged~\footnote{We note that in particular we do not include QED FSR radiation from the leptons, which in the ATLAS analysis are dressed by photon radiation within a cone of $\Delta R= 0.1$; we  consider the impact of this to be beyond the precision of the current comparison, although for future studies this can be straightforwardly accounted for.}, while the SD (DD) cross sections are reduced by $\sim 50\%$ (70\%), and the fractional contributions are modified accordingly. This is again as expected: the veto suppresses the DD contribution most, for which there is a larger potential for radiation in the veto region. The SD and DD are now equally dominant, and the EL contribution is $\sim 10\%$. We note that this is qualitatively similar to the results in Table~\ref{tab:cs_gvbf2}, although not identical to it. In particular, the cross sections including  the veto at the hadron level are $\sim 5\%$ lower and $\sim 10\%$ higher in the SD and DD cases, respectively. Finally, including the survival factor significantly reduces the DD component by a further $\sim 85\%$, while the EL (SD) cases are reduced by $\sim 20\%$ (60\%); that, is the average survival factors are $\sim 0.8,0.6$ and 0.15 in the EL, SD and DD cases, respectively. This is as expected from the discussion at the end of Section~\ref{sec:S2}. That is, the DD component, for which the underlying interaction is entirely inelastic, occurs at relatively low proton--proton impact parameters and is therefore rather sensitive to the impact of MPI once a rapidity veto is applied. For EL and SD production on the other hand, we have at least one elastic photon in the initial--state, and the corresponding interaction is more peripheral. In terms of the invariant mass distributions shown in Fig.~\ref{fig:wwATLAS}, we can see that the fractional contributions are relatively flat, although some trend is observed with increasing mass once the survival factor is included; as discussed above the impact of survival effects is not necessarily constant with respect to the particle kinematics, and this is indeed observed here. The same plot for the lepton pair case discussed below is shown in Fig.~\ref{fig:llATLAS}, and a similar trend is observed.

\renewcommand{\arraystretch}{1.5}
\begin{table}
\begin{center}
\begin{tabular}{|c|c|c|c|c|c|}
\hline
$\sigma$ [fb] ($\sigma_i/\sigma_{\rm tot}$), $W^+ W^-$ &EL  &SD &DD& Total&$f_\gamma^{WW}$  \\
\hline 
No veto, no $S^2$ &0.701 (3.5\%)  & 6.00 (30.3\%) &13.1 (66.2\%)& 19.8&28.2\\
\hline
Veto, no $S^2$ &0.701  (9.2\%) & 3.21 (42.3\%)&3.68 (48.5\%)& 7.59&10.8\\
\hline
Veto, $S^2$ &0.565  (18.6\%) & 1.87 (61.6\%) &0.599 (19.8\%) & 3.03&4.3\\
\hline
\hline
$\langle S^2 \rangle$ &0.81 & 0.58 &0.16 & 0.40&-\\
\hline
\end{tabular}
\end{center}
\caption{Cross section predictions (in fb) for $W^+ W^-$ production at $\sqrt{s}=13$ TeV, from the \texttt{SuperChic 4.1} MC + \texttt{PYTHIA 8.2}. Lepton cuts \eqref{eq:atcuts} applied. Results are shown with and without a rapidity veto~\eqref{eq:veto} applied at the hadron--level, as well as including the survival factor; the `Veto, $S^2$' predictions corresponds to the phenomenologically relevant result, while the rest are given for comparison. The breakdown into El, SD and DD is also given, as well as the corresponding fractional contributions from these. The $f_\gamma^{WW}$ factor \eqref{eq:fgam} is also shown, and the average survival factor, when a veto is imposed, is given in the last row. Theoretical uncertainties not shown, but are discussed in the text.}  \label{tab:wwATLAS}
\end{table}

 \renewcommand{\arraystretch}{1.5}
\begin{table}
\begin{center}
\begin{tabular}{|c|c|c|c|c|c|}
\hline
$\sigma$ [fb] ($\sigma_i/\sigma_{\rm tot}$), $l^+ l^-$ &EL  &SD &DD& Total & $f_\gamma^{ll}$ \\
\hline 
No veto, no $S^2$ &11.3 (9.5\%)  & 50.9 (43.0\%) &56.5 (47.5\%)& 119&10.5\\
\hline
Veto, no $S^2$ &11.3  (13.5\%) & 38.7 (46.0\%)&34.0 (40.5\%)& 84.0&7.4\\
\hline
Veto, $S^2$ &9.61  (24.0\%) & 24.9 (62.5\%) &5.42 (13.5\%) & 39.9&3.5\\
\hline
\hline
$\langle S^2 \rangle$ &0.85& 0.64&0.16& 0.48&-\\
\hline
\end{tabular}
\end{center}
\caption{As in Table~\ref{tab:wwATLAS}, but for lepton pair production. Lepton cuts \eqref{eq:lcuts} are applied, rather than \eqref{eq:atcuts}, with in particular $m_{ll} > 2 \,m_{WW}$ required. The $f_\gamma^{ll}$ factor \eqref{eq:fgam} is also shown, and the average survival factor, when a veto is imposed, is given in the last row. }  \label{tab:llATLAS}
\end{table}

The central prediction which we can compare to the ATLAS data is therefore 3.03 fb. In terms of the theoretical uncertainty, we can include this following the approach discussed in Section~\ref{sec:theorunc}. The uncertainty from higher order corrections is found to be $\pm 0.1$ fb, while that due to the survival factor is $\pm 0.5$ fb (recalling we consider this source to be fully correlated between the different components), and is therefore strongly dominant. Thus our final result is $3.0 \pm 0.5$ fb, which we can now compare with the ATLAS data directly. They report the fiducial cross section of
\be\label{eq:sigat}
\sigma_{\rm meas}^{WW} = 3.13 \pm 0.31 \,({\rm stat.}) \pm 0.28 \,({\rm syst.}) \, {\rm fb}\;,
\ee 
for semi--exclusive $W^+ W^-$ production, that is with the veto \eqref{eq:veto} applied, and with all three components (EL, SD, DD) therefore contributing. We can see that the agreement within uncertainties is very good. Moreover, we can see that it only occurs once the rapidity veto and survival factor are appropriately included in the theoretical prediction; otherwise, the predicted cross section is of course far too high. On the other hand the (conservative) theoretical uncertainty due to the survival factor is non--negligible, being already at the same level as the experimental ones; we will comment further below about how this might be effectively reduced by comparing to lepton pair production.

We note that at the time of the ATLAS analysis~\cite{ATLAS:2020iwi} no full calculation of semi--exclusive $W^+ W^-$ production, accounting consistently for all the effects discussed in this paper, existed. In the absence of this, the result was compared against two distinct and somewhat approximate predictions, which we will now consider in turn. In particular, as we will see, the relative components of EL, SD and DD production are rather different in these cases, and hence this requires further discussion. The first comparison is made by applying a correction to \eqref{eq:sigat} in order to derive a corresponding purely elastic cross section, following the approach of~\cite{Aaboud:2016dkv,Khachatryan:2016mud}. To achieve this, the  fiducial cross section for (same flavour) semi--exclusive lepton pair production is measured in the $m_{ll} > 2 m_{WW}$ region, in order to more closely match the underlying kinematics to that of $W^+ W^-$ production. The ratio of this to the theoretical prediction for purely elastic lepton pair production is then evaluated, i.e.
\be\label{eq:fgam}
f_\gamma^X \equiv \frac{\sigma_{\rm meas}^{X}}{\sigma_{\rm El, theory}^{X}}\;,
\ee
where $X=l^+ l^-$ or $W^+ W^-$ depending on the context, and the elastic cross section does not include the survival factor. With this, and under the assumption that the above ratio is the same between the lepton pair (within the considered mass region) and $W^+ W^-$ cases, one can then use \eqref{eq:fgam} in combination with a theoretical prediction for the purely elastic  $W^+ W^-$ production to compare with \eqref{eq:sigat}. To evaluate the extent to which the above assumption is expected to be valid, in Table~\ref{tab:llATLAS} we show cross section predictions for lepton pair production in the $m_{ll} > 2 m_{WW}$ region and with the same lepton cuts as that considered in the ATLAS analysis to derive this correction factor. We again show results before  and after the veto is applied, and with and without the survival factor included, for illustration. To be precise, we have reweighted the MC results, which only include PI production, by the factors in Table~\ref{tab:lpair} in order to effectively include the non--PI contributions to the SD and DD components. We leave a full MC implementation to future work, but note that the impact of this after applying a veto is at the percent level, and therefore this approach is sufficient at the current level of precision.

We can see that before the veto is applied, the relative contribution from elastic production is predicted to be a factor of $\sim 3$ higher in comparison to the $W^+ W^-$ case. This is due to the differing behaviour of the inelastic (SD and DD) components, and in particular the differing Feynman diagrams that contribute in Fig.~\ref{fig:wwfig} in comparison to Fig.~\ref{fig:llfig}, and similarly in the SD case, as well as the relative role of $Z$--initiated production. In the inelastic case there is inevitably a contribution from non--PI production, and this is more significant in the $W^+ W^-$ case. Once we impose the rapidity veto, the qualitative impact is similar to $W^+ W^-$ production, i.e. the relative elastic component is increased and the DD component is greatly reduced. However, in detail we can see that the average survival factor is somewhat higher in the lepton case for elastic and SD production, due to the differing form of the underlying PI initiated helicity amplitudes (recalling that this process dependence enters in the low $Q_i^2$ region where pure PI production dominates). This effect was observed some time ago for elastic production~\cite{Harland-Lang:2015cta}, and we can see is expected to persist for SD production. On the other hand, for DD production the process dependence is minimal, as here the interaction occurs at rather short distance scales and therefore decouples from survival effects, as discussed further in Section~\ref{sec:S2}. 

The corresponding value of $f_\gamma^{ll}$ is also show in Table~\ref{tab:llATLAS}, and we can see how it reduces as we include the impact of the rapidity veto. We again include theoretical uncertainties on this prediction following the approach discussed in Section~\ref{sec:theorunc}. Including the dominant uncertainty source, due to the survival factor, we find that $f_\gamma^{ll} = 3.5 \pm 0.5$, which is in very good agreement with the value
\be\label{eq:atf}
f_{\gamma}^{ll,{\rm ATLAS}} = 3.59 \pm 0.15
\ee
 determined by ATLAS~\cite{ATLAS:2020iwi}. On the other hand, in Table~\ref{tab:wwATLAS} we show the same predicted $f_\gamma$ factors, but now calculated for the case of $W^+ W^-$ production. We can see that prior to imposing a rapidity veto the predicted $f_\gamma^{WW}$ is significantly larger in this case, again due to the differing Feynman diagrams that contribute in Fig.~\ref{fig:wwfig} in comparison to Fig.~\ref{fig:llfig} discussed above. However once the experimentally relevant veto is imposed, the relative contribution from this is greatly reduced, as we would expect. Nonetheless some difference, driven by this effect and the differing survival factors, remains. The final prediction is a factor of $\sim 20\%$ higher, and is not in agreement with \eqref{eq:atf} within experimental uncertainties. Naively, one might still consider this to be in agreement within theoretical uncertainties, which gives $f_\gamma^{WW} = 4.3 \pm 0.7$, however this omits the fact that the uncertainties in both cases are almost entirely driven by the modelling of the survival factor, which to rather good approximation should be correlated between the two cases. In particular, although the survival factor is not identical in the two cases, being process dependent, one expects that any model variation that results in (say) an increased impact of MPI should do so in both cases. We have seen this explicitly at the bottom of Section~\ref{sec:theorunc}, where the largest uncertainty precisely lies in the larger $Q_i^2$ region, and the process dependence is greatly reduced. Therefore the uncertainty on the ratio of the predicted values of $f_\gamma$ is much smaller. We note that the fractional uncertainty in the current case is slightly larger than in the lepton pair case due to the large contribution from SD and DD production, where the uncertainty, due in particular to the survival factor, is larger.
 
 Indeed, given this fact it is tempting to effectively use the experimental measurement of \eqref{eq:atf} to constrain our modelling of the soft survival factor. In particular, this roughly speaking constrains the variation in the survival factor to be $\sim 30\%$ of the default conservative value, so that the predicted $f_\gamma^{ll}$ is consistent with the data within uncertainties. Assuming complete correlation between the lepton pair and $W^+ W^-$ survival factors, this translates directly across to a reduced uncertainty on our predicted $W^+ W^-$ cross section, from roughly $\pm 0.5$ fb to $\pm 0.2$ fb. Now, the above results are certainly approximate, and in particular the assumption of full correlation between the two cases is not completely correct, due both to the differing components of SD and DD production and the different survival factors themselves. The impact of the former difference can be readily accounted for using standard statistical techniques, while the latter becomes more model dependent. We do not investigate this issue in further detail here, given the size of the current experimental uncertainties, but simply note that the above results will give a good estimate of the exact impact of such a technique, which can (and should) be applied here. We emphasise that this procedure is not the same as assuming the underlying elastic, SD and DD fractions are the same between the two cases, which as discussed above is not correct.
 
Now, returning to the central values of $f_\gamma$, if we were to simply take the value from lepton pair production, and multiply the prediction for elastic $W^+ W^-$ production by this we would arrive at
 \be
 \sigma^{WW}_{f_\gamma} = 3.5 \times 0.701\,{\rm fb} = 2.45\,{\rm fb}\;,
 \ee
 which is somewhat lower than the measured fiducial cross section \eqref{eq:sigat}, albeit not significantly so given the current experimental uncertainties\footnote{We note that in ~\cite{ATLAS:2020iwi} the factor \eqref{eq:atf} is used to multiply the \texttt{Herwig7}~\cite{Bellm:2015jjp} prediction, which gives a somewhat lower value for the corresponding cross section. The  \texttt{Herwig7} prediction for elastic $W^+ W^-$ is in particular $\sim 7\%$ lower than our result, and indeed shows a similar level of suppression with respect to the  \texttt{MadGraph5\_aMC@NLO} prediction quoted in the ATLAS analysis. We have explicitly checked our results here against \texttt{MadGraph5\_aMC@NLO}, and find good agreement, while the reason for this discrepancy in the case of \texttt{Herwig7} is not clear. However for lepton pair production we not that the \texttt{SuperChic} MC predictions are in good agreement with  \texttt{Herwig7}, as can e.g. be seen by comparing the results of~\cite{Harland-Lang:2021ysd} with the quoted results in~\cite{Aaboud:2017oiq}.}. In the future, however, experimental uncertainties will reduce and so this level of approximation will not be sufficient. Moreover, if one aims to use data on semi--exclusive $W^+ W^-$ production to perform an EFT analysis, and hence look for relatively small deviations in e.g. the tails of distributions, then it will be crucial to account for this difference between the $W^+ W^-$ and lepton pair cases. We note that this is not necessarily as straightforward as simply taking the predictions above in order to, say, derive a corresponding elastic $W^+ W^-$ cross section with which an EFT analysis could be performed, as the results in Table~\ref{tab:wwATLAS} are themselves derived assuming the SM. Therefore, some caution is needed here, and a full account of all channels in a complete way may be essential. In this respect, the ability to tag to outgoing protons (or proton) and hence further disentangle the relative (EL, SD, DD) contributions may prove crucial.
 
Finally, we note that a second comparison to the data is presented in~\cite{ATLAS:2020iwi}. Here, the LO collinear $\gamma \gamma \to W^+ W^-$ cross section is evaluated using \texttt{MadGraph5\_aMC@NLO} and then combined with the corresponding elastic and inelastic \texttt{MMHT2015qed}~\cite{Harland-Lang:2019pla} photon PDFs. These are then passed to \texttt{Pythia8} for showering, with MPI turned on only in the DD case. The resultant cross section is $4.3 \pm 0.8$ (scale) fb, which if the EL and SD results are then multiplied by an effective survival factor of 0.82 (taken from the prediction in~\cite{Harland-Lang:2015cta} for the elastic case), gives a cross section of  $3.5 \pm 1.0$ (scale) fb. This is also in good agreement with the ATLAS data, within uncertainties. However, the quoted fractional contributions from EL, SD and DD production are 16\%, 81\% and 3\%, respectively, which is rather different to the results of Table~\ref{tab:wwATLAS}. The reasons for this are multiple. First, the LO collinear prediction omits the contribution from non--PI diagrams in the SD and DD case, and will therefore underestimate these cross sections, as seen in Table~\ref{tab:cs_gvbf2}. Second, for the DD case the use of the default \texttt{Pythia8} treatment of MPI gives a DD cross section (i.e. a survival factor) that is a factor of $\sim 2.5$ lower than the value we predict; we recall that here our direct evaluation of the survival factor is formulated in order to correctly account for the specific DD process under consideration, whereas the default MPI treatment is not specifically designed with this in mind and hence will not necessarily be accurate. This difference is of the same order to (though a little larger than)  the $\pm 50\%$ theoretical uncertainty we have taken for the DD case. Third, in the ATLAS comparison, the dipole recoil scheme described in~\cite{Cabouat:2017rzi}, which as discussed above is the most appropriate for the current situation, is not used. We find that if we do not use the dipole recoil scheme, then the DD component is in particular significantly reduced, by a further factor of $\sim 3$. Fourth, the value of the survival factor taken is not appropriate for the SD case. We find that if we take the above choices, we do indeed reproduce the values quoted in  the ATLAS analysis rather closely. However, this omits the various physical effects listed above, and hence we do not expect the corresponding fractional components (and in particular the very strong suppression of the DD component) to be a reliable prediction.

 \section{Summary and outlook}\label{sec:conc}
 
The production of electroweak (EW) particles with intact protons and/or rapidity gaps in the final state is a key ingredient in the LHC precision physics programme, with unique sensitivity to physics within and beyond the SM. A key process in this category is the case of semi--exclusive $W^+ W^-$ production, which as with VBS scattering more generally, provides a sensitive probe of the gauge structure of SM and of BSM effects that may modify it. The $W^+ W^-$ channel is in particular highly topical in light of the recent  first observation by ATLAS  of semi--exclusive $W^+ W^-$ production~\cite{ATLAS:2020iwi}, at 13 TeV. This is selected by imposing a veto on additional charged tracks in the central detector, in order to isolate the VBS signal, within which the photon--initiated (PI) channel plays a key role.
 
In this paper we have presented the first unified treatment of this process, accounting for all relevant contributing Feynman diagrams and the impact of a rapidity veto on the predicted cross section. We have analysed in detail the issues that arise if one naively applies the so--called Structure Function (SF) approach in the unitary gauge, including the pure PI diagrams alone, and discussed how the dominance (or not) of the PI process may be more appropriately analysed in the EW axial gauge. We have then presented a hybrid approach, which correctly bridges the  region of low squared momentum transfer $Q_i^2$, where the SF approach can be  applied and is given in terms of the low scale proton SFs, with the region of high $Q_i^2$, where the LO quark--initiated calculation is applied. In the latter case this includes the full gauge--invariant set of contributions, and not just the pure PI ones, i.e. due to the underlying $\gamma \gamma \to W^+ W^-$ process. Although the quark--initiated contribution is accounted for at LO, this may straightforwardly be extended to higher order if this is required. 

A further key element in the calculation is the correct  modelling of the `survival factor' probability of no additional particle production due to 
multi--parton interactions (MPI), which must be included when a rapidity veto is applied. This depends sensitively on the final--state kinematics as well as the production process. In particular, for purely elastic production the peripheral nature of the interaction leads to a survival factor that is rather close to unity, and this remains true (albeit to a lesser extent) for single dissociative (SD) production. In the double dissociative (DD) case, on the other hand, the impact of MPI is expected to be significantly larger, and the corresponding survival factor smaller. These effects are fully accounted for in our calculation.

A key consequence of the above discussion is that the total cross section, as well as the relative fractions of elastic, SD and DD production in the  $W^+ W^-$ case depends in general on the full contributing set of Feynman diagrams, and not just PI production, as well as the kinematic dependence of the survival factor. With this in mind, we have compared our predictions to the ATLAS data~\cite{ATLAS:2020iwi}, and find excellent agreement. The predicted contributions from elastic, SD and DD production are on the other hand are rather different from the more approximate theoretical approaches compared to in the analysis, and rather different from the case of lepton pair production. The latter point is particularly important, in light of the fact that it has so far been rather common to use the measured ratio of the fiducial (including elastic, SD and DD production) to the elastic cross sections in the case of lepton pair production in order to extract an exclusive signal in the $W^+ W^-$. This procedure can only be correct if the the fractions of elastic, SD and DD production are the same in the $W^+ W^-$ and lepton pair cases. In this paper we have shown that they are not, and while the effect is relatively mild it is non negligible. This issue may in particular be crucial if the aim is to use such data to look for small deviations from the SM, for example in the context of an EFT analysis.
 
 In summary, in this paper we have presented the results of a new MC implementation of semi--exclusive $W^+ W^-$ production in proton--proton collisions. This is released in the \texttt{SuperChic 4.1} MC, which as well as the case of $W^+ W^-$ production discussed here can generate a range of other processes, as described in~\cite{Harland-Lang:2015cta,Harland-Lang:2018iur}. The code and a user manual can be found at 
\\
\\
{\tt http://projects.hepforge.org/superchic}
\\
\\
The results of our calculation are therefore made available for the community, and we hope will play a key role in future analyses within this very promising area of the LHC precision physics programme.

\section*{Acknowledgments.}

We are very grateful to Marek Tasevsky, who passed the \texttt{SuperChic} LHEs to \texttt{Pythia} in order to produce the results in Section~\ref{sec:ATLAS}.
We thank Valery Khoze, Misha Ryskin and Marek Tasevsky for useful discussions and for passing on comments about the manuscript. We thank Kristin Lowhasser and  Andy Pilkington for useful discussion, in particular with respect to the details of the ATLAS 13 TeV analysis. LHL thanks the Science and Technology Facilities Council (STFC) for support via grant award ST/L000377/1. S. B. acknowledges financial support from STFC

\appendix

\section{Treatment of $\gamma q$ initiated production in hybrid SF approach}\label{app:hybrid}

In this appendix we provide some further details of the implementation of \eqref{eq:rhorep} in the hybrid SF approach. We recall that for the case that  \eqref{eq:qwcut} is satisfied for $i=1$, but not $i=2$, we simply replace in \eqref{eq:sighh}
\be\label{eq:rhorepap}
\rho_{1}^{\mu\mu'}\rho_{2}^{\nu\nu'} M^*_{\mu'\nu'}M_{\mu\nu} \to \frac{Q_1^2}{4\pi \alpha(Q_1^2)} \int  \frac{{\rm d}M_1^2}{Q_1^2} \, \rho_{2}^{\mu\mu'} \sigma_{\mu \mu'}^{1}\;,
\ee
where the integration is as usual performed simultaneously with the other phase space integrals, while for the case that  \eqref{eq:qwcut} is satisfied for $i=2$, but not $i=1$, we simply interchange $1 \leftrightarrow 2$. We will focus on the case given explicitly in the above replacement for concreteness, but everything follows through in exactly the same way for this alternative case. At LO we have
\be\label{eq:siggqap}
\sigma_{\mu \mu'}^{1} = \sum_{j=q,\overline{q}}  f_j(x_{B,1},\mu_F^2)\,A_{\mu}^1 A_{\mu'}^{1*} \;,
\ee
where $A_\mu^1$ is the corresponding $\gamma^* + q \to W^+ W^- + q$ amplitude including all diagrams in Fig.~\ref{fig:wwfigsd}, with a collinear initial--state quark/anti--quark from beam 1, carrying proton momentum fraction $x_{B,1}$. We label the initial--state photon momentum as $q_2$, and the momentum $q_1 = p_q - p_q'$, where $p_q$ ($p_q'$) is the incoming (outgoing) quark momentum. We decompose these as
 \be\label{eq:qdecomp}
 q_1= \xi_1 p_1 + \tilde{x}_1 p_2 + q_{1_\perp}\;,\qquad q_2 = \tilde{x}_2 p_1 + \xi_2 p_2 + q_{2_\perp}\;.
 \ee
 As in~\cite{Harland-Lang:2019eai} we write 
 \be\label{eq:Lexpan}
A_{\mu}^1 A_{\mu'}^{1*}   = -\frac{1}{2}\delta_{T,\mu\mu'}\sum_{\lambda=\pm}  |A_{\lambda}^1|^2  + \epsilon_{0}^\mu\epsilon_{0}^{\mu'}|A_{0}^1|^2 \;,
\ee
where $A_\pm$ is the amplitude corresponding to $\pm$ photon helicities, and $A_0$ corresponds to longitudinal photon. Here
\be\label{eq:projdef}
\epsilon_{0}^\alpha = -\frac{\sqrt{Q^2_2}}{(p_1\cdot q_2)} \left( p_1^\alpha + \frac{q_2 \cdot p_1}{Q^2_2}q_2^\alpha \right)\qquad \qquad
\delta_T^{\mu\mu'} =g^{\mu\mu'} + \frac{q_2^\mu q_2^{\mu'}}{Q^2_2} - \epsilon_{0}^\mu\epsilon_{0}^{\mu'}\;,
\ee
such that these project out the longitudinal and transverse photon helicities in the $\gamma q$ c.m.s. frame. The squared amplitudes can then be straightforwardly calculated via
\be
\sum_{\lambda=\pm}  |A_{\lambda}^1|^2 = \delta_T^{\mu\mu'}A_{\mu}^1 A_{\mu'}^{1*}  \;, \qquad |A_{0}^1|^2  = |\epsilon_{0}^\mu A_{\mu}^1|^2 \;,
\ee
in the usual way. Very similarly to the case of \eqref{eq:rhodr} we have that
\be\label{eq:tdr}
\rho_{\mu\mu'}^2 \delta_T^{\mu\mu'} \approx 2\int \frac{{\rm d}M_2^2}{Q_2^2} \frac{x_{B,2}}{x^2_{2}} \left[\left(z_2 p_{\gamma q}(z_2)+\frac{2\xi_2^2 m_p^2}{Q_2^2}\right)F_2(\xi_2/z_2,Q^2_2)-z_2^2 F_L(\xi_2/z_2,Q^2_2)\right]\;,
\ee
while for the longitudinal case we have
\be\label{eq:0dr}
\rho_{\mu\mu'}^2 \epsilon_{0}^\mu\epsilon_{0}^{\mu'} \approx 4\int \frac{{\rm d}M_2^2}{Q_2^2} \frac{x_{B,2}}{x^2_{2}} \left[\left(1-z_2-\frac{\xi_2^2 m_p^2}{Q_2^2}\right)F_2(\xi_2/z_2,Q^2_2)+\frac{1}{4}z_2^2 F_L(\xi_2/z_2,Q^2_2)\right]\;,
\ee
where $z_2=\xi_2/x_{B,2}$, and we have dropped terms of $O(m_p^2/s, Q_2^2/s)$, which are negligible (as can be confirmed numerically). In reality, for the kinematic region we are limited to, which we recall from \eqref{eq:qwcut} dominantly has $Q_2^2 < 1\,{\rm GeV}^2$, we find that the contribution from the longitudinal photon is very small, but we include it for completeness. These are in principle the expressions we use to  calculate the corresponding cross sections we need. In practice, as discussed in Section~\ref{sec:mc} we use \texttt{MadGraph5\_aMC@NLO}~\cite{Alwall:2014hca,Frederix:2018nkq}  for the corresponding parton--level cross sections. This takes a different basis to \eqref{eq:projdef}, with the transverse polarizations defined to lie orthogonal to the 3--vector $\vec{q}_2$, with no energy component; this only coincides with the above definition in the on--shell limit. As we continue to use the \texttt{MadGraph5\_aMC@NLO} basis, with the scalar photon polarization suitably defined to be consistent with this, in principle we should instead calculate \eqref{eq:tdr} and \eqref{eq:0dr} in this basis. However, we have checked numerically that the difference between doing this and taking the simpler analytic results of \eqref{eq:tdr} and \eqref{eq:0dr} is negligible. We in particular recall that the above approach is (dominantly) only applied when $Q_2^2 < 1\,{\rm GeV}^2$, i.e. rather close to the on--shell limit where these two bases will coincide. Therefore, for speed of implementation we continue to use these analytic result in the \texttt{SuperChic} release.

Finally, it is instructive to recall the connection between the above results and the on--shell $Q_2^2 \to 0$ limit, as would be applied for a collinear calculation. In this case the longitudinal contribution can be dropped, as this is subleading in this limit.  We then start with \eqref{eq:sighhf}, which we repeat for clarity:
 \be\label{eq:sighhfap}
 \sigma_{pp} = \frac{1}{2s}  \int  {\rm d}x_1 {\rm d}x_2\,{\rm d}^2 q_{1_\perp}{\rm d}^2 q_{2_\perp
} {\rm d \Gamma} \,\alpha(Q_1^2)\alpha(Q_2^2)\frac{1}{\tilde{\beta}} \frac{\rho_1^{\mu\mu'}\rho_2^{\nu\nu'} M^*_{\mu'\nu'}M_{\mu\nu}}{Q_1^2Q_2^2}\delta^{(4)}(q_1+q_2 - p_X)\;.
 \ee
We can then change variables to give 
\be
{\rm d}q_{2_\perp}^2 {\rm d}x_1 {\rm d}x_2 = \frac{\tilde{\beta}}{1-\xi_1} {\rm d}Q_2^2 {\rm d}\xi_1 {\rm d}\xi_2\;,
\ee
where the integration variables are as in \eqref{eq:sighhf}. We in addition have that 
 \be
\frac{{\rm d}M_1^2}{Q_1^2}  {\rm d} \xi_1  {\rm d}^2 q_{1_\perp}= \frac{16\pi^3}{x_{B,1}}  (1-\xi_1){\rm d} \tilde{\xi}_1  {\rm d}\Gamma_{q'}\;,
\ee 
where $\Gamma_{q'}= {\rm d}^3 p_{q'} / [2 E_{q'} (2\pi)^3]$ is the phase space integral with respect to the outgoing quark, and $\tilde{\xi}$ is the momentum fraction of the incoming quark.

We can then change variables from $M_2$ to $z_2$ (at fixed $Q_2^2$) to give 
 \be\label{eq:rhophotap}
 \frac{1}{\alpha(\mu^2)}\int  \frac{{\rm d} Q^2_2}{Q^2_2}\alpha(Q^2_2)^2 \rho_{\mu\mu'}^2 \delta_T^{\mu\mu'} \approx  \frac{4\pi}{\xi_2}   f_{\gamma/p}^{\rm PF}(\xi_2,\mu^2)\;,
\ee
as in \eqref{eq:rhophot}. We identify in the usual way
\be
{\rm d}\hat{\sigma}(\gamma q \to l^+ l^- + q)=\frac{1}{2 M_{q\gamma}^2} {\rm d \Gamma}{\rm d \Gamma_{q'}}   \frac{1}{2}\sum_{\lambda=\pm}|A_{\lambda}^1|^2 (2\pi)^4 \delta^{(4)} (\tilde{\xi}_1 p_1 + q_2 - p_{ll} - p_q') \;,
\ee
and note that the argument of the delta function is simply equal to $q_1 + q_2 - p_X$ as in \eqref{eq:sighhfap}, while $M_{q \gamma}^2 =  x_{B,1} x_2 s$ (the on--shellness of the outgoing quark gives $x_{B,1} = \xi_1'$). Putting the above together we arrive at 
\be
\sigma_{pp} \approx\sum_{q,\overline{q}}  \int  {\rm d}\xi_1 {\rm d}\xi_2\,q(\xi_1,\mu^2)f_{\gamma/p}^{\rm PF}(\xi_2,\mu^2)\hat{\sigma}(\gamma q \to l^+ l^- + q)\;,
\ee
where we have relabelled the dummy variable $ \tilde{\xi}_1 \to \xi_1$ in the last step to match the usual notation.

\bibliography{references}{}
\bibliographystyle{h-physrev}

\end{document}